\documentstyle{article}  
\newlength{\extralineskip}
\newcommand{\beq}{\begin{equation}}
\newcommand{\eeq}{\end{equation}}
\newcommand{\bd}{\begin{displaymath}}
\newcommand{\ed}{\end{displaymath}}

\def\bea{\begin{eqnarray}}
\def\eea{\end{eqnarray}}

\def\ba{\beq\new\begin{array}{c}}
\def\ea{\end{array}\eeq}

\def\inbar{\,\vrule height1.5ex width.4pt depth0pt}
\def\IC{\relax\hbox{$\inbar\kern-.3em{\rm C}$}}
\def\IR{\relax{\rm I\kern-.18em R}}

\def\IN{\relax{\rm I\kern-.15em N}}

\makeatletter
\newdimen\normalarrayskip              
\newdimen\minarrayskip                 
\normalarrayskip\baselineskip
\minarrayskip\jot
\newif\ifold             \oldtrue            \def\new{\oldfalse}
\def\arraymode{\ifold\relax\else\displaystyle\fi} 
\def\@arrayskip{\ifold\baselineskip\z@\lineskip\z@
     \else
     \baselineskip\minarrayskip\lineskip2\minarrayskip\fi}
\def\@arrayclassz{\ifcase \@lastchclass \@acolampacol \or
\@ampacol \or \or \or \@addamp \or
   \@acolampacol \or \@firstampfalse \@acol \fi
\edef\@preamble{\@preamble
  \ifcase \@chnum
     \hfil$\relax\arraymode\@sharp$\hfil
     \or $\relax\arraymode\@sharp$\hfil
     \or \hfil$\relax\arraymode\@sharp$\fi}}
\def\@array[#1]#2{\setbox\@arstrutbox=\hbox{\vrule
     height\arraystretch \ht\strutbox
     depth\arraystretch \dp\strutbox
     width\z@}\@mkpream{#2}\edef\@preamble{\halign \noexpand\@halignto
\bgroup \tabskip\z@ \@arstrut \@preamble \tabskip\z@ \cr}%
\let\@startpbox\@@startpbox \let\@endpbox\@@endpbox
  \if #1t\vtop \else \if#1b\vbox \else \vcenter \fi\fi
  \bgroup \let\par\relax
  \let\@sharp##\let\protect\relax
  \@arrayskip\@preamble}

\begin{document}
\thispagestyle{empty}

\begin{center}
{\huge \bf Quantum Spin-1/2 Antiferromagnetic Chains and Strongly Coupled 
Multiflavor Schwinger Models}\\
\vskip 0.6 truein
{\bf F. Berruto$^{(a)}$, G. Grignani$^{(a)}$, G. W. Semenoff$^{(b)}$ and \underline{P. Sodano$^{(a)}$}
\protect\footnote{Lectures delivered by P. Sodano at the 
International Conference on ``Mathematical Methods in Modern Theoretical Physics" Tbilisi, Georgia, September 1998.}}
\vskip 0.3truein
$(a)${\it Dipartimento di Fisica and Sezione
I.N.F.N., Universit\`a di Perugia, Via A. Pascoli I-06123 Perugia,
Italy}\\
\vskip 1.truecm
$(b)$ {\it Department of Physics and Astronomy\\University of British 
Columbia\\
6224 Agricultural Road\\Vancouver, British Columbia, Canada V6T 1Z1}\\
\end{center}
\vskip 1.truecm
\begin{center}
{\large \bf Abstract}
\end{center}

We review the correspondence between strongly coupled lattice multiflavor Schwinger models and $SU({\cal N})$ 
antiferromagnetic chains. We show that finding the low lying states of the gauge models is equivalent to solving an $SU({\cal N})$ 
Heisenberg antiferromagnetic chain. For the two-flavor lattice Schwinger model the massless excitations correspond to gapless states of the 
Heisenberg chain, while the massive states are created by fermion transport in the ground state of the spin chain. Our 
analysis shows explicitly how spinons may arise in lattice gauge theories. 
 
\vskip 0.3truein
\section{Introduction}

Though it is far from the scaling regime, the strong coupling limit is 
often used to study the qualitative properties of gauge field theories~\cite{b1}. 
Two important features of the spectrum of non-Abelian gauge theories appear 
there. The strong coupling limit exhibits confinement in a rather natural way. 
Furthermore with certain numbers of flavors and colors of dynamical quarks, it 
is straightforward to show that  they also exhibit chiral symmetry breaking. 
This has motivated several more quantitative investigations of gauge theories 
using strong coupling techniques and there have been attempts to compute the mass spectrum of 
realistic models such as quantum chromodynamics (QCD)~\cite{b2,bqcd}. 

It has been recognized for some time that the strong coupling limit of lattice 
gauge theory with dynamical fermions is related to certain quantum spin systems. 
This is particularly true in the hamiltonian formalism and had already appeared in some of the earliest 
analyses of chiral symmetry breaking in the strong coupling limit~\cite{b3}. On the other side, it has been noted 
that there are several similarities between some condensed matter systems with lattice 
fermions and lattice gauge theory systems, particularly in their strong coupling limit. 
For example, it is well known that the quantum spin-1/2 Heisenberg antiferromagnet is 
equivalent to the strong coupling limit of either $U(1)$ or $SU(2)$ lattice gauge theory~\cite{b4a,b4b,b4c,b4d,b4e}. 

Moreover, the staggered fermions which are used to put the Dirac operator on a lattice 
resemble ordinary lattice fermions used in tight binding models in condensed matter physics 
when the latter have an half-filled band and are placed in a background $U(1)$ magnetic 
field $\pi$ ($mod$ $2\pi$) (1/2 of a flux quantum) through every plaquette of the lattice. 
This is an old result for $d=2+1$~\cite{b5} where it was already recognized in the first work on the 
Azbel-Wannier-Hofstaeder problem and has since been discussed in the context of the 
so-called flux phases of the Hubbard model. It is actually true for all $d\ge 2+1$~\cite{b6}, $i.e.$ 
staggered fermions with the minimal number of flavors are identical to a simple nearest neighbor 
hopping problem with a background magnetic field which has 1/2 flux quanta per plaquette. 
In the condensed matter context, the magnetic flux can be produced by a condensate, as in the 
flux phases. As an external magnetic field, for ordinary lattice spacing it is as yet an 
experimentally inaccessible flux density. However, it could be achieved in analog experiments 
where macroscopic arrays of Josephson junctions, for example, take the place of atoms at lattice sites. 

In~\cite{semen} it was shown that the strong coupling limit of d-dimensional $QED$ with $2^d/2^{[d/2]}$ flavors of 
fermions can be mapped onto the $S=1/2$ quantum Heisenberg antiferromagnet in $d-1$ dimensions. 
The chiral symmetry breaking phase transition corresponds to a transition between the flux phase and the 
conventional N\'eel ordered phase of an antiferromagnet. 
        
In this lectures we shall discuss the correspondence between lattice gauge theories and spin systems by 
analyzing, as a concrete example, the mapping  of the strongly coupled multiflavor lattice 
Schwinger models $-$ $i.e.$ $QED_2$ with massless fermions~\cite{schw} $-$ into the antiferromagnetic 
quantum spin-1/2 Heisenberg chains. In the one-flavor Schwinger model, the anomaly is realized in the lattice 
theory via spontaneous breaking of a residual chiral symmetry~\cite{b10}. In the multiflavor models, at variance 
with the one-flavor case, the chiral symmetry is 
explicitly broken by the staggered fermions, and the non-zero vacuum expectation value of a pertinent condensate is the only 
relic on the lattice of the chiral anomaly in the continuum~\cite{b8,b9}.

In section 2 we introduce the antiferromagnetic quantum spin-1/2 chains; we review the Bethe ansatz~\cite{b6b} 
solution of the Heisenberg chain~\cite{b6c} and compare it with the exact diagonalization of 4, 6 and 8 sites chains. 
Furthermore we generalize the analysis to the spin-1/2 $SU({\cal N})$ antiferromagnetic 
Heisenberg chains~\cite{b7,b8,b9}.

Section 3 is devoted to a detailed investigation of the two-flavor lattice Schwinger 
model in the strong coupling limit using the hamiltonian formalism and staggered fermions. 
We show that the problem of finding the low lying states is equivalent to solving the Heisenberg 
antiferromagnetic spin chain. The massless excitations of the gauge model correspond to 
gapless states of the Heisenberg chain, while the massive excitations are created by fermion transport 
in the ground state of the spin chain and we compute their masses in terms of vacuum expectation values 
(VEV's) of spin-spin correlation functions~\cite{b7,b8,b9,b10}. 

In section 4 we study the generic ${\cal N}$-flavor lattice Schwinger models and their mapping 
into spin-1/2 quantum $SU({\cal N})$ antiferromagnetic spin chains. The analysis strictly 
parallels what we found in section 3, but some rather surprising difference arises depending on if 
${\cal N}$ is even or odd.

\section{Antiferromagnetic spin chains}

The title of this section deserves an entire book to be properly developed, and in fact some books, 
reviews and long articles about spin chains do already exist~\cite{b31}. 
Spin chains interested mathematical physicists for their exact solvability, field theorists for the possibility to test their methods 
by comparation with exact results and condensed matter physicists for the possibility to understand the spectra of real physical systems. 
It is our purpose to report a self-contained analysis of antiferromagnetic spin chains, useful to understand section 3 and 4 where we shall show 
that the multiflavor lattice Schwinger models in the strong coupling limit are effectively described by the 
spin-$1/2$ $SU({\cal N})$ antiferromagnetic Heisenberg chains. 

A common tool to study spin systems is spin-wave theory. Spin-wave theory was developed by Anderson, Kubo and others~\cite{b32} in the 1950s 
and it is a very powerful approach to quantum magnetic systems in dimensions greater than one, predicting long-range order and gapless Goldstone bosons. 
The situation remained clouded for magnetic  chains because it was known that long-range order could not occur in one dimension~\cite{b39}, but the Bethe ansatz 
predicts massless excitations. There is ``quasi-long-range-order" corresponding to a power law decay of spin-spin correlators and there are gapless excitations 
which are not true Goldstone bosons. 

Quantum spin chains have been extensively studied in the literature, starting from the seminal paper by Hans Bethe~\cite{b6b} in 1931 for the spin-$1/2$ 
case, where he introduced an ansatz to write down the eigenfunctions of the Heisenberg Hamiltonian, describing a chain with periodic boundary conditions. 
Seven years later Hulthen~\cite{b34} was able to compute the ground state energy of the antiferromagnetic Heisenberg chain. We had to wait until 1984 
to know exactly the spectrum of this model, when Faddeev and Takhtadzhyan~\cite{b6c} analyzed the model using the algebraic Bethe ansatz 
and showed that the only one-particle excitation is a doublet of spin-$1/2$ quantum 
excitations with gapless dispersion relation. This excitation is a kink rather than an ordinary particle and is called spinon. All the eigenstates of the 
antiferromagnetic Heisenberg Hamiltonian contain an even number of kinks, nevertheless the kinks are localizable objects and one can consider 
their scattering. There are no bound states of kinks in this model. 

The one dimensional spin systems are not only unusual because they are exactly solvable but also for their incompatibility with long range order. 
Actually solvability and absence of long range order are deeply related concepts; systems with spontaneously broken symmetries are more difficult to 
describe and resist analytical solutions. All isotropic antiferromagnetic quantum spin chains with short range interactions 
exhibit quantum spin-liquid ground states $-$ $i.e.$ states with short range antiferromagnetic correlations and no order. Moreover these 
chains have strange quantum elementary excitations above these ground states, that are not ordinary spin waves but are usually called spinons $-$ neutral 
spin-$1/2$ kinks. 

As R. B. Laughlin~\cite{b35} points out:``spinons are not ``like" anything familiar to most of us, but are instead an important and beautiful instance of 
fractional  quantization, the physical phenomenon in which particles carrying pieces of a fundamental quantum number, such as charge or spin, are created 
as a collective motion of many conventional particles obeying quantum mechanical laws. The fractional quantum number of the spinon is its spin. It is 
fractional because the particles out of which the magnetic states are constructed are spin flips, which carry spin 1."

One significant result of our analysis~\cite{b7,b8,b9} has been to show explicitly how spinons appear in lattice gauge theories. More precisely we showed that the massless 
excitations of the two-flavor Schwinger model coincide with the spinons of the spin-$1/2$ antiferromagnetic Heisenberg model.\\

Section (\ref{baa}) is devoted to review the Bethe ansatz solution of the antiferromagnetic Heisenberg chain. We shall compare the exact solution 
given in \cite{b6c} with a study of finite size chains of 4,6, and 8 sites in section (2.2). An original result presented is the thermodynamic 
limit coefficient of the states with $N-2$ domain walls composing the ground state. 

The spin-spin correlation function 
$\sum_{x}<g.s.|\vec{S}_{x}\cdot \vec{S}_{x+2}|g.s.>$ originally computed by M. Takahashi~\cite{b38}, is discussed in section (\ref{ssc}). The vacuum expectation 
value of the square of the vector $\vec{V}=\sum_{x}\vec{S}_{x}\wedge \vec{S}_{x+1}$ is also computed. 

In section (\ref{sunc}) we introduce the spin-$1/2$ $SU({\cal N})$ antiferromagnetic Heisenberg chains. To provide an intuitive picture we study 
the $SU(3)$ two sites chain and we find the ground state. A short review of the literature on $SU({\cal N})$ antiferromagnetic Heisenberg chains 
is provided.

\subsection{The Bethe Ansatz solution of the antiferromagnetic Heisenberg chain}
\label{baa}

In this section the Bethe ansatz solution of the antiferromagnetic 
Heisenberg chain~\cite{b6c} is discussed in detail. Moreover we shall compare the exact solution given by L.D. Faddeev and L.A. Takhtadzhyan in~\cite{b6c} 
with a study of finite size chains of 4, 6 and 8 sites. We show that already these very small finite systems exhibit spectra 
that match very well with the thermodynamic limit solution. We suggest to the reader interested in the subject the references 
\cite{b31}.

The Bethe ansatz is a method of solution of a number of quantum field theory and statistical mechanics 
models in two space-time dimensions. This method was first suggested by Bethe~\cite{b6b} in 1931 from which takes its name. 
Historically one can call this formulation the Coordinate Bethe ansatz to distinguish it from the modern formulation 
known as Algebraic Bethe ansatz. The eigenfunctions of some (1+1)-dimensional Hamiltonians can be constructed imposing 
periodic boundary conditions which lead to a system of equations for the permitted values of momenta. These are 
known as the Bethe equations which are also useful in the thermodynamic limit. The energy of the ground state may be 
calculated in this limit and its excitations can be investigated. 

We review here the method applyied to the study of 
the antiferromagnetic Heisenberg chain~\cite{b6c}. In particular it is shown that there is only one excitation with spin-$1/2$ 
which is a kink: physical states have only an even number of kinks, therefore they always have an integer spin.
The one dimensional isotropic Heisenberg model describes a system of 
$N$ interacting spin-$\frac{1}{2}$ particles. 
The Hamiltonian of the model is
\begin{equation}
H_{J}=J\sum_{x=1}^{N}(\vec{S}_{x}\cdot \vec{S}_{x+1}-\frac{1}{4})\quad .
\label{heis}
\end{equation}
where $J>0$ ($J<0$ would describe a ferromagnet) and the spin operators 
have the following form
\begin{equation}
\vec{S}_{x}=1_{1}\otimes 1_{2}\otimes \ldots \otimes 
\frac{\vec{\sigma}_{x}}{2} 
\otimes \ldots \otimes 1_{N}\quad .
\end{equation}
They act nontrivially only on the Hilbert space of the $x^{th}$ site. 
Periodic boundary conditions are assumed. 

The Hamiltonian (\ref{heis}) is invariant under global rotations in the 
spin space, generated by
\begin{equation}
\vec{S}=\sum_{x=1}^{N}\vec{S}_{x}\quad .
\end{equation}
Due to the periodic boundary conditions, under translations generated by 
the operator $\hat{T}$, 
\begin{equation}
\hat{T}\quad : \vec{S}_{x} \longrightarrow \vec{S}_{x-1} 
\end{equation}
the Hamiltonian is invariant and $[\vec{S},\hat{T}]=0$.

In order to diagonalize $H_{J}$ it is convenient to use an eigenfunction basis of operators commuting with 
$H_{J}$, so obviously $\vec{S}^2$, $S^z$ and also $\hat{T}$.
Let us sketch the Coordinate Bethe ansatz technique. One has to introduce the ``false vacuum"
\begin{equation}
|\Omega>=\prod_{x=1}^{N}\otimes |\uparrow>_{x}
\end{equation}
with
\begin{eqnarray}
S_{x}^{+} |\uparrow>_{x}&=&0\\
S_{x}^{3} |\uparrow>_{x}&=&\frac{1}{2}|\uparrow>_{x}
\end{eqnarray}
where
\begin{equation}
S_{x}^{\pm}=S_{x}^{1}\pm S_{x}^{2}
\end{equation}
\begin{eqnarray}
S^{3}|\Omega>&=&\frac{N}{2}|\Omega>\\
\vec{S}^{2}|\Omega>&=&\frac{N}{2}(\frac{N}{2}+1)|\Omega>\\
\hat{T}|\Omega>&=&|\Omega>\quad .
\end{eqnarray}
All the other basis vectors have $S^{3}<\frac{N}{2}$ and one can get them by properly acting on $|\Omega>$ with 
the lowering operators $S_{x}^{-}$. Let us start from the generic state with $S^{3}=\frac{N}{2}-1$, with $N-1$ spins up and 
$M=1$ spin down
\begin{equation}
|M=1>=\sum_{x=1}^{N}\phi_{x}|x>\quad with \quad |x>=S_{x}^{-}|\Omega>
\end{equation}
where the coefficients $\phi_{x}$ must be such that $|M=1>$ is a translationally and rotationally invariant state
\begin{equation}
\hat{T}|M=1>=\sum_{x=1}^{N}\phi_{x}|x-1>=\sum_{x=1}^{N}\phi_{x+1}|x>=\mu \sum_{x=1}^{N}\phi_{x}|x>
\label{tm}
\end{equation}
and from Eq.(\ref{tm}) one gets
\begin{equation}
\phi_{x+1}=\mu \phi_{x}
\end{equation}
\begin{eqnarray}
\phi_{x+1}&=&\mu ^{x} \phi_{1}\quad \quad x\neq N \\
\phi_{1}&=&\mu \phi_{N}=\mu^{N} \phi_{1}\longrightarrow \mu^N =1\quad .
\end{eqnarray}
Setting $\phi_{1}=1$ one has $\phi_{x}=\mu ^{x-1}$. There are N possible values for $\phi_{x}$, 
corresponding to the N roots of the unity. One of these roots corresponds to a state with $S=\frac{N}{2}$ 
\begin{equation}
\sum_{x=1}^{N}|x>=S^{-}|\Omega>\quad\longrightarrow \quad \mu=1
\end{equation}
while the other $N-1$ roots have $S=\frac{N}{2}-1$.

The generic case with M spins down is more complicated. Let us consider the case $M=2$ to understand 
what happens 
\begin{equation}
|M=2>=\sum_{x_{1}<x_{2}=1}^{N}\phi(x_{1},x_{2})|x_{1},x_{2}>\quad with \quad |x_{1},x_{2}>=S_{x_{1}}^{-}S_{x_{2}}^{-}|\Omega>\quad .
\end{equation}      
By requiring the translational invariance of the state $|M=2>$ one has
\begin{equation}
\hat{T}|M=2>=T|M=2>
\end{equation}
and for $x_{2}<N$  one has
\begin{equation}
\phi(x_{1}+1,x_{2}+1)=T \phi(x_{1},x_{2})
\label{ee1}
\end{equation}
that would easily give
\begin{equation}
\phi(x_{1},x_{2})=\mu_{1}^{x_{1}-1} \mu_{2}^{x_{2}-1}\quad ,\quad T=\mu_{1} \mu_{2}
\label{trial}
\end{equation}
but due to the periodic boundary conditions one has to find coefficients $\phi(x_{1},x_{2})$ that satisfy not only Eq.(\ref{ee1}) 
but also 
\begin{equation}
\phi(1,x+1)=T\phi(x,N)\quad .
\label{ee2}
\end{equation}
Eq.(\ref{ee2}) is no more satisfied by (\ref{trial}). Bethe proposed the following ansatz for the coefficients $\phi(x_{1},x_{2})$
\begin{equation}
\phi(x_{1},x_{2})=A_{1,2}\mu_{1}^{x_{1}-1}\mu_{2}^{x_{2}-1}+A_{2,1}\mu_{2}^{x_{1}-1}\mu_{1}^{x_{2}-1}\quad .
\label{ba}
\end{equation}
Eq.(\ref{ee1},\ref{ee2}) are satisfied if the following equations hold
\begin{eqnarray}
A_{12}&=&A_{21}\mu_{1}^{N}
\label{cc1}\\
A_{21}&=&A_{12}\mu_{2}^{N}\quad .
\label{cc2}
\end{eqnarray}
By imposing to $|M=2>$ to be an highest weight state
\begin{equation}
S^{+}|M=2>=0
\end{equation}
taking into account Eq.(\ref{cc1},\ref{cc2}) and introducing the following change of variables
\begin{equation}
\mu_{\alpha}=\frac{\lambda_{\alpha}-\frac{i}{2}}{\lambda_{\alpha}-\frac{i}{2}} 
\end{equation}
one gets the so called Bethe ansatz equations
\begin{equation}
(\frac{\lambda_{\alpha}-\frac{i}{2}}{\lambda_{\alpha}+\frac{i}{2}})^{N}=-
\prod_{\beta=1}^{2}\frac{\lambda_{\alpha}-\lambda_{\beta}-i}{ \lambda_{\alpha}-\lambda_{\beta}+i}\quad .
\end{equation}
In the general case of M spins flipped one gets
\begin{equation}
(\frac{\lambda_{\alpha}-\frac{i}{2}}{\lambda_{\alpha}+\frac{i}{2}})^{N}=-
\prod_{\beta=1}^{M}\frac{\lambda_{\alpha}-\lambda_{\beta}-i}{ \lambda_{\alpha}-\lambda_{\beta}+i}\quad .
\label{baeq}
\end{equation}

The energy and the momentum of a given state with $M$ spins down can be expressed in terms of the parameters $\lambda_{\alpha}$
\begin{eqnarray}
E_{M}&=&\sum_{\alpha=1}^{M}\epsilon_{\alpha}=-\frac{J}{2}\sum_{\alpha=1}^{M}\frac{1}{\lambda_{\alpha}^{2}+\frac{1}{4}}\\
P_{M}&=&i\ln T=\sum_{\alpha=1}^{M}p_{\alpha}=i\sum_{\alpha=1}^{M}\ln \frac{\lambda_{\alpha}-\frac{i}{2}}{\lambda_{\alpha}+\frac{i}{2}}\quad .
\end{eqnarray}
Energy and momentum are thus additive as if there were M independent particles and the $\lambda_{\alpha}$ must satisfy the Bethe ansatz equations 
(\ref{baeq}) in order for $E_{M}$ and $P_{M}$ to be eigenvalues of the Hamiltonian and momentum operators. 

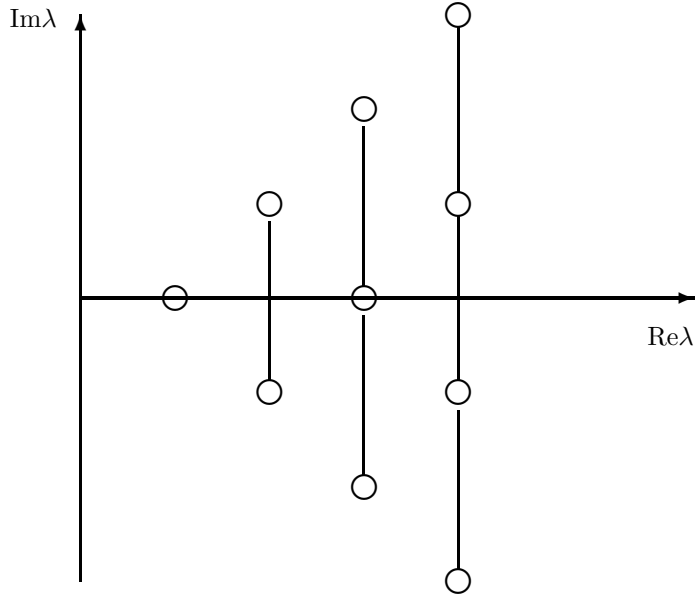
\begin{figure}[htb]
\begin{center}
\setlength{\unitlength}{0.00041700in}%
\begingroup\makeatletter\ifx\SetFigFont\undefined
\def\x#1#2#3#4#5#6#7\relax{\def\x{#1#2#3#4#5#6}}%
\expandafter\x\fmtname xxxxxx\relax \def\y{splain}%
\ifx\x\y   
\gdef\SetFigFont#1#2#3{%
  \ifnum #1<17\tiny\else \ifnum #1<20\small\else
  \ifnum #1<24\normalsize\else \ifnum #1<29\large\else
  \ifnum #1<34\Large\else \ifnum #1<41\LARGE\else
     \huge\fi\fi\fi\fi\fi\fi
  \csname #3\endcsname}%
\else
\gdef\SetFigFont#1#2#3{\begingroup
  \count@#1\relax \ifnum 25<\count@\count@25\fi
  \def\x{\endgroup\@setsize\SetFigFont{#2pt}}%
  \expandafter\x
    \csname \romannumeral\the\count@ pt\expandafter\endcsname
    \csname @\romannumeral\the\count@ pt\endcsname
  \csname #3\endcsname}%
\fi
\fi\endgroup
\begin{picture}(8712,7514)(2701,-7718)
\thicklines
\put(6001,-5161){\circle{300}}
\put(7201,-1561){\circle{300}}
\put(7201,-6361){\circle{300}}
\put(7201,-3961){\circle{300}}
\put(8401,-2761){\circle{300}}
\put(8401,-361){\circle{300}}
\put(8401,-5161){\circle{300}}
\put(8401,-7561){\circle{300}}
\put(6001,-2761){\circle{300}}
\put(4801,-3961){\circle{300}}
\put(3601,-7561){\vector( 0, 1){7200}}
\put(10801,-4561){\makebox(0,0)[lb]{\smash{\SetFigFont{10}{12.0}{rm}Re$\lambda$}}}
\put(3601,-3961){\vector( 1, 0){7800}}
\multiput(6001,-2986)(0.00000,-7.98817){254}{\line( 0,-1){  3.994}}
\multiput(7201,-1786)(0.00000,-7.98817){254}{\line( 0,-1){  3.994}}
\multiput(7201,-4186)(0.00000,-7.98817){254}{\line( 0,-1){  3.994}}
\multiput(8401,-511)(0.00000,-8.00000){263}{\line( 0,-1){  4.000}}
\multiput(8401,-2911)(0.00000,-8.00000){263}{\line( 0,-1){  4.000}}
\multiput(8401,-5386)(0.00000,-7.98817){254}{\line( 0,-1){  3.994}}
\put(2701,-511){\makebox(0,0)[lb]{\smash{\SetFigFont{10}{12.0}{rm}Im$\lambda$}}}
\end{picture}
\end{center}
\caption{Strings for \protect $L=0,\frac{1}{2},1,\frac{3}{2}$}
\label{strings}
\end{figure}

The solution of the antiferromagnetic Heisenberg chain is 
reduced to the solution of the system of the $M$ algebraic equations 
(\ref{baeq}). This, in general, is not an easy task.
It can be shown~\cite{b6c}, however, that, in the thermodynamic limit 
$N\to\infty$, the complex parameters $\lambda$ have the form
\begin{equation}
\lambda_{l}=\lambda_{j,L}+il\quad,\quad l=-L,-L+1,\dots,L-1,L;
\label{complex}
\end{equation}
where $L$ is a non-negative integer or half-integer, $\lambda_{j,L}$
is the real part of the solution of (\ref{baeq}) and we shall 
define shortly the set of allowed values for the integer index $j$.
The $\lambda$'s that, for a given $\lambda_{j,L}$, are obtained varying $l$ 
between  $[-L,L]$ by integer steps, form a string of length $2L+1$,
see fig.(\ref{strings}). 
This arrangement of $\lambda$'s in the complex plane is called the 
``string hypothesis" \cite{b6c}. 
In the following we shall verify 
that, even on finite size systems, the ``string hypothesis" 
is very well fulfilled. 

In a generic Bethe state with $M$ spins down, 
there are $M$ solutions to (\ref{baeq}), which can be grouped
according to the length of their strings.
Let us denote  by $\nu_{L}$ the number of strings of
length $2L+1$, $L=0,\frac{1}{2},\ldots$; strings of the same length
are obtained by changing the real parts, $\lambda_{j,L}$, of the $\lambda$'s 
in (\ref{complex}); as a consequence $\quad j=1,\ldots,\nu_{L}$. 
If one denotes the total number of strings by $q$ one has
\begin{equation}
q=\sum_{L}\nu _{L}\quad,\quad M=\sum_{L} (2L+1)\nu _{L}\quad .
\label{constraint}
\end{equation}

The set of integers $(M,q,\{\nu_{L}\})$ constrained 
by (\ref{constraint}), characterizes Bethe 
states up to the fixing of the $q$ numbers $\lambda_{j,L}$; this set 
is called the ``configuration''. 
Varying $M$, $q$ and $\nu_L$, one is able to construct all the
$2^N$ eigenstates of an Heisenberg antiferromagnetic chain of $N$ 
sites~\cite{b6c}.
The energy and momentum of the Bethe's state, corresponding to a 
given configuration $-$ within exponential accuracy as
$N\rightarrow \infty$ $-$ consist of $q$ summands representing 
the energy and momentum of separate strings. 
For the parameters $\lambda_{j,L}$ of the given configuration, 
taking the logarithm of (\ref{baeq}) 
the following system of equations is obtained in the thermodynamic limit
\begin{equation}
2N\arctan \frac{\lambda_{j,L_{1}}}{L_{1}+\frac{1}{2}}=2\pi 
Q_{j,L_{1}}+\sum_{L_{2}}
\sum_{k=1}^{\nu_{L_{2}}}\Phi_{L_1 L_2}(\lambda_{j,L_{1}}-
\lambda_{k,L_{2}})\quad, 
\label{tbaeq}
\end{equation}
where
\begin{equation}  
\Phi_{L_1 L_2}(\lambda)=2\sum_{L=|L_1 -L_2|\neq 0}^{L_1 +L_2}
(\arctan \frac{\lambda}{L}+\arctan \frac{\lambda}{L+1})\quad .
\end{equation}

Integer and half integer numbers $Q_{j,L}$ parametrize the 
branches of the arcotangents and, consequently, the 
possible solutions of the system of Eqs.(\ref{tbaeq}). In ref.\cite{b6c}
it was shown that the $Q_{j,L}$  are limited as
\begin{equation}
-Q_{L}^{max}\leq Q_{1,L}<Q_{2,L}<\ldots<Q_{\nu_{L},L}\leq Q_{L}^{max}\quad 
\label{vacan}
\end{equation}
with $Q_{L}^{max}$ given by
\begin{equation}
Q_{L}^{max}=\frac{N}{2}-\sum_{L'}J(L,L')\nu_{L'}-\frac{1}{2}
\label{qmax}
\end{equation}
and
\begin{equation}
 J(L_1 ,L_2)=\left\{\begin{array}{ll}
 2{\rm min}(L_1 ,L_2)+1 &\mbox{if $L_1 \neq L_2$}\\ 
 2L_1+\frac{1}{2} &\mbox{if $L_1 =L_2$}\quad .
\end{array}
\right.
\label{jll}
\end{equation}
The admissible values for the numbers $Q_{j,L}$ are called 
the ``vacancies'' and their number for every $L$ is denoted by $P_{L}$
\begin{equation}
P_{L}=2Q_{L}^{max}+1\quad .
\label{pelle}
\end{equation}

The main hypothesis formulated in~\cite{b6c} is that to every 
admissible collection of $Q_{j,L}$ there corresponds a unique 
solution of the system of equations (\ref{tbaeq}). 
The solution always provides, in a multiplet, the state with the highest
value of the third spin component $S^3$. 

Let us now consider some simple example.
The simplest configuration has only strings of length 1,
$i.e.$ all the $\lambda$'s are real. 
The singlet associated to this configuration
\begin{equation}
M=q=\nu_0 =\frac{N}{2}\quad,\quad \nu_{L}=0\quad,\quad L>0\quad ,
\end{equation}
is the ground state.
The vacancies of the strings of length 1, $i.e.$ the 
admissible values of $Q_{j,0}$, due to 
eqs.(\ref{vacan},\ref{qmax},\ref{jll}), belong to the segment 
\begin{equation}
-\frac{N}{4}+\frac{1}{2}\leq Q_{j,0}\leq \frac{N}{4}-\frac{1}{2}\quad .
\label{igs}
\end{equation}
Therefore they are $N/2$.
All these vacancies must then be used to find the $N/2$ strings of length 1.
As a consequence this state is uniquely specified and no degeneracy 
is possible.

Next we consider the configuration that provides a singlet
with 1 string of length 2 and all the others of length 1:
\begin{equation}
M=\frac{N}{2}\quad,\quad q=\frac{N}{2}-1\quad,\quad \nu_{0}=
\frac{N}{2}-2 \quad,\quad \nu_{\frac{1}{2}}=1\quad,\quad \nu_{L}
=0\quad,\quad L>\frac{1}{2}\ .
\label{c1}
\end{equation}
For the strings of length 1 the 
number of vacancies is again ${N}/{2}$; 
for the string of length 2 there is one vacancy and the only 
admissible $Q_{j,1}$ equals 0. 
Thus, since the number of strings of length 1 is $\nu_{0}=
\frac{N}{2}-2$, there are two vacancies for which Eqs.(\ref{tbaeq}) 
have no solution; they are called ``holes'' and are denoted $Q_{1}^{(h)}$ 
and $Q_{2}^{(h)}$. 
This configuration is determined by two parameters: 
the positions of two ``holes" which vary independently in the interval 
(\ref{igs}). 

There is another state with only 2 holes:
the triplet corresponding to the configuration
\begin{equation}
M=q=\nu_{0}=\frac{N}{2}-1\quad,\quad \nu_{L}=0\quad,\quad L>0
\label{c2}
\end{equation}
The number of vacancies for the strings of length 1 equals $\frac{N}{2}+1$, 
while $\nu_{0}=\frac{N}{2}-1$.

The excitations determined by the configurations 
(\ref{c1},\ref{c2}) belong to the configuration class called in~\cite{b6c}
$\cal{M}_{AF}$. The class $\cal{M}_{AF}$ is characterized as follows: 
the number of strings of length 1 in each configuration 
belonging to this class
differs by a finite quantity from ${N}/{2}$, $\nu_0=\frac{N}{2}-k_0$ 
where $k_0$
is a positive finite constant, so that 
the number of strings of length greater than 1 is finite. From (\ref{pelle}) 
one then has
\begin{eqnarray}
P_0&=&\frac{N}{2}+k_0-2\sum_{L>0}\nu_L\label{p0}\\
P_L&=&2k_0-2\sum_{L'>0}J(L,L')\nu_{L'}\ ,\quad L>0
\end{eqnarray}
so that
\begin{equation}
P_0\ge \frac{N}{2}\ ,\quad P_L<2 k_0\ ,\quad L>0\ .
\end{equation}
From (\ref{p0}) follows that the number of holes for the strings of length 1 
is always even and equals 2 only for the singlet and the triplet 
excitations discussed above.
One can imagine the class $\cal{M}_{AF}$ as a
``sea" of strings of length 1 with a finite number of 
strings of length greater than 1 immersed into it. 
It was proven in~\cite{b6c} that, in the 
thermodynamic limit, the states belonging to 
$\cal{M}_{AF}$ have finite energy and momentum with 
respect to the antiferromagnetic vacuum, whereas 
each of the states which 
corresponds to a configuration not included in the class $\cal{M}_{AF}$ has 
an infinite energy relative to the antiferromagnetic vacuum.

Let us now sketch the computation of the thermodynamic limit 
ground state energy. Eqs.(\ref{tbaeq}) for the ground state have the form
\begin{equation}
\arctan 2\lambda_{j}=\frac{\pi Q_{j}}{N}+\frac{1}{N}
\sum_{k=1}^{{N}/{2}}\arctan (\lambda_{j} -\lambda_{k})\quad .
\label{gstbae}
\end{equation}
Taking the thermodynamic limit $N\rightarrow \infty$, one has 
\begin{equation}
\frac{Q_{j}}{N}\rightarrow x\quad,\quad -\frac{1}{4}\leq x\leq 
\frac{1}{4}\quad,\quad \lambda_{j}\rightarrow \lambda(x)\quad ,
\end{equation}
and Eqs.(\ref{gstbae}) can be rewritten in the form
\begin{equation}
\arctan 2\lambda(x)=\pi x+\int_{-\frac{1}{4}}^{\frac{1}{4}}
\arctan(\lambda(x)-\lambda(y)) dy\quad .
\label{2gsbae}
\end{equation}
Upon introducing the density of the numbers $\lambda(x)$ 
in the interval $d\lambda$ 
\begin{equation}
\rho(\lambda)=\frac{1}{\frac{d\lambda(x)}{dx}|_{x=x(\lambda)}}
\label{den}
\end{equation}
and differentiating Eqs.(\ref{2gsbae}), one gets
\begin{equation}
\rho(\lambda)=\frac{1}{2\pi}\int_{-\infty}^{\infty}
\frac{e^{-\frac{1}{2}|\xi|}}{1+e^{-|\xi|}}e^{-i\lambda |\xi|}
d\xi =\frac{1}{2\cosh\pi\lambda}\quad .
\label{den1}
\end{equation}
The density $\rho(\lambda)$ introduced in this way is normalized to $1/2$.
It is now easy to compute the energy and the momentum of the ground state 
\begin{equation}
E_{g.s.}=\sum_{\alpha =1}^{\frac{N}{2}}\epsilon_{\alpha}
=N\int_{-\infty}^{\infty}\epsilon(\lambda)\rho(\lambda)d\lambda=
-\frac{J N}{4}\int_{-\infty}^{\infty}d\lambda
\frac{1}{\left(\lambda^2+\frac{1}{4}
\right)\cosh\pi\lambda}=-JN\ln 2
\label{gse}
\end{equation}
\begin{equation}
P_{g.s.}=\sum_{\alpha=1}^{\frac{N}{2}}p_{\alpha}=N
\int_{-\infty}^{\infty}p(\lambda)
\rho(\lambda)d\lambda=-\frac{N}{2}
\int_{-\infty}^{\infty}d\lambda\frac{\pi}{\cosh\pi\lambda}=
\frac{N}{2}\pi\quad ({\rm mod}\quad 2\pi)\quad .
\label{gsm}
\end{equation}
According to Eq.(\ref{gsm}), $P_{g.s.}=0\ ({\rm mod} 2\pi)$ 
for $\frac{N}{2}$ even, and 
$P_{g.s.}=\pi\ ({\rm mod}\  2\pi)$ for $\frac{N}{2}$ odd. 
The ground state, as expected, is a singlet, in fact the spin $S$ 
is given by
\begin{equation}
S=\frac{N}{2}-\sum_{\alpha=1}^{N/2}1=\frac{N}{2}-
N\int_{-\infty}^{\infty}\rho(\lambda) d\lambda=0\quad .
\end{equation}

Let us analyze the triplet described by Eq.(\ref{c2}); 
Eqs.(\ref{tbaeq}) take the form
\begin{equation}
\arctan 2\lambda_{j}=\frac{\pi Q_{j}}{N}+\frac{1}{N}
\sum_{k=1}^{\frac{N}{2}-1}\arctan (\lambda_{j} -\lambda_{k})
\label{trbae}
\end{equation}
where now the numbers $Q_{j}$ lie in the segment 
$[-\frac{N}{4},\frac{N}{4}]$ and have two holes, $Q_1 ^{(h)}$ and 
$Q_2 ^{(h)}$ with $Q_1 ^{(h)} <Q_2 ^{(h)}$. Taking the 
thermodynamic limit one gets
\begin{equation}
\frac{Q_1 ^{(h)}}{N}\rightarrow x_1\quad,\quad 
\frac{Q_2 ^{(h)}}{N}\rightarrow x_2\quad,\quad \frac{Q_j}{N}
\rightarrow x+\frac{1}{N}(
\theta(x-x_1)+\theta(x-x_2))
\end{equation}
where $\theta (x)$ is the Heaviside function. Eqs.(\ref{trbae}) become 
\begin{equation}
\arctan 2\lambda(x)=\pi x+\frac{\pi}{N}(\theta(x-x_1)+\theta(x-x_2)) +
 \int_{-\frac{1}{4}}^{\frac{1}{4}}\arctan(\lambda(x)-\lambda(y)) dy\quad .
\label{2trbae}
\end{equation}
Eq.(\ref{2trbae}) gives, for this triplet, the density of 
$\lambda$, $\rho(\lambda)=\frac{d \lambda}{dx}$ 
\begin{equation}
\rho_{t}(\lambda)=\rho(\lambda)+\frac{1}{N}(\sigma (\lambda -
\lambda_1)-\sigma (\lambda -\lambda_2))
\end{equation}
where $\rho (\lambda)$ is given in Eq.(\ref{den1}) and  
\begin{equation}
\sigma(\lambda )=-\frac{1}{2\pi }\int_{-\infty}^{\infty}\frac{1}{1+e^{-|\xi|}}
e^{-i\lambda \xi} d\xi \quad .
\label{densig}
\end{equation}
$\lambda_1$ and $\lambda_2$ are the parameters of the holes, 
$\lambda_i=\lambda(x_{i})$, $i=1,2$.
The energy and the momentum of this state measured from the 
ground state are now easily computed
\begin{equation}
\epsilon_{T}(\lambda_1,\lambda_2)=
N\int_{-\infty}^{\infty}\epsilon(\lambda) 
(\rho_{t}(\lambda)-\rho(\lambda))d\lambda =
\epsilon(\lambda_1)+\epsilon(\lambda_2)
\label{tre}
\end{equation}
\begin{equation}
p_{T}(\lambda_1 ,\lambda_2)=N\int_{-\infty}^{\infty}p(\lambda) 
(\rho_{t}(\lambda)-\rho(\lambda))d\lambda=
p(\lambda_1)+p(\lambda_2)\quad (mod\quad 2\pi)
\end{equation}
where
\begin{equation}
\epsilon(\lambda)=\int_{-\infty}^{\infty}\epsilon(\mu)
\sigma(\lambda -\mu)d\mu=J\frac{\pi}{2\cosh \pi\lambda}
\label{et}
\end{equation}
\begin{equation}
p(\lambda)=\int_{-\infty}^{\infty}p(\mu)\sigma(\lambda -\mu)
d\mu=\arctan \sinh \pi \lambda -\frac{\pi}{2},\quad -\pi
\leq p(\lambda) \leq 0\quad .
\label{pt}
\end{equation}
From Eqs.(\ref{et},\ref{pt}) one gets
\begin{equation}
\epsilon=-\frac{J\pi}{2}\sin p\quad .
\label{disrel}
\end{equation}
The momentum $p_T(\lambda_1,\lambda_2)$ varies over 
the interval $[0,2\pi)$, when $\lambda_1$ and $\lambda_2$ run 
independently over the whole real axis.
The spin of this state can be computed by the formula
\begin{equation}
S=-\int_{-\infty}^{\infty}(\sigma(\lambda -\lambda_1)+
\sigma(\lambda -\lambda_2))d\lambda=1\quad .
\end{equation}

Let us finally analize the singlet excitation 
characterized by the configuration (\ref{c1}). 
Denoting by $\lambda_S$ 
the only number among the $\lambda_{j,{1}/{2}}$ 
which characterizes the string of length 2 and by $\lambda_{j}$ 
the numbers $\lambda_{j,0}$ 
for the strings of length 1, Eqs.(\ref{tbaeq}) read
\begin{eqnarray}
\arctan 2\lambda_{j}&=&\frac{\pi Q_j}{N}+\frac{1}{N}
\Phi(\lambda_j -\lambda_S)+\frac{1}{N}
\sum_{k=1}^{\frac{N}{2}-2}\arctan (\lambda_j -\lambda_k)\\
\arctan \lambda_S&=&\frac{1}{N}
\sum_{j=1}^{\frac{N}{2}-2}\Phi(\lambda_S -\lambda_j)
\end{eqnarray}
with
\begin{equation}
\Phi(\lambda)=\arctan 2\lambda+\arctan \frac{2}{3}\lambda\quad .
\end{equation}
The $\frac{N}{2}-2$ numbers $Q_{j}$ vary in the segment 
$[-\frac{N}{4}+\frac{1}{2},\frac{N}{4}-\frac{1}{2}]$; 
among them there are the two holes $Q_{1}^{(h)}$ and 
$Q_{2}^{(h)}$. Taking the thermodynamic limit one finds 
the density of $\lambda$'s for the singlet
\begin{equation}
\rho(\lambda_{S})=\rho(\lambda)+\frac{1}{N}
(\sigma(\lambda -\lambda_1)+\sigma(\lambda -\lambda_2)+
\omega(\lambda -\lambda_S))
\label{dens}
\end{equation}
where $\rho$ and $\sigma$ were given in Eqs.(\ref{den1}, 
\ref{densig}) and where 
\begin{equation}
\omega(\lambda)=-\frac{1}{2\pi}\int_{-\infty}^{\infty}
e^{-\frac{1}{2}|\xi|-i\lambda \xi} d\xi=
-\frac{2}{\pi(1+4\lambda^2)}\quad .
\end{equation}
In \cite{b6c} it was demonstrated that the 
string parameter $\lambda_S$ is uniquely determined by 
the $\lambda$'s parametrizing the two holes
\begin{equation}
\lambda_S=\frac{\lambda_{1}^{(h)}+\lambda_{2}^{(h)}}{2}\quad .
\end{equation}
In \cite{b6c} it was also proved the remarkable fact 
that the string of length 2 does not contribute to the 
energy and momentum of the excitation, 
so that the singlet and the triplet have the same dispersion relations
\begin{eqnarray}
\epsilon_S(\lambda_1,\lambda_2)&=&\epsilon_{T}(\lambda_1,\lambda_2)
=\epsilon(\lambda_1)+\epsilon(\lambda_2) \\
p_S(\lambda_1 ,\lambda_2)&=&p_T(\lambda_1 ,\lambda_2)=
p(\lambda_1)+p(\lambda_2)\quad ({\rm mod}\quad 2\pi)\quad .
\end{eqnarray}
The spin of this excitation is, of course, zero
\begin{equation}
S=-2-\int_{-\infty}^{\infty}(2\sigma(\lambda)+\omega(\lambda)) d\lambda=0
\end{equation}
The only difference between the state 
whose configuration is given in Eq.(\ref{c2}) and the state of
Eq.(\ref{c1}) is the spin. 

To summarize,
the finite energy 
excitations of the antiferromagnetic Heisenberg chain are only 
those belonging to the class ${\cal M_{AF}}$ and are described by
scattering states of an even number of quasiparticles or kinks. 
The momentum $p$ of these kinks runs over half the Brillouin 
zone $-\pi\le p\le 0$, the 
dispersion relation is $\epsilon(p)=\frac{J\pi}{2}\sin p$, Eq.(\ref{disrel}),
and the spin of a kink is $1/2$. The singlet and the triplet excitations
described above are the only states composed of two kinks, the spins of the 
kinks being parallel for the triplet and antiparallel for the singlet.
For vanishing total momentum all the states belonging to ${\cal M_{AF}}$
have the same energy of the ground state so that they are gapless excitations.
Since the eigenstates of $H_{J}$ always contain an even number of kinks,
the dispersion relation is determined by a set of two-parameters: the
momenta of the even number of kinks whose scattering provides the excitation. 
There are no bound states of kinks.

\subsection{Finite size antiferromagnetic Heisenberg chains}
\label{fsac}
Let us now turn to the computation of the spectrum of 
finite size quantum antiferromagnetic chains 
by exact diagonalization. We shall see that already 
for very small chains, the spectrum is well described  
by the Bethe ansatz solution in the thermodynamic limit. 
Furthermore, an intuitive picture 
of the ground state and of the lowest lying excitations
of the strongly coupled two-flavor 
lattice Schwinger model emerges, due to the mapping of the gauge model onto the spin chain $-$ see section 3. 
 
The states of an antiferromagnetic chain are 
classified according to the quantum numbers of spin, 
third spin component, energy and momentum $|S, S^3,E,p>$. 
For a 4 site chain the 
momenta allowed for the states are: $0,
\frac{\pi}{2},\frac{3\pi}{2}\ {\rm mod}\  2\pi $. The ground state is
\begin{equation}
|g.s.>=|0,0,-3J,0>=\frac{1}{\sqrt{12}}(2|\uparrow
\downarrow\uparrow\downarrow>+2|\downarrow\uparrow\downarrow\uparrow>-
|\uparrow\uparrow\downarrow \downarrow> -|\uparrow
\downarrow \downarrow\uparrow>-|\downarrow \downarrow\uparrow\uparrow>-
|\downarrow\uparrow\uparrow\downarrow>)\quad .
\label{gs4}
\end{equation}
This state is $P$-parity even.
In fact, by the definition of $P$-parity given in Eq.(\ref{par}),
the $P$-parity inverted state (\ref{gs4}) is
obtained by reverting the order of the spins in each vector $|\dots>$
appearing in (\ref{gs4}), e.g. $|\downarrow\downarrow\uparrow\uparrow>
\buildrel P\over\longrightarrow|\uparrow\uparrow\downarrow \downarrow>$.

The $\lambda$'s associated to the ground state (solution of the Bethe 
ansatz equations (\ref{baeq})) are 
$\lambda_1=-\frac{1}{2\sqrt{3}}$ and $\lambda_2=\frac{1}{2\sqrt{3}}$. 
There is also an excited singlet 
\begin{equation}
|0,0,-J,\pi>=\frac{1}{\sqrt{4}}(|\downarrow \downarrow
\uparrow\uparrow>-|\downarrow\uparrow\uparrow\downarrow>
-|\uparrow\downarrow \downarrow\uparrow>+ |\uparrow
\uparrow\downarrow \downarrow>)\quad .
\label{sex1}
\end{equation}
It is $P$-even, so that it is a $S^{P}=0^{+}$ excitation, with
the same quantum numbers (the isospin is replaced by the spin)
of the lowest lying singlet excitation of the
strongly coupled Schwinger model discussed by Coleman~\cite{b46}. 
The state (\ref{sex1}) also coincides with the excited 
singlet described by the configuration (\ref{c1}). It has only two 
complex $\lambda$'s which arrange 
themselves in a string approximately of length 2, 
$\lambda_1=-\lambda_2=i\sqrt{\frac{\sqrt{481}-17}{8}}$ and 
there are two holes with $Q_{1}^{(h)}=-\frac{1}{2}$ and 
$Q_{2}^{(h)}=\frac{1}{2}$. 

There are also three excited triplets, whose highest weight states are 
\begin{eqnarray}
|1,1,-J,\frac{\pi}{2}>&=&\frac{1}{\sqrt{4}}
(|\downarrow\uparrow\uparrow\uparrow>+i|\uparrow\downarrow\uparrow\uparrow>-
|\uparrow\uparrow\downarrow\uparrow>-i|\uparrow\uparrow\uparrow\downarrow>)\\
|1,1,-2J,\pi>&=&\frac{1}{\sqrt{4}}(|\downarrow
\uparrow\uparrow\uparrow>-|\uparrow\downarrow\uparrow\uparrow>+
|\uparrow\uparrow\downarrow\uparrow>-|\uparrow\uparrow\uparrow\downarrow>)
\label{tr63}\\
|1,1,-J,\frac{3\pi}{2}>&=&\frac{1}{\sqrt{4}}(|\downarrow
\uparrow\uparrow\uparrow>-i|\uparrow\downarrow\uparrow\uparrow>-
|\uparrow\uparrow\downarrow\uparrow>+i|\uparrow\uparrow\uparrow\downarrow>)\quad .
\label{6tri}
\end{eqnarray}
Among these, only the non-degenerate state with the lowest energy has a
well defined $P$-parity (\ref{tr63}). It is a $S^P=1^{-}$
like the lowest lying triplet of the two-flavor strongly coupled Schwinger
model. The degenerate states can be always combined
to form a $P$-odd state. 

We thus see that within the states in a given 
configuration there is always a representative state with well defined parity,
the others are degenerate and can be used to construct state of well defined 
energy and parity. Moreover the parity of the representative states
(with respect to the parity of the ground state) is
the same of the one of the lowest-lying
Schwinger model excitations in strong coupling.

\begin{figure}[htb]
\begin{center}
\setlength{\unitlength}{0.00041700in}%
\begingroup\makeatletter\ifx\SetFigFont\undefined
\def\x#1#2#3#4#5#6#7\relax{\def\x{#1#2#3#4#5#6}}%
\expandafter\x\fmtname xxxxxx\relax \def\y{splain}%
\ifx\x\y   
\gdef\SetFigFont#1#2#3{%
  \ifnum #1<17\tiny\else \ifnum #1<20\small\else
  \ifnum #1<24\normalsize\else \ifnum #1<29\large\else
  \ifnum #1<34\Large\else \ifnum #1<41\LARGE\else
     \huge\fi\fi\fi\fi\fi\fi
  \csname #3\endcsname}%
\else
\gdef\SetFigFont#1#2#3{\begingroup
  \count@#1\relax \ifnum 25<\count@\count@25\fi
  \def\x{\endgroup\@setsize\SetFigFont{#2pt}}%
  \expandafter\x
    \csname \romannumeral\the\count@ pt\expandafter\endcsname
    \csname @\romannumeral\the\count@ pt\endcsname
  \csname #3\endcsname}%
\fi
\fi\endgroup
\begin{picture}(10224,7224)(1189,-6973)
\thicklines
\put(6001,-3361){\circle{450}}
\put(7801,-2161){\circle{450}}
\put(6676,-5986){\circle{450}}
\put(2401,-6961){\vector( 0, 1){7200}}
\put(1201,-961){\vector( 1, 0){10200}}
\put(2401,-2161){\line( 1, 0){7200}}
\put(2401,-2161){\line( 1, 0){7200}}
\put(2401,-3361){\line( 1, 0){3600}}
\put(2401,-4561){\line( 1, 0){ 75}}
\put(2251,-2161){\line( 1, 0){150}}
\put(2251,-3361){\line( 1, 0){150}}
\put(2251,-4561){\line( 1, 0){150}}
\put(4201,-661){\line( 0,-1){1500}}
\put(6001,-661){\line( 0,-1){2700}}
\put(7801,-661){\line( 0,-1){1500}}
\put(9601,-661){\line( 0,-1){300}}
\multiput(2626,-961)(-8.91855,14.86426){14}{\makebox(13.3333,20.0000){\SetFigFont{7}{8.4}{rm}.}}
\put(2513,-766){\line(-1, 0){224}}
\multiput(2289,-766)(-8.91855,-14.86426){14}{\makebox(13.3333,20.0000){\SetFigFont{7}{8.4}{rm}.}}
\multiput(2176,-961)(8.91855,-14.86426){14}{\makebox(13.3333,20.0000){\SetFigFont{7}{8.4}{rm}.}}
\put(2289,-1156){\line( 1, 0){224}}
\multiput(2513,-1156)(8.91855,14.86426){14}{\makebox(13.3333,20.0000){\SetFigFont{7}{8.4}{rm}.}}
\multiput(9826,-961)(-8.91855,14.86426){14}{\makebox(13.3333,20.0000){\SetFigFont{7}{8.4}{rm}.}}
\put(9713,-766){\line(-1, 0){224}}
\multiput(9489,-766)(-8.91855,-14.86426){14}{\makebox(13.3333,20.0000){\SetFigFont{7}{8.4}{rm}.}}
\multiput(9376,-961)(8.91855,-14.86426){14}{\makebox(13.3333,20.0000){\SetFigFont{7}{8.4}{rm}.}}
\put(9489,-1156){\line( 1, 0){224}}
\multiput(9713,-1156)(8.91855,14.86426){14}{\makebox(13.3333,20.0000){\SetFigFont{7}{8.4}{rm}.}}
\put(4201,-2161){\circle{450}}
\put(5851,-2311){\framebox(300,300){}}
\put(1726,-2161){\makebox(0,0)[lb]{\smash{\SetFigFont{10}{12.0}{rm}-1}}}
\put(2326,-4711){\framebox(300,300){}}
\multiput(6901,-5311)(-8.91855,14.86426){14}{\makebox(13.3333,20.0000){\SetFigFont{7}{8.4}{rm}.}}
\put(6788,-5116){\line(-1, 0){224}}
\multiput(6564,-5116)(-8.91855,-14.86426){14}{\makebox(13.3333,20.0000){\SetFigFont{7}{8.4}{rm}.}}
\multiput(6451,-5311)(8.91855,-14.86426){14}{\makebox(13.3333,20.0000){\SetFigFont{7}{8.4}{rm}.}}
\put(6564,-5506){\line( 1, 0){224}}
\multiput(6788,-5506)(8.91855,14.86426){14}{\makebox(13.3333,20.0000){\SetFigFont{7}{8.4}{rm}.}}
\put(6601,-6736){\framebox(300,300){}}
\put(2401,-2161){\line( 1, 0){5400}}
\put(2401,-4561){\line( 1, 0){7200}}
\put(9601,-961){\line( 0,-1){3600}}
\put(9451,-4711){\framebox(300,300){}}
\put(11101,-1561){\makebox(0,0)[lb]{\smash{\SetFigFont{14}{16.8}{rm}p}}}
\put(1726,-3361){\makebox(0,0)[lb]{\smash{\SetFigFont{10}{12.0}{rm}-2}}}
\put(1726,-4561){\makebox(0,0)[lb]{\smash{\SetFigFont{10}{12.0}{rm}-3}}}
\put(7201,-5461){\makebox(0,0)[lb]{\smash{\SetFigFont{10}{12.0}{rm}QUINTET}}}
\put(7201,-6061){\makebox(0,0)[lb]{\smash{\SetFigFont{10}{12.0}{rm}TRIPLETS}}}
\put(7201,-6661){\makebox(0,0)[lb]{\smash{\SetFigFont{10}{12.0}{rm}SINGLETS}}}
\put(3976,-361){\makebox(0,0)[lb]{\smash{\SetFigFont{10}{12.0}{rm}$\pi /2$}}}
\put(5851,-361){\makebox(0,0)[lb]{\smash{\SetFigFont{10}{12.0}{rm}$\pi$}}}
\put(7576,-361){\makebox(0,0)[lb]{\smash{\SetFigFont{10}{12.0}{rm}$3\pi /2$}}}
\put(9451,-361){\makebox(0,0)[lb]{\smash{\SetFigFont{10}{12.0}{rm}$2\pi$}}}
\put(1201,-61){\makebox(0,0)[lb]{\smash{\SetFigFont{20}{24.0}{rm}E/J}}}
\end{picture}
\end{center}
\caption{Four sites chain spectrum}
\label{4spectrum}
\end{figure}
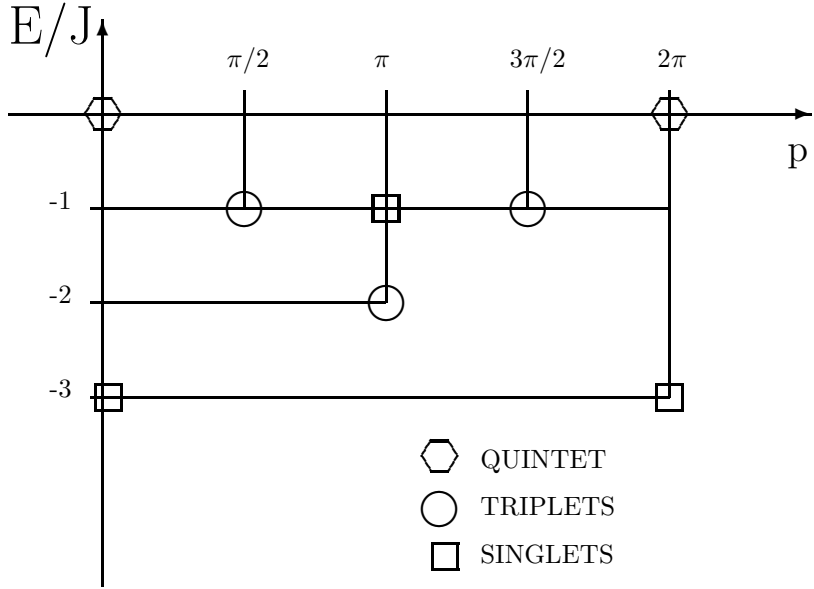

All the triplets in (\ref{6tri}) have one real $\lambda$ and two holes; 
they can be associated with the family of triplets (\ref{c2}).
 In table (\ref{tiqn}) we summarize the triplet $\lambda$'s and $Q^{(h)}$'s.
\begin{table}[htbp]
\begin{center}
\caption{Triplet internal quantum numbers }\label{tiqn}
\vspace{.1in}
\begin{tabular}{|lccc|}
\hline
TRIPLET&$\lambda$&$Q_{1}^{(h)}$&$Q_{2}^{(h)}$\rule{0in}{4ex}\\[2ex] \hline
$|1,1,-J,\frac{\pi}{2}>$& $\frac{1}{2} $&$-1$&$0$\rule{0in}{4ex}\\[2ex] \hline
$|1,1,-2J,\pi>$&$0$&$-1$&$1$\rule{0in}{4ex}\\[2ex] \hline
$|1,1,-J,\frac{3\pi}{2}>$&$-\frac{1}{2}$&$0$&$1$\rule{0in}{4ex}\\[2ex] \hline
\end{tabular}
\end{center}
\end{table}
The spectrum exhibits also a quintet, whose highest weight state is 
\begin{equation}
|2,2,0,0>=|\uparrow\uparrow\uparrow\uparrow>
\end{equation}
We report in fig.(\ref{4spectrum}) the spectrum of the 4 sites chain. 

\begin{figure}[htb]
\begin{center}
\setlength{\unitlength}{0.00050000in}%
\begingroup\makeatletter\ifx\SetFigFont\undefined
\def\x#1#2#3#4#5#6#7\relax{\def\x{#1#2#3#4#5#6}}%
\expandafter\x\fmtname xxxxxx\relax \def\y{splain}%
\ifx\x\y   
\gdef\SetFigFont#1#2#3{%
  \ifnum #1<17\tiny\else \ifnum #1<20\small\else
  \ifnum #1<24\normalsize\else \ifnum #1<29\large\else
  \ifnum #1<34\Large\else \ifnum #1<41\LARGE\else
     \huge\fi\fi\fi\fi\fi\fi
  \csname #3\endcsname}%
\else
\gdef\SetFigFont#1#2#3{\begingroup
  \count@#1\relax \ifnum 25<\count@\count@25\fi
  \def\x{\endgroup\@setsize\SetFigFont{#2pt}}%
  \expandafter\x
    \csname \romannumeral\the\count@ pt\expandafter\endcsname
    \csname @\romannumeral\the\count@ pt\endcsname
  \csname #3\endcsname}%
\fi
\fi\endgroup
\begin{picture}(12087,8499)(451,-8173)
\thicklines
\put(12001,-6661){\circle{450}}
\put(4801,-5461){\circle{450}}
\put(8401,-5461){\circle{450}}
\put(3001,-4861){\circle{450}}
\put(10201,-4861){\circle{450}}
\put(1201,-2761){\circle{450}}
\put(12001,-2761){\circle{450}}
\put(6601,-2161){\circle{450}}
\put(4801,-1786){\circle{450}}
\put(8401,-1786){\circle{450}}
\put(10726,-7186){\circle{450}}
\put(1201,-8161){\vector( 0, 1){8400}}
\put(3001,-61){\line( 0,-1){300}}
\put(4801,-61){\line( 0,-1){300}}
\put(6601,-61){\line( 0,-1){300}}
\put(8401,-61){\line( 0,-1){300}}
\put(10201,-61){\line( 0,-1){300}}
\put(12001,-61){\line( 0,-1){300}}
\put(976,-2161){\line( 1, 0){5625}}
\put(6601,-361){\line( 0,-1){1800}}
\put(976,-1261){\line( 1, 0){9225}}
\put(3001,-361){\line( 0,-1){900}}
\put(10201,-361){\line( 0,-1){900}}
\put(976,-1561){\line( 1, 0){5625}}
\put(4801,-361){\line( 0,-1){1425}}
\put(8401,-361){\line( 0,-1){1425}}
\put(976,-3061){\line( 1, 0){7425}}
\put(976,-2761){\line( 1, 0){225}}
\put(4801,-1786){\line( 0,-1){1275}}
\put(8401,-1786){\line( 0,-1){1275}}
\put(3001,-1261){\line( 0,-1){2700}}
\put(6601,-2161){\line( 0,-1){1800}}
\put(10201,-1261){\line( 0,-1){2700}}
\put(976,-3961){\line( 1, 0){9225}}
\put(976,-4861){\line( 1, 0){9225}}
\put(3001,-3961){\line( 0,-1){900}}
\put(10201,-3961){\line( 0,-1){900}}
\put(976,-5761){\line( 1, 0){225}}
\put(976,-5461){\line( 1, 0){7425}}
\put(8401,-3061){\line( 0,-1){2400}}
\put(4801,-3061){\line( 0,-1){2400}}
\put(976,-6661){\line( 1, 0){225}}
\put(976,-7861){\line( 1, 0){5625}}
\put(6601,-3961){\line( 0,-1){3900}}
\put(1201,-6661){\circle{450}}
\put(12001,-361){\line( 0,-1){6300}}
\put(1276,-7711){\makebox(0,0)[lb]{\smash{\SetFigFont{7}{8.4}{rm}$-(\sqrt{13}+5)/2$}}}
\put(1201,-6661){\line( 1, 0){10800}}
\put(1201,-5761){\line( 1, 0){10800}}
\put(1201,-2761){\line( 1, 0){10800}}
\put(6451,-8011){\framebox(300,300){}}
\put(6451,-1711){\framebox(300,300){}}
\put(1051,-5911){\framebox(300,300){}}
\put(11851,-5911){\framebox(300,300){}}
\put(2851,-4111){\framebox(300,300){}}
\put(10051,-4111){\framebox(300,300){}}
\multiput(6826,-3961)(-7.24632,12.07721){17}{\makebox(11.1111,16.6667){\SetFigFont{7}{8.4}{rm}.}}
\put(6713,-3766){\line(-1, 0){224}}
\multiput(6489,-3766)(-7.24632,-12.07721){17}{\makebox(11.1111,16.6667){\SetFigFont{7}{8.4}{rm}.}}
\multiput(6376,-3961)(7.24632,-12.07721){17}{\makebox(11.1111,16.6667){\SetFigFont{7}{8.4}{rm}.}}
\put(6489,-4156){\line( 1, 0){224}}
\multiput(6713,-4156)(7.24632,12.07721){17}{\makebox(11.1111,16.6667){\SetFigFont{7}{8.4}{rm}.}}
\multiput(5026,-3061)(-7.24632,12.07721){17}{\makebox(11.1111,16.6667){\SetFigFont{7}{8.4}{rm}.}}
\put(4913,-2866){\line(-1, 0){224}}
\multiput(4689,-2866)(-7.24632,-12.07721){17}{\makebox(11.1111,16.6667){\SetFigFont{7}{8.4}{rm}.}}
\multiput(4576,-3061)(7.24632,-12.07721){17}{\makebox(11.1111,16.6667){\SetFigFont{7}{8.4}{rm}.}}
\put(4689,-3256){\line( 1, 0){224}}
\multiput(4913,-3256)(7.24632,12.07721){17}{\makebox(11.1111,16.6667){\SetFigFont{7}{8.4}{rm}.}}
\multiput(8626,-3061)(-7.24632,12.07721){17}{\makebox(11.1111,16.6667){\SetFigFont{7}{8.4}{rm}.}}
\put(8513,-2866){\line(-1, 0){224}}
\multiput(8289,-2866)(-7.24632,-12.07721){17}{\makebox(11.1111,16.6667){\SetFigFont{7}{8.4}{rm}.}}
\multiput(8176,-3061)(7.24632,-12.07721){17}{\makebox(11.1111,16.6667){\SetFigFont{7}{8.4}{rm}.}}
\put(8289,-3256){\line( 1, 0){224}}
\multiput(8513,-3256)(7.24632,12.07721){17}{\makebox(11.1111,16.6667){\SetFigFont{7}{8.4}{rm}.}}
\multiput(3226,-1261)(-7.24632,12.07721){17}{\makebox(11.1111,16.6667){\SetFigFont{7}{8.4}{rm}.}}
\put(3113,-1066){\line(-1, 0){224}}
\multiput(2889,-1066)(-7.24632,-12.07721){17}{\makebox(11.1111,16.6667){\SetFigFont{7}{8.4}{rm}.}}
\multiput(2776,-1261)(7.24632,-12.07721){17}{\makebox(11.1111,16.6667){\SetFigFont{7}{8.4}{rm}.}}
\put(2889,-1456){\line( 1, 0){224}}
\multiput(3113,-1456)(7.24632,12.07721){17}{\makebox(11.1111,16.6667){\SetFigFont{7}{8.4}{rm}.}}
\multiput(10426,-1261)(-7.24632,12.07721){17}{\makebox(11.1111,16.6667){\SetFigFont{7}{8.4}{rm}.}}
\put(10313,-1066){\line(-1, 0){224}}
\multiput(10089,-1066)(-7.24632,-12.07721){17}{\makebox(11.1111,16.6667){\SetFigFont{7}{8.4}{rm}.}}
\multiput(9976,-1261)(7.24632,-12.07721){17}{\makebox(11.1111,16.6667){\SetFigFont{7}{8.4}{rm}.}}
\put(10089,-1456){\line( 1, 0){224}}
\multiput(10313,-1456)(7.24632,12.07721){17}{\makebox(11.1111,16.6667){\SetFigFont{7}{8.4}{rm}.}}
\put(1201,-61){\line(-2,-3){300}}
\put(901,-511){\line( 1, 0){600}}
\put(1501,-511){\line(-2, 3){300}}
\put(1201,-61){\line( 0, 1){  0}}
\put(12001,-136){\line(-2,-3){300}}
\put(11701,-586){\line( 1, 0){600}}
\put(12301,-586){\line(-2, 3){300}}
\put(12001,-136){\line( 0, 1){  0}}
\put(10576,-8011){\framebox(300,300){}}
\multiput(7951,-7861)(-7.24632,12.07721){17}{\makebox(11.1111,16.6667){\SetFigFont{7}{8.4}{rm}.}}
\put(7838,-7666){\line(-1, 0){224}}
\multiput(7614,-7666)(-7.24632,-12.07721){17}{\makebox(11.1111,16.6667){\SetFigFont{7}{8.4}{rm}.}}
\multiput(7501,-7861)(7.24632,-12.07721){17}{\makebox(11.1111,16.6667){\SetFigFont{7}{8.4}{rm}.}}
\put(7614,-8056){\line( 1, 0){224}}
\multiput(7838,-8056)(7.24632,12.07721){17}{\makebox(11.1111,16.6667){\SetFigFont{7}{8.4}{rm}.}}
\put(7726,-6961){\line(-2,-3){300}}
\put(7426,-7411){\line( 1, 0){600}}
\put(8026,-7411){\line(-2, 3){300}}
\put(7726,-6961){\line( 0, 1){  0}}
\put(601,-361){\vector( 1, 0){11925}}
\put(976,-1786){\line( 1, 0){7425}}
\put(2851, 14){\makebox(0,0)[lb]{\smash{\SetFigFont{12}{14.4}{rm}$\pi/3$}}}
\put(4651, 14){\makebox(0,0)[lb]{\smash{\SetFigFont{12}{14.4}{rm}$2\pi/3$}}}
\put(6526, 14){\makebox(0,0)[lb]{\smash{\SetFigFont{12}{14.4}{rm}$\pi$}}}
\put(8251, 14){\makebox(0,0)[lb]{\smash{\SetFigFont{12}{14.4}{rm}$4\pi/3$}}}
\put(10051, 14){\makebox(0,0)[lb]{\smash{\SetFigFont{12}{14.4}{rm}$5\pi/3$}}}
\put(11851, 89){\makebox(0,0)[lb]{\smash{\SetFigFont{12}{14.4}{rm}$2\pi$}}}
\put(451, 14){\makebox(0,0)[lb]{\smash{\SetFigFont{17}{20.4}{rm}E/J}}}
\put(12226,-1036){\makebox(0,0)[lb]{\smash{\SetFigFont{17}{20.4}{rm}p}}}
\put(8101,-7261){\makebox(0,0)[lb]{\smash{\SetFigFont{9}{10.8}{rm}SEVENTHPLET}}}
\put(8101,-7945){\makebox(0,0)[lb]{\smash{\SetFigFont{9}{10.8}{rm}QUINTETS}}}
\put(11026,-7261){\makebox(0,0)[lb]{\smash{\SetFigFont{9}{10.8}{rm}TRIPLETS}}}
\put(11026,-7945){\makebox(0,0)[lb]{\smash{\SetFigFont{9}{10.8}{rm}SINGLETS}}}
\put(1201,-1486){\makebox(0,0)[lb]{\smash{\SetFigFont{7}{8.4}{rm}$-(5-\sqrt{13})/2$}}}
\put(1276,-1711){\makebox(0,0)[lb]{\smash{\SetFigFont{7}{8.4}{rm}$-(7-\sqrt{17})/2$}}}
\put(1201,-2011){\makebox(0,0)[lb]{\smash{\SetFigFont{7}{8.4}{rm}$-1$}}}
\put(1276,-1186){\makebox(0,0)[lb]{\smash{\SetFigFont{7}{8.4}{rm}$-1/2$}}}
\put(1276,-2536){\makebox(0,0)[lb]{\smash{\SetFigFont{7}{8.4}{rm}$-(5-\sqrt{5})/2$}}}
\put(1501,-2986){\makebox(0,0)[lb]{\smash{\SetFigFont{7}{8.4}{rm}$-3/2$}}}
\put(1276,-3886){\makebox(0,0)[lb]{\smash{\SetFigFont{7}{8.4}{rm}$-2$}}}
\put(1276,-4786){\makebox(0,0)[lb]{\smash{\SetFigFont{7}{8.4}{rm}$-5/2$}}}
\put(1276,-5386){\makebox(0,0)[lb]{\smash{\SetFigFont{7}{8.4}{rm}$-(\sqrt{17}+7)/4$}}}
\put(1501,-5686){\makebox(0,0)[lb]{\smash{\SetFigFont{7}{8.4}{rm}$-3$}}}
\put(1501,-6511){\makebox(0,0)[lb]{\smash{\SetFigFont{7}{8.4}{rm}$-(\sqrt{5}+5)/2$}}}
\end{picture}
\end{center}
\caption{Six sites chain spectrum}
\label{sixspectrum}
\end{figure}
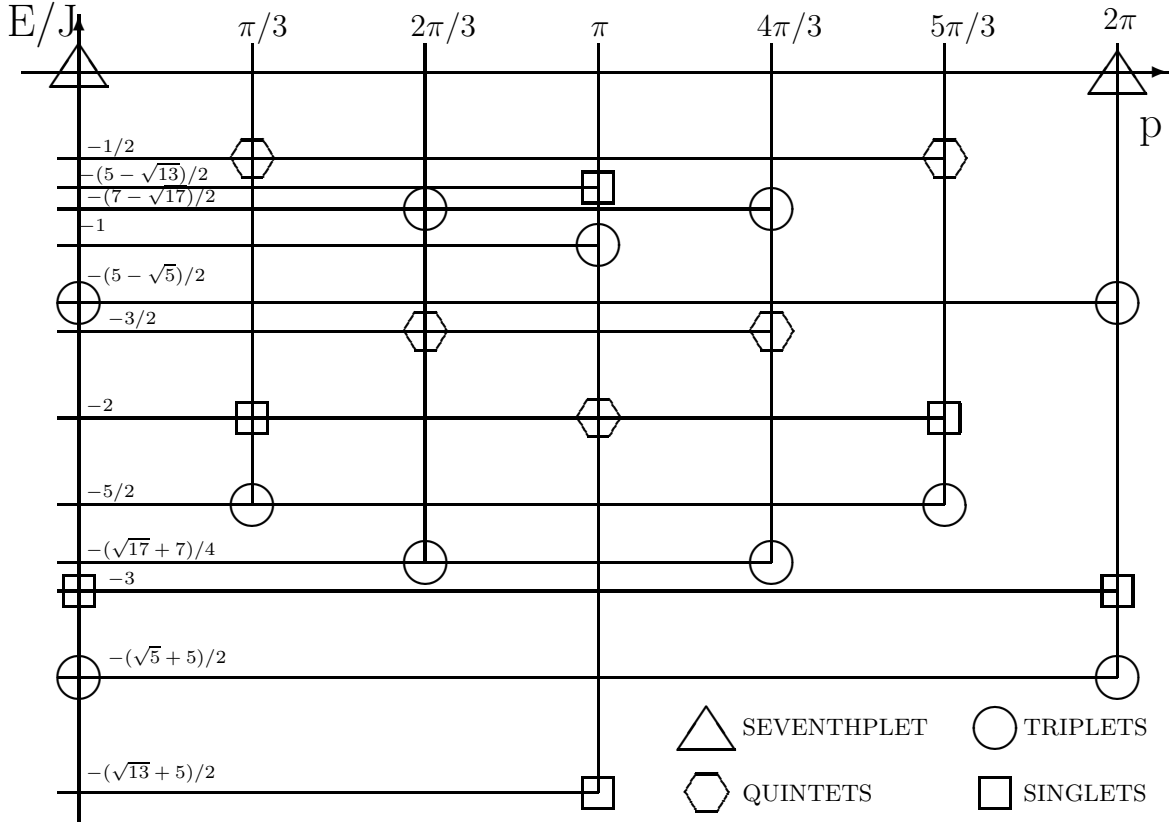

Let us analize the spectrum of the 6 site antiferromagnetic chain. 
The momenta allowed for 
the states are now $0,\frac{\pi}{3},\frac{2\pi}{3},\pi,\frac{4\pi}{3},
\frac{5\pi}{3}\ {\rm mod}\  2\pi$. 
The ground state is
\begin{eqnarray}
|g.s.>&=&|0,0,-\frac{J}{2}(5+\sqrt{13}),\pi>=
\frac{1}{\sqrt{26-6\sqrt{13}}}
\{|\downarrow\uparrow\downarrow\uparrow\downarrow\uparrow>-
|\uparrow\downarrow\uparrow\downarrow\uparrow\downarrow>\nonumber\\
&+&\frac{1-\sqrt{13}}{6}(|\uparrow\uparrow\downarrow
\uparrow\downarrow\downarrow>-|\uparrow\downarrow\uparrow
\downarrow\downarrow\uparrow>+
|\downarrow\uparrow\downarrow\downarrow\uparrow\uparrow>-|
\uparrow\downarrow\downarrow\uparrow\uparrow\downarrow>
+|\downarrow\downarrow\uparrow\uparrow\downarrow\uparrow>- |
\downarrow\uparrow\uparrow\downarrow\uparrow\downarrow>\nonumber\quad \quad \\
&-&|\downarrow\downarrow\uparrow\downarrow\uparrow\uparrow>+|
\downarrow\uparrow\downarrow\uparrow\uparrow\downarrow>-
|\uparrow\downarrow\uparrow\uparrow\downarrow\downarrow>+|
\downarrow\uparrow\uparrow\downarrow\downarrow\uparrow>
-|\uparrow\uparrow\downarrow\downarrow\uparrow\downarrow>+|
\uparrow\downarrow\downarrow\uparrow\downarrow\uparrow>)\nonumber\quad \quad \\
&+&\frac{4-\sqrt{13}}{3}(|\uparrow\uparrow\uparrow\downarrow
\downarrow\downarrow>-|\uparrow\uparrow\downarrow\downarrow\downarrow\uparrow>+
|\uparrow\downarrow\downarrow\downarrow\uparrow\uparrow>
-|\downarrow\downarrow\downarrow\uparrow\uparrow\uparrow>+|
\downarrow\downarrow\uparrow\uparrow\uparrow\downarrow>-
|\downarrow\uparrow\uparrow\uparrow\downarrow\downarrow>)\}
\quad \quad \quad .
\label{6gs}
\end{eqnarray}
This state is odd under $P$-parity.
The spectrum of the six sites chain is reported in 
fig.(\ref{sixspectrum}). There are 9 triplets in the spectrum. 
In~\cite{b45} 
it was already pointed out that the number of lowest 
lying triplets for a finite system with N sites is 
$N(N+2)/8$, so for $N=6$ there are 6 lowest lying triplet states. 
In order to identify these 6 states among the 9 that 
are exhibited by the spectrum of fig.(\ref{sixspectrum}), 
it is necessary to compute their $\lambda$'s and their $Q$'s.
In this way in fact, We can find out which are the triplets characterized by 
two holes and thus belonging to the triplet of type (\ref{c2}). In 
table (\ref{6tiqn}) we report the internal quantum numbers 
of the lowest lying triplets. The $Q^{(h)}$'s vary in the segment 
$[-\frac{3}{2},\frac{3}{2}]$.
The highest weight state of the triplet of zero 
momentum and energy $-(J/2)(5+\sqrt{5})$ reads
\begin{eqnarray}
|0,0,-\frac{J}{2}(5+\sqrt{5}),0>&=&\frac{1}{\sqrt{45-15\sqrt{5}}}
\{\frac{-3+\sqrt{5}}{2}(
|\downarrow\downarrow\uparrow\uparrow\uparrow\uparrow>+
|\downarrow\uparrow\uparrow\uparrow\uparrow\downarrow>+
|\uparrow\uparrow\uparrow\uparrow\downarrow\downarrow>+
|\uparrow\uparrow\uparrow\downarrow\downarrow\uparrow>\nonumber\\
&+&|\uparrow\uparrow\downarrow\downarrow\uparrow\uparrow>+
|\uparrow\downarrow\downarrow\uparrow\uparrow\uparrow>)\nonumber\\
&+&(|\downarrow\uparrow\downarrow\uparrow\uparrow\uparrow>+
|\uparrow\downarrow\uparrow\uparrow\uparrow\downarrow>+
|\downarrow\uparrow\uparrow\uparrow\downarrow\uparrow>+
|\uparrow\uparrow\uparrow\downarrow\uparrow\downarrow>+
|\uparrow\uparrow\downarrow\uparrow\downarrow\uparrow>+
|\uparrow\downarrow\uparrow\downarrow\uparrow\uparrow>)\nonumber\\
&+&(1-\sqrt{5})(|\downarrow\uparrow\uparrow\downarrow
\uparrow\uparrow>+|\uparrow\uparrow\downarrow\uparrow\uparrow\downarrow>+
|\uparrow\downarrow\uparrow\uparrow\downarrow\uparrow>)\}\quad\quad .
\label{tpo}
\end{eqnarray}
One can get the triplet of energy $-(J/2)(5-\sqrt{5})$ 
from (\ref{tpo}) by changing $\sqrt{5}\rightarrow -\sqrt{5}$.
As can be explicitly checked from (\ref{tpo}),
the two non-degenerate triplets of zero momentum are then $P$-parity even, 
namely they have opposite parity 
with respect to that of the  ground state, as it happens for the 
lowest lying triplet excitations of the two-flavor 
Schwinger model. 
For what concerns the degenerate triplets of momenta 
$\pi/3$ and $5\pi/3$ (or $2\pi/3$ and $4\pi/3$) they do not have definite 
$P$-parity, but it is always possible to take a 
linear combination of them with parity opposite to the ground state.  
\begin{table}[htbp]
\begin{center}
\caption{Triplet internal quantum numbers }\label{6tiqn}
\vspace{.1in}
\begin{tabular}{|lcccc|}
\hline
TRIPLET& $\lambda_1$ & $\lambda_2 $ & $Q_{1}^{(h)}$ & 
$Q_{2}^{(h)}$ \rule{0in}{4ex}\\[2ex] \hline

$|1,1,-\frac{5+\sqrt{5}}{2}J,0>$ & $-\sqrt{\frac{5-2\sqrt{5}}{20}}$ 
& $\sqrt{\frac{5-2\sqrt{5}}{20}}$ & $-\frac{1}{2}$ & $\frac{1}{2}$ 
\rule{0in}{4ex}\\[2ex] \hline

$|1,1,-\frac{5-\sqrt{5}}{2}J,0>$ & $-\sqrt{\frac{5+2\sqrt{5}}{20}}$ 
& $\sqrt{\frac{5+2\sqrt{5}}{20}}$ & $-\frac{3}{2}$ & $\frac{3}{2}$     
\rule{0in}{4ex}\\[2ex] \hline

$|1,1,-\frac{5}{2}J,\frac{\pi}{3}>$ & $-\frac{\sqrt{3}+\sqrt{\pi}}{8}$ 
& $-\frac{\sqrt{3}-\sqrt{\pi}}{8}$ & $-\frac{3}{2}$ & $\frac{1}{2}$     
\rule{0in}{4ex}\\[2ex] \hline

$|1,1,-\frac{7+\sqrt{17}}{4}J,\frac{2\pi}{3}>$ & $\frac{-2\sqrt{3}-
\sqrt{-2+2\sqrt{17}}}{2+2\sqrt{17}}$ & 
$\frac{-2\sqrt{3}+\sqrt{-2+2\sqrt{17}}}{2+2\sqrt{17}}$ 
& $-\frac{3}{2}$ & $-\frac{1}{2}$     
\rule{0in}{4ex}\\[2ex] \hline

$|1,1,-\frac{7+\sqrt{17}}{4}J,\frac{4\pi}{3}>$ 
& $\frac{2\sqrt{3}-\sqrt{-2+2\sqrt{17}}}{2+2\sqrt{17}}$ & 
$\frac{2\sqrt{3}+\sqrt{-2+2\sqrt{17}}}{2+2\sqrt{17}}$ 
& $\frac{1}{2}$ & $\frac{3}{2}$     
\rule{0in}{4ex}\\[2ex] \hline

$|1,1,-\frac{5}{2}J,\frac{5\pi}{3}>$ & $\frac{\sqrt{3}-
\sqrt{\pi}}{8}$ & $\frac{\sqrt{3}+\sqrt{\pi}}{8}$ 
& $-\frac{1}{2}$ & $\frac{3}{2}$\rule{0in}{4ex}\\[2ex] \hline
\end{tabular}
\end{center}
\end{table} 

The remaining three triplets in fig.(\ref{sixspectrum}) 
have no real $\lambda$'s and are characterized by a 
string of length 2 and four holes 
for $Q=-\frac{3}{2},-\frac{1}{2},\frac{1}{2},\frac{3}{2}$, $i.e.$ do 
not belong to the type (\ref{c2}). More precisely, two 
triplets have a string approximately of length 2, due to the finite size 
of the system, while the triplet of momentum $\pi$ has a 
string exactly of length 2. In table (\ref{26tiqn}) we summarize the quantum numbers of 
these triplets. 

\begin{table}[htbp]
\begin{center}
\caption{Four holes triplet internal quantum numbers }\label{26tiqn}
\vspace{.1in}
\begin{tabular}{|lcc|}
\hline
TRIPLET&$\lambda_1$&$\lambda_2$\rule{0in}{4ex}\\[2ex] \hline

$|1,1,-\frac{7-\sqrt{17}}{4}J,\frac{2\pi}{3}>$ 
& $\frac{2\sqrt{3}-i\sqrt{2+2\sqrt{17}}}{-2+2\sqrt{17}}$ 
& $\frac{2\sqrt{3}+i\sqrt{2+2\sqrt{17}}}{-2+2\sqrt{17}}$ 
\rule{0in}{4ex}\\[2ex] \hline

$|1,1,-J,\pi>$ & $-\frac{i}{2}$ & $\frac{i}{2}$ 
\rule{0in}{4ex}\\[2ex] \hline

$|1,1,-\frac{7-\sqrt{17}}{4}J,\frac{4\pi}{3}>$ 
& $\frac{2\sqrt{3}+i\sqrt{2+2\sqrt{17}}}{2-2\sqrt{17}}$ 
& $\frac{2\sqrt{3}-i\sqrt{2+2\sqrt{17}}}{2-2\sqrt{17}}$ 
\rule{0in}{4ex}\\[2ex]\hline
\end{tabular}
\end{center}
\end{table} 

In fig.(\ref{sixspectrum}) it is shown that the spectrum 
exhibits five singlet states. The lowest lying state
at momentum $\pi$ is the ground state. Then there are 
three excited singlets characterized by the 
configuration with two holes (\ref{c1}), $i.e.$ they 
have one real $\lambda$ and a string of length almost 2.
In table (\ref{esin}) we summarize their quantum numbers. 
Among these singlets, those which are
not degenerate, have   
$P$-parity equal to that of the
ground state (odd) as it happens in the two-flavor Schwinger 
model. The non-  singlet in fact reads
\begin{eqnarray}
|0,0,-3J,0>&=&\frac{1}{\sqrt{12}}\{|\uparrow\uparrow\downarrow
\uparrow\downarrow\downarrow>+|\uparrow\downarrow\uparrow
\downarrow\downarrow\uparrow>+
|\downarrow\uparrow\downarrow\downarrow\uparrow\uparrow>+
|\uparrow\downarrow\downarrow\uparrow\uparrow\downarrow>+
|\downarrow\downarrow\uparrow\uparrow\downarrow\uparrow>+
|\downarrow\uparrow\uparrow\downarrow\uparrow\downarrow>\nonumber\quad \quad\\
&-&|\downarrow\downarrow\uparrow\downarrow\uparrow\uparrow>-
|\downarrow\uparrow\downarrow\uparrow\uparrow\downarrow>-
|\uparrow\downarrow\uparrow\uparrow\downarrow\downarrow>-
|\downarrow\uparrow\uparrow\downarrow\downarrow\uparrow>-
|\uparrow\uparrow\downarrow\downarrow\uparrow\downarrow>-
|\uparrow\downarrow\downarrow\uparrow\downarrow\uparrow>\}\quad .
\label{spo}
\end{eqnarray}
The degenerate singlets are again not eigenstates of the $P$-parity, 
but it is always possible to take a linear combination of 
them with the a $P$-parity 
that coincides with that of the representative state (\ref{spo})
of the configuration.
\begin{table}[htbp]
\begin{center}
\caption{Singlet internal quantum numbers }\label{esin}
\vspace{.1in}
\begin{tabular}{|lcccc|}
\hline
SINGLET&$\lambda$ & $\lambda_S $ & $Q_1^{(h)}$ 
& $Q_2^{(h)}$\rule{0in}{4ex}\\[2ex] \hline

$|0,0,-3J,0>$ & 0 & 0 & $-1$ & 1 \rule{0in}{4ex}\\[2ex] \hline

$|0,0,-2J,\frac{\pi}{3}>$ & $-\frac{\sqrt{3}+2\sqrt{6}}{14}$ 
& $\frac{-2+3\sqrt{2}}{\sqrt{3}(4+\sqrt{2})}$ & $0$ 
& 1 \rule{0in}{4ex}\\[2ex] \hline

$|0,0,-2J,\frac{5\pi}{3}>$ & $\frac{\sqrt{3}+2\sqrt{6}}{14} $ 
&  $\frac{2-3\sqrt{2}}{\sqrt{3}(4+\sqrt{2})} $& $-1$ 
& 0 \rule{0in}{4ex}\\[2ex]\hline
\end{tabular}
\end{center}
\end{table}

The remaining singlet $|0,0,-\frac{5-\sqrt{13}}{2} J,\pi>$ 
it is not of the type (\ref{c1}). It is characterized by 
a string approximately of length 3 with $\lambda_{1,1}=
i\sqrt{\frac{5+2\sqrt{13}}{12}}$, $\lambda_{2,1}=0$ 
and $\lambda_{3,1}=-i\sqrt{\frac{5+2\sqrt{13}}{12}}$.

Even in finite systems very small like the 4 and 6 sites chains, 
the ``string hypothesis" is a very good approximation and 
it allows us to classify and distinguish among states with the same spin.   

The ground state of the antiferromagnetic Heisenberg 
chain with N sites is a linear combination of all the 
$\left( \begin{array}{c} N \\ 
\frac{N}{2} 
\end{array} \right)$ states with $\frac{N}{2}$ spins up and $\frac{N}{2}$ 
spins down. 
These states group themselves into sets with the same coefficient 
in the linear combination  according to the fact that 
the ground state is translationally invariant (with momentum 0 ($\pi$) 
for $\frac{N}{2}$ even (odd)), it 
is an eigenstate of $P$-parity and it is 
invariant under the exchange of up with down spins. 
The states belonging to the same set 
have the same number of domain walls, which ranges from 
$N$, for the two N\'eel states, to 2 for the states with $\frac{N}{2}$ 
adjacent spins up and $\frac{N}{2}$ adjacent spins down.  

The ground state of the 8 sites chain is 

\begin{equation}
|g.s.>= \frac{1}{ \sqrt{\cal{N}} }
(|\psi_{8} >+\alpha |\psi_6 ^{(1)} >+\beta |\psi_6^{(2)} >
+\gamma |\psi_4 ^{(1)}>+\delta |\psi_4 ^{(2)}>+\epsilon
|\psi_4 ^{(3)} >+\zeta |\psi_2>)
\label{8gs}
\end{equation} 
where 
\begin{eqnarray}
& &|\psi_{8}> = |\uparrow\downarrow\uparrow\downarrow\uparrow
\downarrow\uparrow\downarrow>+
|\downarrow\uparrow\downarrow\uparrow\downarrow\uparrow
\downarrow\uparrow>\\
& &|\psi_6 ^{(1)}> = |\uparrow\uparrow\downarrow\uparrow
\downarrow\uparrow\downarrow\downarrow>+
|\downarrow\downarrow\uparrow\downarrow\uparrow\downarrow
\uparrow\uparrow>+translated\quad states \\
& &|\psi_6^{(2)}>= |\uparrow\uparrow\downarrow\uparrow
\downarrow\downarrow\uparrow\downarrow>+
|\downarrow\downarrow\uparrow\downarrow\uparrow\uparrow
\downarrow\uparrow>+translated\quad states\\
& &|\psi_4 ^{(1)}>= |\uparrow\uparrow\downarrow\downarrow
\uparrow\uparrow\downarrow\downarrow>+translated\quad states\\
& &|\psi_4 ^{(2)}> = |\uparrow\uparrow\uparrow\downarrow
\downarrow\downarrow\uparrow\downarrow>+
|\downarrow\downarrow\downarrow \uparrow\uparrow\uparrow
\downarrow\uparrow>+translated\quad states \\
& &|\psi_4 ^{(3)}> = |\uparrow\uparrow\downarrow\downarrow
\downarrow\uparrow\uparrow\downarrow>+
|\downarrow\downarrow\uparrow\uparrow\uparrow\downarrow
\downarrow\uparrow>+translated\quad states\\
& &|\psi_2> = |\uparrow\uparrow\uparrow\uparrow\downarrow
\downarrow\downarrow\downarrow>+translated\quad states\ .
\label{8states}
\end{eqnarray}
By direct diagonalization one gets
\begin{eqnarray}
\alpha&=&-0.412773\\
\beta&=&0.344301\\
\gamma&=&0.226109\\
\delta&=&-0.087227\\
\epsilon&=&0.136945\\
\zeta&=&0.018754\\
\cal{N}&=&2+16\alpha^2+8\beta^2+4\gamma^2+16\delta^2
+16\epsilon^2+8\zeta^2=6.30356\quad .
\end{eqnarray}

The energy of the ground state is 
\begin{equation}
E_{g.s.}=-5.65109J\quad .
\label{egs8}
\end{equation}
Eq.(\ref{egs8}) differs only by $1.8\%$ from the thermodynamic 
limit expression $E_{g.s.}=-8\ln 2=-5.54518$. Moreover 
also the correlation function of distance 2 Eq.(\ref{corrd2}) 
computed for the 8 sites chain is $G(2)=0.1957N$, value which is $7\%$ 
higher than the exact answer Eq.(\ref{corrd2n}).

In the analysis of finite size systems we were able to find 
the coefficient $\beta$ of the first set of states containing
$N-2$ domain walls in 
the ground state. These states 
are obtained interchanging two adjacent spins in the N\'eel states. 
The $\beta$ is for a generic chain of $N$-sites
\begin{equation}
\beta=\frac{N+2E_{g.s.}}{N}=1-2\ln 2\quad .
\end{equation}    
\subsection{Spin-spin correlators}
\label{ssc}
The explicit computation of spin-spin correlations is far from being trivial since the correlator 
$G(r)=<g.s.|\vec{S}_{0}\cdot \vec{S}_{r}|g.s.>$ is not known for arbitrary lattice separations $r$.
For $r=2$ it was computed by M. Takahashi \cite{b38} 
in his perturbative analysis of the half filled Hubbard model in one dimension. For $r>2$ no exact numerical 
values of $G(r)$ are known. In \cite{b47} were given two representations of $G(r)$, while in \cite{b48,b49} the exact asymptotic 
($r\rightarrow \infty$) expression of $G(r)$ was derived. 

In order to explicitly compute the second order energies Eq.(\ref{senes}) and 
Eq.(\ref{senet}) one has to evaluate the correlation function
\begin{equation}
G(2)=\frac{1}{N}\sum_{x=1}^{N}<g.s.|\vec{S}_{x}\cdot \vec{S}_{x+2}|g.s>
\label{corrd2}
\end{equation}
which has been exactly computed in~\cite{b38} and is given by
\begin{equation}
G(2)=\frac{1}{4}(1-16\ln 2+9\zeta (3))=0.1820\quad . 
\label{corrd2n}
\end{equation}
In the following we shall show how the knowledge of this correlator 
allows one to compute explicitly the first three ``emptiness formation 
probabilities'', used in Ref.~\cite{b47} in the study of the Heisenberg
chain correlators, $G(r)$. The isotropy of the Heisenberg model implies that
\begin{equation}
\sum_{x=1}^{N}<g.s.|\vec{S}_{x}\cdot \vec{S}_{x+2}|g.s.>=3
\sum_{x=1}^{N}<g.s.|S^{3}_{x}\cdot S^3_{x+2}|g.s.>\quad .
\label{e1}
\end{equation}
Let us introduce the probability $P_3$ for finding three adjacent spins
in a given position in the Heisenberg antiferromagnetic vacuum. 
Taking advantage of the isotropy of the Heisenberg model 
ground state and of its
translational invariance, it is easy to see that the correlator (\ref{e1}) 
can be written in terms of the $P_3$'s as
\begin{equation}
\sum_{x=1}^{N}<g.s.|S^{3}_{x}\cdot S^3_{x+2}|g.s>=N~
\frac{1}{4}~2~(~P_3(\uparrow\uparrow\uparrow)+P_3(\uparrow\downarrow\uparrow)-
P_3(\uparrow\uparrow\downarrow)-P_3(\downarrow\uparrow\uparrow)~)\quad .
\label{e2}
\end{equation}
The factor 2 appears in (\ref{e2}) due again to the isotropy of the 
Heisenberg model: the probability of  
a configuration and of the configuration rotated by $\pi$ around the 
chain axis, are the same.  

In \cite{b47} the so called ``emptiness formation probability'' $P(x)$
was introduced.
\begin{equation}
P(x)=<g.s.|\prod_{j=1}^{x}P_{j}|g.s.>\quad ,
\label{e3}
\end{equation}
where 
\begin{equation}
P_{j}=\frac{1}{2}(\sigma_{j}^{3}+1)
\label{e4}
\end{equation}
and $\sigma_{j}^{3}$ is the Pauli matrix.
$P(x)$ determines the probability of finding $x$ adjacent spins up in the 
antiferromagnetic vacuum.
One gets
\begin{eqnarray}
P(\uparrow\uparrow\uparrow)&=&P(3)\\
P(\uparrow\downarrow\uparrow)&=&P(1)-2P(2)+P(3)\\
P(\downarrow\uparrow\uparrow)&=&P(\uparrow\uparrow\downarrow)=P(2)-P(3)
\end{eqnarray}
so that Eq.(\ref{corrd2}) reads
\begin{equation}
G(2)=2P(3)-2P(2)+\frac{1}{2}P(1)\quad .
\label{corrd22}
\end{equation}
Using the exact  value the correlator $G(2)$ computed in~\cite{b38}
from (\ref{corrd22}) and from the known values of $P(2)$
and $P(1)$ given in \cite{b47}
\begin{eqnarray}
P(1)&=&\frac{1}{2}\label{p1}\\
P(2)&=&\frac{1}{3}(1-\ln 2)\label{p2}\\
\label{p1p2}
\end{eqnarray}
one gets
\begin{eqnarray}
P(3)&=&\frac{1}{3}(1-7\ln 2)+\frac{9}{8} \zeta(3)\label{p3}\quad .
\label{efp3}
\end{eqnarray}
For the general emptiness formation probability $P(x)$ of the antiferromagnetic
Heisenberg chain,
an integral representation was given in~\cite{b47},
but, to our knowledge, the exact value of $P(3)$ (\ref{efp3}) was not
known.

We illustrate now the computation of a spin-spin correlator which appears in the mass spectrum of the two-flavor 
lattice Schwinger model
\begin{eqnarray}
& &<g.s.|\vec{V}\cdot \vec{V}|g.s.>=\sum_{x,y=1}^{N}(<g.s.|(\vec{S}_{x}\cdot \vec{S}_{y})(\vec{S}_{x+1}\cdot \vec{S}_{y+1})|g.s.>\nonumber\\
&-&<g.s.|(\vec{S}_{x}\cdot \vec{S}_{y+1})(\vec{S}_{x+1}\cdot \vec{S}_{y})|g.s.>)
-\sum_{x=1}^{N}<g.s.|\vec{S}_{x}\cdot \vec{S}_{x+1}|g.s.>\quad \quad .
\label{vv}
\end{eqnarray}
It is possible to extract a numerical value from Eq.(\ref{vv}) only within the random phase approximation \cite{b38,b50}. 
For this purpose it is first convenient 
to rewrite the unconstrained sum over the sites x and y as a sum where all the four spins involved in the VEV's lie on different sites,
\begin{eqnarray}
<g.s.|\vec{V}\cdot \vec{V}|g.s.>&=&\sum_{\stackrel{y\neq x}{ y\neq x\pm 1}}(<g.s.|(\vec{S}_{x}\cdot \vec{S}_{y})(\vec{S}_{x+1}\cdot \vec{S}_{y+1})|g.s.>
\nonumber\\
&-&<g.s.|(\vec{S}_{x}\cdot \vec{S}_{y+1})(\vec{S}_{x+1}\cdot \vec{S}_{y})|g.s.>)+\frac{3}{8}N\nonumber\\
&-&\frac{1}{2}\sum_{x=1}^{N}<g.s.|\vec{S}_{x}\cdot \vec{S}_{x+1}|g.s.>-\sum_{x=1}^{N}
<g.s.|\vec{S}_{x}\cdot \vec{S}_{x+2}|g.s.>\quad \quad 
\label{vvv}
\end{eqnarray}
and then factorize the four spin operators in Eq.(\ref{vvv}) as 
\begin{eqnarray}
<g.s.|\vec{V}\cdot \vec{V}|g.s.>&=&N\sum_{r=2}^{\infty}(<g.s.|\vec{S}_{0}\cdot \vec{S}_{r}|g.s.>^2 -<g.s.|\vec{S}_{0}\cdot \vec{S}_{r+1}|g.s.>
<g.s.|\vec{S}_{1}\cdot \vec{S}_{r}|g.s.>)\nonumber\\
&+&\frac{3}{8}N-\frac{1}{2}\sum_{x=1}^{N}<g.s.|\vec{S}_{x}\cdot \vec{S}_{x+1}|g.s.>-\sum_{x=1}^{N}<g.s.|\vec{S}_{x}\cdot \vec{S}_{x+2}|g.s.>\quad .
\label{rpavv}
\end{eqnarray}
Of course, Eq.(\ref{rpavv}) provides an answer larger than the exact result; 
terms such as $<(\ldots)(\ldots)>$ yield negative contributions 
which are eliminated once one factorizes them in the form $<(...)><(...)>$. 
This is easily checked also by direct computation on finite size systems. 

The spin-spin correlation functions G(r) are exactly known for $r=1,2$. For the spin-spin correlation functions $G(r)$ up to a distance of 
$r=30$ the results are reported in table (\ref{30c})  \cite{b48,b51}. 
\begin{table}[htbp]
\begin{center}
\caption{Spin-spin correlation functions}\label{30c}
\vspace{.1in}
\begin{tabular}{|rc|rc|}\hline
$r$   &   $G(r)$ & $r$   &  $G(r)$ \rule{0in}{4ex}\\[2ex] \hline 
 
 1    & -0.4431              & 16    &   0.0305 \rule{0in}{4ex}\\
  
 2    & 0.1821               & 17    &  -0.0296 \\

 3    & -0.1510              & 18    &   0.0274 \\
 
 4    & 0.1038               & 19    &  -0.0267 \\
 
 5    & -0.0925              & 20    &   0.0249 \\
 
 6    & 0.0731               & 21    &   -0.0242 \\
 
 7    & -0.0671              & 22    &   0.0228 \\

 8    & 0.0567              & 23    &   -0.0223 \\
 
 9    & -0.0532              & 24    &   0.0211 \\

 10   & 0.0465              & 25    &   -0.0206 \\
 
 11   &-0.0442              & 26    &   0.0196 \\
 
 12   & 0.0395              & 27    &   -0.0193 \\
 
 13   & -0.0379              & 28    &   0.0183 \\
 
 14   & 0.0344              & 29    &   -0.0181 \\
 
 15   & -0.0332              & 30    &   0.0172 \\[2ex] \hline
 \end{tabular}
 \end{center}
 \end{table}

 For $r>30$, one may write~\cite{b48}

 \begin{eqnarray}
 G(r)&=&\frac{3}{4}\sqrt{\frac{2}{\pi^3}}\frac{1}{r\sqrt{g(r)}}[ 1-\frac{3}{16}g(r)^2+\frac{156\zeta(3)-73}{384}g(r)^3+O(g(r)^4)-\nonumber \\
 & &\frac{0.4}{2r}((-1)^r +1+O(g(r))+O(\frac{1}{r^2}) ]
 \label{l1}
 \end{eqnarray}
 with $g(r)$ satisfying
 \begin{equation}
 g(r)=\frac{1}{C(r)}(1+\frac{1}{2}g(r)\ln (g(r)))
 \label{l2}
 \end{equation}
 and 
 \begin{equation}
 C(r)=\ln(2\sqrt{2\pi}e^{\gamma +1} r)\quad .
 \label{l3}
 \end{equation}
 Eq.(\ref{l2}) may be solved by iteration. To the lowest order in $\frac{1}{C}$ one finds
 \begin{equation}
 g(r)\approx \frac{1}{C(r)}-\frac{1}{C(r)^2}\ln C(r)\quad .
 \label{lg}
 \end{equation}
 Inserting (\ref{lg}) in Eq.(\ref{l1}) leads to 
\begin{equation}
G(r)\approx \sqrt{2}{\pi^3}\frac{1}{r}\sqrt{C(r)} [1+\frac{1}{4C(r)}\ln C(r)]+O(\frac{1}{C(r)^2})\quad .
\label{sslu}
\end{equation}
Inserting Eq.(\ref{sslu}) in (\ref{rpavv}), one finally gets  
\begin{equation}
<g.s.|\vec{V}\cdot \vec{V}|g.s.>=0.3816N
\label{vvapp}
\end{equation}

Last, we report the following exact three spin correlators that have been used in the determination of the mass spectrum of the two-flavor 
lattice Schwinger model
\begin{eqnarray}
<g.s.|\sum_{x=1}^{N}S^{3}_{x}S^{3}_{x+1}S^{3}_{x+2}|g.s.>&=&0\\
<g.s.|\sum_{x=1}^{N}S^{+}_{x}S^{-}_{x+1}S^{3}_{x+2}|g.s.>&=&0\\
<g.s.|\sum_{x=1}^{N}S^{-}_{x}S^{+}_{x+1}S^{3}_{x+2}|g.s.>&=&0\\
<g.s.|\sum_{x=1}^{N}S^{+}_{x}S^{3}_{x+1}S^{-}_{x+2}|g.s.>&=&0\\  
<g.s.|\sum_{x=1}^{N}S^{-}_{x}S^{3}_{x+1}S^{+}_{x+2}|g.s.>&=&0
\end{eqnarray}

So, on a spin singlet, not only the VEV of $\sum_{x=1}^{N}S^{3}_{x}$ is zero, but also every VEV with an odd number of $S^{3}$.

\subsection{$SU({\cal N})$ quantum antiferromagnetic chains}
\label{sunc}

It is our purpose to introduce now spin-$1/2$ antiferromagnetic Heisenberg chains where  ``spins" are generators of the $SU({\cal N})$ 
group. In the limit ${\cal N}=2$ one has the usual antiferromagnetic Heisenberg chain discussed in the previous section. 
An $U({\cal N})$ spin-$1/2$ quantum antiferromagnetic chain is described by the Hamiltonian
\begin{equation}
H_{J}^{U({\cal N})}=J\sum_{x=1}^{N}S_{ab}(x)S_{ba}(x+1)
\label{hun}
\end{equation}
where $S_{ab}(x)$ $a,b=1,\ldots,{\cal N}$ are the generators of $U({\cal N})$ satisfying the Lie algebra
\begin{equation}
\left[S_{ab}(x),S_{cd}(y)\right]=(S_{ad}(x)\delta_{bc}- S_{cb}(x)\delta_{ad})\delta_{xy}
\end{equation}
and they can be conventionally represented by fermion bilinear operators
\begin{equation}
S_{ab}(x)=\psi_{ax}^{\dagger}\psi_{bx}-\frac{\delta_{ab}}{2}\quad .
\end{equation}
The representation of the algebra on each site is fixed by specifying the fermion number occupation
\begin{equation}
\rho(x)=\sum_{a=1}^{\cal N}S_{aa}(x)\quad .
\end{equation}
By fulfilling the global neutrality condition $\sum_{x=1}^{N}\rho(x)=0$, one may choose
\begin{equation}
\sum_{a=1}^{\cal N}\psi^{\dagger}_{ax}\psi_{ax}=\left\{\begin{array}{cc}
m&x\quad $even$\\
{\cal N}-m&x\quad $odd$
\end{array} \right.
\label{nnn}
\end{equation}
or viceversa the opposite choice for $x$ even-odd. 
Eq.(\ref{nnn}) restricts on each site the Hilbert space to a representation with Young tableau of $m$ rows for $x$ even and ${\cal N}-m$ rows for 
$x$ odd. 
For each site $x$ $\rho(x)$ is the generator of the $U(1)$ subgroup of $U({\cal N})$. 

Let us use the basis $T^{\alpha}=(T^{\alpha})^{*}$, 
$\alpha=1,\ldots,{\cal N}^{2}-1$, 
of the Lie algebra of $SU({\cal N})$ in the fundamental representation such that $tr(T^{\alpha}T^{\beta})=\delta ^{\alpha \beta}/2$ and 
$\left[T^{\alpha},T^{\beta}\right]=if^{\alpha \beta \gamma}T^{\gamma}$, where $f^{\alpha \beta \gamma}$ are the structure constants. By 
means of
\begin{equation}
T_{ab}^{\alpha} T_{cd}^{\alpha}=\frac{1}{2}\delta_{ad}\delta_{bc}-\frac{1}{2{\cal N}} \delta_{ab}\delta_{cd} 
\end{equation}
and redefining the group generators
\begin{equation}
S^{\alpha}_x=\psi_{ax}^{\dagger}T_{ab}^{\alpha}\psi_{bx}
\label{spisch}
\end{equation}
one can rewrite the Hamiltonian (\ref{hun}) as
\begin{equation}
H_{J}^{U({\cal N})}=J\sum_{x=1}^{N}\rho(x)\rho(x+1)+ H_{J}^{SU({\cal N})}
\label{hun2}
\end{equation}
where
\begin{equation}
H_{J}^{SU({\cal N})}=J\sum_{x=1}^{N}S^{\alpha}_x  S^{\alpha}_{x+1}
\label{sunn}
\end{equation}
is the Hamiltonian of an $SU({\cal N})$ quantum antiferromagnet. 
From Eq.(\ref{hun2}) it is clear that by fixing $\rho(x)$ $H_{J}^{U({\cal N})}$ is reduced to $H_{J}^{SU({\cal N})}$. 

The representations of the $SU({\cal N})$ spins relevant for the relationship of the spin Hamiltonians (\ref{sunn})  with 
the Schwinger models are different when ${\cal N}$ is even or odd. When ${\cal N}$ is even we shall consider the representation on each site 
with Young tableau of one column and ${\cal N}/2$ rows so that $\rho(x)=0$. When ${\cal N}$ is odd we shall consider on one sublattice 
the representation with Young tableau of one column and $({\cal N}+1)/2$ rows and of one column and $({\cal N}-1)/2$ rows on the other sublattice. 

As an explicit example let us consider the case ${\cal N}=3$. The generators $T^{\alpha}$ of the $SU(3)$ group in Eq.(\ref{spisch}) 
are the Gell Mann matrices. By denoting $u$,$d$ and $s$ the three flavors, the eight spin operators $S^{\alpha}_{x}$ read
\begin{eqnarray}
S_{x}^{1}&=&\psi_{ux}^{\dagger}\psi_{dx}+\psi_{dx}^{\dagger}\psi_{ux}\\  
S_{x}^{2}&=&-i\psi_{ux}^{\dagger}\psi_{dx}+i\psi_{dx}^{\dagger}\psi_{ux}\\ 
S_{x}^{3}&=&\psi_{ux}^{\dagger}\psi_{ux}-\psi_{dx}^{\dagger}\psi_{dx}\\
S_{x}^{4}&=&\psi_{ux}^{\dagger}\psi_{sx}+\psi_{sx}^{\dagger}\psi_{ux}\\  
S_{x}^{5}&=&-i\psi_{ux}^{\dagger}\psi_{sx}+i\psi_{sx}^{\dagger}\psi_{ux}\\     
S_{x}^{6}&=&\psi_{dx}^{\dagger}\psi_{sx}+\psi_{sx}^{\dagger}\psi_{dx}\\   
S_{x}^{7}&=&-i\psi_{dx}^{\dagger}\psi_{sx}+i\psi_{sx}^{\dagger}\psi_{dx}\\    
S_{x}^{8}&=&\frac{1}{\sqrt{3}}(\psi_{ux}^{\dagger}\psi_{ux}+\psi_{dx}^{\dagger}\psi_{dx}-2\psi_{sx}^{\dagger}\psi_{sx})
\end{eqnarray}
Let us find the ground state of the Hamiltonian (\ref{sunn}) for a chain of two sites. Taking the representations with one particle 
on site 1 and two particles on site 2, the ground state with energy $E_{g.s.}=-16J/3$ reads
\begin{equation}
|g.s.>=\frac{1}{\sqrt{3}}(|u\begin{array}{c} d\\ s \end{array}>-|d\begin{array}{c} u\\ s \end{array}>+|s\begin{array}{c} u\\ d \end{array}>)
\label{hsgs1}   
\end{equation}
The state (\ref{hsgs1}) is a singlet of $SU(3)$, $i.e.$ it is annihilated by the Casimir $\vec{S}^{2}=(\vec{S}_{1}+ \vec{S}_{2})^{2}=
\sum_{\alpha=1}^{8}\left[(S_{1}^{\alpha})^{2}+(S_{2}^{\alpha})^{2}+2S_{1}^{\alpha}S_{2}^{\alpha}\right]$. If one chooses the representation 
with two particles on site  1 and one particle on site 2, the ground state degenerate with (\ref{hsgs1}) is
\begin{equation}
|g.s.>'=\frac{1}{\sqrt{3}}(|\begin{array}{c} d\\ s \end{array}u>-|\begin{array}{c} u\\ s \end{array}d>+|\begin{array}{c} u\\ d \end{array}s>)
\label{hsgs2}   
\end{equation}
Diagonalizing the translation operator $\hat{T}$ with the Hamiltonian (\ref{sunn}), one gets two degenerate ground states
\begin{equation}
|G.S.>^{\pm}=\frac{|g.s.>\pm|g.s.>'}{\sqrt{2}}
\label{GSPM}
\end{equation}
In Eq.(\ref{GSPM}) the linear combination with the $+$ has momentum zero and the one with the $-$ has momentum $\pi$. 
By studying this very simple example, one can infer that the thermodynamic limit $N\rightarrow \infty$ analysis for 
a generic $SU({\cal N})$ chain is very involved.
 
Unfortunately, no analysis with a level of completeness such as the one given in~\cite{b6c} for the $SU(2)$ case does exist 
for a generic symmetry group $SU({\cal N})$ with spins in the representation in which we are interested. In~\cite{b52} $SU({\cal N})$ antiferromagnetic models 
were solved for a particular spin representation such that the Hamiltonian (\ref{sunn}) becomes
\begin{equation}
H_{J}^{SU({\cal N})}=J\sum_{x=1}^{N}P_{x x+1}
\label{sunnp}
\end{equation}
with $P_{x x+1}$ the operator that permutes whatever objects occupy sites $x$ and $x+1$. The spectrum of Eq.(\ref{sunnp}) 
was shown to exhibit massless excitations. 

A large ${\cal N}$ expansion approach has been performed in~\cite{b53} for an $SU({\cal N})$ antiferromagnetic chain characterized 
by spins living in a representation with Young tableaux of one row on one sublattice and ${\cal N}-1$ rows on the 
other sublattice. In the case of Young tableaux of one column it was found that the ground state is twofold degenerate 
and breaks translational and parity symmetry. Moreover the elementary excitations are massive non relativistic solitons in 
the large ${\cal N}$ limit with a mass of $O({\cal N})$. 

In~\cite{b54} the Lieb-Shultz-Mattis theorem~\cite{b55} was generalized to $SU({\cal N})$ spin chains. The theorem proves 
that a half-integer-S spin chain with essentially any reasonably local Hamiltonian respecting 
translational and rotational symmetry either has zero gap or else has degenerate ground states spontaneously breaking 
translational and parity invariance. In~\cite{b55} it was proved the existence of a unique ground state of the $SU({\cal N})$ 
chains where ${\cal N}$ is even for spins in the antisymmetric ${\cal N}$ tensor representation. Under the assumption of a 
unique ground state  which must be a $SU({\cal N})$ singlet, an infinitesimal energy gap was found for all the representations 
of $SU({\cal N})$ whose Young tableaux contain a number of boxes not divisible by ${\cal N}$. Of course the degeneracy of the ground state 
trivially implyes a zero gap. 

Using the Lieb-Shultz-Mattis theorem we can state that when ${\cal N}$ is even the ground state of the Hamiltonian (\ref{sunn}) is 
unique and there are gapless excitations for the spin representation with a Young tableau of one column and ${\cal N}/2$ rows. 
When ${\cal N}$ is odd the ground state is twofold degenerate, as we illustrated with the simple example of the two site $SU(3)$ chain. 
It is doubtful if there exist gapless excitations in this case and at present time we are investigating this problem.

\section{The two-flavor lattice Schwinger model}
The one-flavor lattice Schwinger model has been studied in~\cite{b10}. The solution of the strong coupling problem is identical to 
solving a particular type of Ising system with long range interaction. 
We showed that the mass of the elementary excitations and the chiral condensate could be computed reliably from an extrapolation to 
weak coupling using Pad\'e approximants. Our analysis improved previous ones~\cite{b58} by taking careful account of all discrete symmetries 
of the continuum theory. We do not discuss here the one-flavor model, since it is mapped into the classical Ising spin model and 
it is out of the scope of these lectures.
  
We study the $SU(2)$-flavor lattice Schwinger model in the hamiltonian formalism using staggered fermions. 
The existence of the continuum internal isospin symmetry makes the model much more interesting than the one-flavor case; the 
spectrum is extremely richer, exhibiting also massless excitations and the chiral symmetry breaking pattern is 
completely different from the one-flavor case. We shall demonstrate~\cite{b7,b8}
 that the strong coupling limit of the two-flavor 
lattice Schwinger model is mapped onto an interesting quantum spin model $-$ the one-dimensional spin-$1/2$
 quantum Heisenberg antiferromagnet. 
The ground state of the antiferromagnetic chain has been known since many years~\cite{b6b} and its energy 
was computed in~\cite{b34}; the complete spectrum as been determined by Faddeev and Takhtadzhyan~\cite{b6c} using the algebraic 
Bethe ansatz. 

The two-flavor lattice Schwinger model with non-zero fermion mass $m$ has been analysed in~\cite{b64b} in the limit 
of heavy fermions $m\gg e^{2}$; good agreement with the continuum theory has been found.

There are by now many hints at a correspondence between quantized gauge theories and quantum spin models, aimed at analyzing new phases relevant 
for condensed matter systems~\cite{b4a,b4b,b4c,b4d,b4e}. Recently Laughlin has argued that there is an analogy between the spectral data of gauge 
theories and strongly correlated electron systems~\cite{b58b}. Moreover,
 certain spin ladders have been shown to be related to the two-flavor Schwinger model \cite{b65}. 
 
The correspondence between the SU(2) flavor Schwinger model and the quantum Heisenberg antiferromagnetic chain provides a concrete 
computational scheme in which the issue of the correspondence between quantized gauge theories and quantum spin models may be 
investigated. Because of dimensionality of the coupling constant in (1+1)-dimensions the infrared behavior is governed by the strong coupling limit, 
and it is tempting to conjecture the existence of an exact correspondence between the infrared limits of the Heisenberg and two-flavor Schwinger models. 
We shall derive~\cite{b7,b8} results which support this conjecture. For example the gapless modes in the spectra have 
identical quantum numbers; within the accurancy of the strong coupling limit, the gapped mode of the two-flavor Schwinger model was also 
identified in the spectrum of the Heisenberg model.

In this section we present a complete study~\cite{b7,b8,b9} of the strong coupling limit of the two-flavor lattice Schwinger model. 
We firstly compute explicitly the masses 
of the excitations to the second order in the strong coupling expansion; this 
computation needs the knowledge of the spin-spin correlators of the quantum Heisenberg antiferromagnetic chain. 
The continuum massless two-flavor Schwinger model does exhibit neither an isoscalar $\left<\bar\psi\psi\right>$ nor an isovector 
$\left< \bar\psi \sigma^a\psi\right>$ chiral condensate, since this is forbidden by the Coleman theorem~\cite{b39}. 
On the lattice these fermion 
condensates are zero to all the orders in the strong coupling expansion~\cite{b8}. The pertinent non-zero chiral condensate is 
$<\overline{\psi}_{L}^{(2)}\overline{\psi}_{L}^{(1)}\psi_{R}^{(1)}\psi_{R}^{(2)}>$ and we computed its lattice expression up to the 
second order in the strong coupling expansion~\cite{b8}. It should be noticed that, in abscence of gauge fields, the chiral condensate is zero, is 
different from zero only when the fermions are coupled to gauge fields. This can be viewed as the manifestation of the 
chiral anomaly in this model. 

\subsection{The model}
\label{2fcl}

The action of the $1+1$-dimensional electrodynamics with two charged 
Dirac spinor fields is
\begin{equation}
S = \int d^{2} x\left[\sum_{a=1}^{2} \overline{\psi}_{a}
(i\gamma_{\mu}\partial^{\mu}+\gamma_{\mu}
A^{\mu})\psi_{a}-\frac{1}{4e^{2}_{c}}F_{\mu\nu}F^{\mu\nu}\right]
\label{action}
\end{equation}
The theory has an internal $SU_L(2)\otimes SU_R(2)$-flavor isospin symmetry; 
the Dirac fields are 
an isodoublet whereas the electromagnetic field is an isosinglet.
It is well known that in $1+1$ dimensions there is no spontaneous breakdown 
of continuous internal symmetries, unless there are anomalies or the Higgs 
phenomenon occurs. Neither mechanism is possible in the two-flavor Schwinger 
model for the $SU_L(2)\otimes SU_R(2)$-symmetry: isovector currents do not 
develop anomalies and there are no gauge 
fields coupled to the isospin currents. The particles belong then to 
isospin multiplets. For what concerns the $U(1)$ gauge symmetry there is an Higgs phenomenon~\cite{b60}.

The action is invariant under the symmetry 
\beq
SU_{L}(2)\otimes SU_{R}(2)\otimes U_{V}(1) \otimes U_{A}(1) \nonumber
\eeq
The group generators act on the fermion isodoublet to give
\begin{eqnarray}
SU_{L}(2) &:& \psi_{a}(x)\longrightarrow (
e^{i\theta_{\alpha}\frac{\sigma^{\alpha}}{2}P_{L}})_{ab}\ \psi_{b}(x)\   , \  
\overline{\psi_{a}}(x)\longrightarrow \overline{\psi_{b}}\ (x)
(e^{-i\theta_{\alpha}\frac{\sigma^{\alpha}}{2}P_{R}})_{ba}\label{s1} \\
SU_{R}(2) &:& \psi_{a}(x)\longrightarrow 
(e^{i\theta_{\alpha}\frac{\sigma^{\alpha}}{2}P_{R}})_{ab}\ \psi_{b}(x)\   ,\   
\overline{\psi_{a}}(x)\longrightarrow \overline{\psi_{b}}(x)\ 
(e^{-i\theta_{\alpha}\frac{\sigma^{\alpha}}{2}P_{L}})_{ba} \\
U_{V}(1) &:& \psi_{a}(x)\ \longrightarrow 
(e^{i\theta(x){\bf 1}})_{ab}\ \psi_{b}(x)\  ,\  
\psi_{a}^{\dagger}(x)\longrightarrow \psi_{b}^{\dagger}(x)\ 
(e^{-i\theta(x){\bf 1}})_{ba} \\
U_{A}(1) &:& \psi_{a}(x)\longrightarrow (e^{i\alpha 
\gamma_{5}{\bf 1}})_{ab}\ \psi_{b}(x)\  ,\  
\psi_{a}^{\dagger}(x)\longrightarrow \psi_{b}^{\dagger}(x)\ 
(e^{-i\alpha \gamma_{5}{\bf 1}})_{ba}
\quad ,
\label{s4} 
\end{eqnarray}
where $\sigma^{\alpha}$ are the Pauli matrices, $\theta_{\alpha}$, $\theta(x)$ and $\alpha$ are real coefficients and
\beq
P_{L}=\frac{1}{2}(1-\gamma_{5})\ ,\ P_{R}=\frac{1}{2}(1+\gamma_{5})\quad .
\eeq
At the classical level the symmetries (\ref{s1}$-$\ref{s4}) lead to 
conservation laws for the isovector, vector and axial currents
\begin{eqnarray}
j_{\alpha}^{\mu}(x)_{R}&=&\overline{\psi}_{a}(x)\gamma^{\mu}P_{R}
(\frac{\sigma_{\alpha}}{2})_{ab}\psi_{b}(x)
\label{ca}\\
j_{\alpha}^{\mu}(x)_{L}&=&\overline{\psi}_{a}(x)\gamma^{\mu}P_{L}
(\frac{\sigma_{\alpha}}{2})_{ab}\psi_{b}(x)
\label{cb}\\
j^{\mu}(x)&=&\overline{\psi}_{a}(x)\gamma^{\mu}{\bf 1}_{ab}\psi_{b}(x)
\label{cc}\\
j^{\mu}_{5}(x)&=&\overline{\psi}_{a}(x)\gamma^{\mu}\gamma ^{5}
{\bf 1}_{ab}\psi_{b}(x)
\end{eqnarray}
It is well known that at the quantum level the vector and axial currents 
cannot be simultaneously conserved, 
due to the anomaly phenomenon~\cite{b23}. If the regularization is gauge 
invariant, so that 
the vector current is conserved, then the axial current acquires the 
anomaly which breaks the $U_{A}(1)$-symmetry 
\beq
\partial_{\mu} j_5^{\mu}(x)=2\frac{e_{c}^{2}}{2\pi}\epsilon_{\mu \nu}
F^{\mu \nu}(x)
\label{anomaly}
\eeq
The isoscalar and isovector chiral condensates are zero due to the 
Coleman theorem \cite{b39}; in fact, they would break not only the 
$U_A(1)$ symmetry 
of the action, but also the continuum internal symmetry 
$SU_L(2)\otimes SU_R(2)$ down to $SU_V(2)$. 
There is, however,  a $SU_L(2)\otimes SU_R(2)$ invariant operator, 
which is non-invariant under the $U_A(1)$-symmetry;
it can acquire a non-vanishing VEV without violating 
Coleman's theorem and consequently 
may be regarded as a good order parameter for the 
$U_A(1)$-breaking.
Its expectation value is given by~\cite{b67,b57}
\begin{equation} 
<F>\equiv
<\overline{\psi}_{L}^{(2)}\overline{\psi}_{L}^{(1)}
\psi_{R}^{(1)}\psi_{R}^{(2)}>=(\frac{e^{\gamma}}{4\pi})^{2} 
\frac{2}{\pi} e_{c}^{2}\quad .
\label{chico}
\end{equation}
It describes a process in which two right movers are anihilated
and two left movers are created. 
Note that $F$, being quadrilinear in the fields,
is actually invariant under chiral rotations of $\pi/2$, namely under the 
discrete axial symmetry
\begin{equation}
\psi_a(x)\to \gamma^5\psi_a(x)\quad\bar\psi_a(x)\to-\bar\psi_a(x)\gamma_5\ \ .
\label{disax}
\end{equation}
As a consequence, this part of the chiral symmetry group is not broken by 
the non-vanishing VEV of $F$ (\ref{chico}).

The lattice theory faithfully reproduces the 
pattern of symmetry breaking of the continuum theory; this happens even if 
on the lattice the $SU(2)$-flavor symmetry is not protected 
by the Coleman theorem. The isoscalar and isovector chiral 
condensates are zero also on the lattice,
whereas the operator $F$ acquires a non-vanishing VEV due to the 
coupling of left and right movers induced by the gauge field.
The continuous axial symmetry is broken explicitly by 
the staggered fermion, but the discrete axial symmetry (\ref{disax}) remains.

The action (\ref{action}) may be presented 
in the usual abelian bosonized form \cite{b68}. Setting
\beq
:\overline{\psi}_{a}\gamma^{\mu}\psi_{a}:=\frac{1}{\sqrt{\pi}}\epsilon^{\mu 
\nu}\partial_{\nu}\Phi_{a}\  ,\  a=1,2\quad ,
\eeq
the electric charge density and the action read
\beq
j_{0}=:\psi^{\dagger}_{1}\psi_{1}+\psi^{\dagger}_{2}\psi_{2}:=
\frac{1}{\sqrt{\pi}}\partial_{x}(\Phi_{1}+\Phi_{2})
\label{chde}
\eeq
\beq
S=\int d^{2}x \left[\frac{1}{2}\partial_{\mu}\Phi_{1}\partial^{\mu}\Phi_{1}
+\frac{1}{2}\partial_{\mu}\Phi_{2}\partial^{\mu}\Phi_{2}-
\frac{e_{c}^{2}}{2\pi}(\Phi_{1}+\Phi_{2})^{2}\right] \quad .
\eeq
By changing the variables to
\begin{eqnarray}
\Phi_{+}&=&\frac{1}{\sqrt{2}}(\Phi_{1}+\Phi_{2})\\
\Phi_{-}&=&\frac{1}{\sqrt{2}}(\Phi_{1}-\Phi_{2})\quad ,
\end{eqnarray}
one has
\beq
S=\int d^{2}x \left(\frac{1}{2}\partial_{\mu}\Phi_{+}
\partial^{\mu}\Phi_{+}+\frac{1}{2}\partial_{\mu}\Phi_{-}
\partial^{\mu}\Phi_{-}-
\frac{e_{c}^{2}}{\pi}\Phi_{+}^{2}\right)\quad .
\label{boa}
\eeq

The theory describes two scalar fields, one massive and one massless. 
$\Phi_{+}$ is an isosinglet as 
evidenced from Eq.(\ref{chde}); its mass $m_{S}=\sqrt{\frac{2}{\pi}}e_{c}$ 
comes from the anomaly Eq.(\ref{anomaly})~\cite{b60}. Local electric charge 
conservation is 
spontaneously broken, but no Goldstone boson appears because the 
Goldstone mode may be gauged away.
$\Phi_{-}$ represents an isotriplet; it has rather involved nonlinear 
transformation properties under a general 
isospin transformation. All three isospin currents can be written in 
terms of $\Phi_{-}$ but only the third component has a simple 
representation in terms of $\Phi_-$; namely
\begin{eqnarray}
j_{\mu} ^{3} (x) = : \overline{\psi} _{a} (x) \gamma_{\mu} 
(\frac{\sigma^{3}}{2})_{ab} \psi_{b} (x):=\quad \quad \quad 
\quad \quad \nonumber\\
   = \frac{1}{2}:\overline{\psi}_{1}(x)\gamma_{\mu}\psi_{1}(x)-
\overline{\psi}_{2}(x)\gamma_{\mu}\psi_{2}(x):=(2\pi)^{\frac{1}{2}}
\epsilon^{\mu \nu}
\partial_{\nu}\Phi_{-}\quad .
\end{eqnarray}
The other two isospin currents $j_{\mu}^{1}(x)$ and $j_{\mu}^{2}(x)$ 
are nonlinear 
and nonlocal functions of $\Phi_{-}$ \cite{b68}; a more symmetrical 
treatment of the bosonized form of the isotriplet currents is available 
within the framework of non abelian bosonization~\cite{b68bis}. For the multiflavor 
Schwinger model this approach has been carried out in~\cite{b68tris}, providing results in 
agreement with~\cite{b68}.

The excitations are most conveniently classified in terms of the 
quantum numbers of $P$-parity and $G$-parity; 
$G$-parity is related to the charge conjugation $C$ by
\beq
G=e^{i\pi \frac{\sigma^{2}}{2}} C\quad .
\eeq
$\Phi_{-}$ is a $G$-even pseudoscalar, while $\Phi_{+}$ is a $G$-odd pseudoscalar
\begin{eqnarray}
\Phi_{-}\  &:&\  I^{PG}=1^{-+}\\
 \Phi_{+}\  &:&\  I^{PG}=0^{--}\ \ .
\end{eqnarray}
The massive meson $\Phi_{+}$ is stable by $G$ conservation since the action 
(\ref{boa}) is invariant under $\Phi_{+}\longrightarrow -\Phi_{+}$.

In the massive $SU(2)$ Schwinger model $-$ when the mass of the fermion 
$m$ is small compared to $e^2$ (strong coupling) $-$ 
Coleman \cite{b68} showed that - in addition to the triplet $\Phi_-$ 
($I^{PG}=1^{-+}$)
the low-energy spectrum exhibits a singlet 
$I^{PG}=0^{++}$  lying on 
top of the triplet $\Phi_-$. In this limit the gauge theory is 
mapped to a sine-Gordon model and  the low lying excitations  
are soliton-antisoliton states. When $m\rightarrow 0$, these
soliton-antisoliton states become massless \cite{b69}; in this 
limit, the analysis of the many body wave functions, carried out in 
ref.\cite{b69}, hints to the existence of  
a whole class of massless states with positive G-parity; 
P-parity however cannot be determined with the procedure developed
in \cite{b69}. 
These are not the only excitations of the model:
way up in mass there is the 
isosinglet $I^{PG}=0^{--}$, ($\Phi_+$), already discussed in 
ref.~\cite{b68}. 
The model exhibits also triplets, 
whose mass $-$ of order $m_{S}$ or greater $-$ stays finite~\cite{b69}; 
among the triplets there is a G-even state
\footnote{K. Harada private communication.}.

The Hamiltonian, gauge constraint and non-vanishing (anti-)commutators
of the continuum two-flavor Schwinger model are
\begin{eqnarray}
H=\int dx&\left[\frac{e^2}{2}E^2(x)+\sum_{a=1}^2
\psi^{\dagger}_a (x)\alpha\left(i\partial_x +eA(x)\right)\psi_a(x)\right]
\label{ham1}\\ 
&\partial_x E(x)\ +\sum_{a=1}^2 \psi^{\dagger}_a
(x)\psi_a (x)\sim 0\label{ga1}\\
&\left[ A(x),E(y)\right]=i\delta(x-y) ~,
 \left\{\psi_a(x),\psi_b^{\dagger}(y)\right\}=\delta_{ab}\delta(x-y)\ \ .
\label{commu1}
\end{eqnarray}
A lattice Hamiltonian, constraint and (anti-) commutators reducing to 
(\ref{ham1},\ref{ga1},\ref{commu1}) 
in the naive continuum limit are 
\begin{eqnarray}
H_{S}=\frac{e^{2}a}{2}\sum_{x=1}^N E_{x}^{2}&-&\frac{it}{2a}\sum_{x=1}^N
\sum_{a=1}^2 \left(\psi_{a,x+1}^{\dag}e^{iA_{x}}\psi_{a,x}
-\psi_{a,x}^{\dag}e^{-iA_{x}}\psi_{a,x+1}\right)\label{hamilton}\nonumber\\
E_{x}-E_{x-1}&+&\psi_{1,x}^{\dag}\psi_{1,x}+\psi_{2,x}^{\dag}\psi_{2,x}-1\sim
0\ ,
\label{gauss}\\
\left[ A_x,E_y\right]=i\delta_{x,y}~&,&
\left\{\psi_{a,x},\psi_{b,y}^{\dagger}\right\}=\delta_{ab}\delta_{xy}\ \ .
\nonumber
\end{eqnarray}
The fermion fields are defined on the sites, $x=1,...,N$, 
the gauge and electric fields, $ A_{ x}$ and
 $E_{x}$,  on the links $[x; x + 1]$, $N$ is an even integer 
and, when $N$ is finite it is convenient to impose periodic boundary conditions.  When $N$ is finite, the continuum limit is the 
two-flavor Schwinger model on a circle \cite{b62}.
The coefficient $t$ of the hopping term in (\ref{hamilton})
plays the role of the lattice light speed. In the naive continuum limit,
$e_L=e_c$ and $t=1$. 

The Hamiltonian and gauge constraint exhibit the discrete symmetries 
\begin{itemize}
\item{}Parity P: 
\begin{equation}
A_{x}\longrightarrow -A_{-x-1},\ E_{x}\longrightarrow -E_{-x-1},\ 
\psi_{a,x}\longrightarrow (-1)^{x}\psi_{a,-x},\ \psi_{a,x}^{\dag}\longrightarrow 
(-1)^{x}\psi_{a,-x}^{\dag}
\label{par}
\end{equation}

\item{}Discrete axial symmetry $\Gamma$: 
\begin{equation}
A_{x}\longrightarrow A_{x+1},\ E_{x}\longrightarrow E_{x+1},\ 
\psi_{a,x}\longrightarrow \psi_{a,x+1},\ \psi_{a,x}^{\dag}\longrightarrow 
\psi_{a,x+1}^{\dag}
\label{chir}
\end{equation}

\item{}Charge conjugation C:
\begin{equation}
A_{x}\longrightarrow -A_{x+1},\  E_{x}\longrightarrow -E_{x+1},\ 
\psi_{a,x}\longrightarrow \psi^{\dag}_{a,x+1},\ \psi_{a,x}^{\dag}\longrightarrow 
\psi_{a,x+1}
\label{char}
\end{equation}

\item{}G-parity:
\begin{eqnarray}
A_{x}\longrightarrow -A_{x+1},\  E_{x}\longrightarrow -E_{x+1}\nonumber\\
\psi_{1,x}\longrightarrow \psi^{\dag}_{2,x+1},\ \psi_{1,x}^{\dag}
\longrightarrow \psi_{2,x+1}\\
\psi_{2,x}\longrightarrow -\psi^{\dag}_{1,x+1},\ \psi_{2,x}^{\dag}
\longrightarrow -\psi_{1,x+1}\ .\nonumber
\end{eqnarray}
\end{itemize}

The lattice two-flavor Schwinger model is equivalent to a one 
dimensional quantum Coulomb gas on the lattice with two kinds of particles. To see this one can fix the 
gauge, $A_{x} = A$ (Coulomb gauge). Eliminating the non-constant electric 
field and using the gauge constraint, one obtains the effective Hamiltonian
\begin{eqnarray}
H_{S}&=&H_u+H_p
\equiv\left[\frac{e^{2}_{L}}{2 N}E^{2}+\frac{e^{2}_{L}a}{2}
\sum_{x,y}\rho(x) V(x-y)\rho(y)\right]+\nonumber\\
&+&\left[
-\frac{it}{2a}\sum_{x}\sum_{a=1}^{2}(\psi_{a,x+1}^{\dag}e^{iA}\psi_{a,x}-\psi_{a,x}^{\dag}e^{-iA}
\psi_{a,x+1})\right]\ ,
\label{hs}
\end{eqnarray}
where the charge density is
\begin{equation}
\rho(x)=\psi^{\dag}_{1,x}\psi_{1,x}+\psi^{\dag}_{2,x}\psi_{2,x}-1\ \ ,
\label{cd1}
\end{equation}
and the potential
\begin{equation}
V(x-y)=\frac{1}{N}
\sum^{N-1}_{n=1} e^{i 2\pi n (x-y)/N}\frac{1}{4\sin^2\frac{\pi n}{N}}
\label{pote}
\end{equation}
is the Fourier transform of the inverse laplacian on the lattice
for non zero momentum.
The constant 
modes of the gauge field decouple in the thermodynamic limit 
$ N \longrightarrow \infty $.

\subsection{The strong coupling limit and the antiferromagnetic Heisenberg Hamiltonian}
\label{2fhm}

In the thermodynamic limit the Schwinger Hamiltonian (\ref{hs}), 
rescaled by the
 factor ${e_{L}^{2}a}/{2}$, reads 
 \begin{equation}
H=H_{0}+\epsilon H_{h}
\end{equation}
with
\begin{eqnarray}
H_{0}&=&\sum_{x>y}\left[\frac{(x-y)^{2}}{N}-(x-y)\right]\rho(x)\rho(y)\quad ,
\label{hu}\\
H_{h}&=&-i(R-L)
\label{hp}
\end{eqnarray}
and $\epsilon=t/e_{L}^{2}a^{2}$.
In Eq.(\ref{hp}) the right $R$ and left $L$ hopping operators are defined ($L=R^{\dagger}$) as
\begin{equation}
R=\sum_{x=1}^{N}R_{x}=\sum_{x=1}^{N}\sum_{a=1}^{2}R_{x}^{(a)}= 
\sum_{x=1}^{N} \sum_{a=1}^2 \psi_{a,x+1}^{\dag}e^{iA}\psi_{a,x}\quad .
\end{equation}
On a periodic chain the commutation relation 
\begin{equation}
[R,L]=0\quad 
\end{equation}
is satisfied.

We shall consider the strong coupling perturbative expansion 
where the Coulomb 
Hamiltonian (\ref{hu}) is the unperturbed Hamiltonian and the hopping 
Hamiltonian 
(\ref{hp}) the perturbation.
Due to Eq.(\ref{cd1}) every configuration with one particle per site 
has zero energy, so that the ground state of the Coulomb 
Hamiltonian (\ref{hu}) is $2^N$ times degenerate. The degeneracy of 
the ground state can be removed only at the second perturbative order 
since the first order is trivially zero. 

At the second order the 
lattice gauge theory is effectively described by the antiferromagnetic 
Heisenberg Hamiltonian. 
The vacuum energy $-$ at order $\epsilon^2$ $-$ reads  
\begin{equation}
E^{(2)}_{0}=<H_{h}^{\dagger}\frac{\Pi}{E_{0}^{(0)}-H_{0}}H_{h}>
\label{secorder}
\end{equation}
where the expectation values are defined on the degenerate subspace and 
$\Pi$ is the operator projecting on a set orthogonal to the states with one 
particle per site. 
Due to the vanishing of the charge density on the ground states of $H_{0}$, 
the commutator
\begin{equation}
[H_0, H_h]=H_h
\label{comm1}
\end{equation}
holds on any linear combination of the degenerate ground states. 
Consequently, from Eq.(\ref{secorder}) one finds
\begin{equation}
E^{(2)}_{0}=-2<RL>\quad .
\label{secorder2}
\end{equation}
On the ground state the combination $R L$ can be written in terms of the 
Heisenberg Hamiltonian.
By introducing the Schwinger spin operators
\begin{equation}
\vec{S}_{x}=\psi_{a,x}^{\dag}\frac{\vec{\sigma}_{ab}}{2}\psi_{b,x}
\end{equation}
the Heisenberg Hamiltonian $H_{J}$ reads
\begin{eqnarray}
H_{J}&=&\sum_{x=1}^{N}\left(\vec{S}_x
\cdot \vec{S}_{x+1}-\frac{1}{4}\right)=\nonumber\\ 
&=&\sum_{x=1}^{N} \left( -\frac{1}{2} 
L_{x}R_{x}-\frac{1}{4}\rho(x) \rho(x+1) \right)
\end{eqnarray}
and, on the degenerate subspace, one has 
\begin{equation}
<H_{J}>=\left<\sum_{x=1}^{N}\left(\vec{S}_x\cdot \vec{S}_{x+1}-\frac{1}{4}
\right)\right>
=\left<\sum_{x=1}^{N}\left(-\frac{1}{2}L_{x}R_{x}\right)\right>\quad .
\label{mainequation}
\end{equation}
Taking into account that products of $L_x$ and $R_y$ at different points 
have vanishing expectation values on the ground states,
and using Eq.(\ref{mainequation}), Eq.(\ref{secorder2}) reads
\begin{equation}
E^{(2)}_{0}=4<H_{J}>\quad .
\label{secorder3}
\end{equation}
The ground state of $H_{J}$ singles out the correct vacuum, on which to 
perform the perturbative expansion. 
In one dimension $H_{J}$ is exactly diagonalizable \cite{b6b,b31}. 
In the spin model a flavor 1 particle on a site 
can be represented by a spin up, a flavor 2 particle by a spin down. 
The spectrum of $H_{J}$ exhibits $2^N$ eigenstates; among these, the spin 
singlet with lowest energy is the non degenerate ground state $|g.s.>$. 

We shall construct the strong coupling perturbation theory of the 
two-flavor Schwinger model using $|g.s.>$ as the unperturbed ground state. 
$|g.s.>$ is invariant under translations by one lattice site, which amounts
to invariance under discrete chiral transformations. As a consequence,
at variance with 
the one-flavor model~\cite{b10}, chiral symmetry cannot be 
spontaneously broken even in the infinite coupling limit. 
 
$|g.s.>$ has zero charge density on each site and zero electric flux 
on each link 
\begin{equation}
\rho(x)|g.s.>=0\quad ,\quad E_{x}|g.s.>=0 \quad \quad (x=1,...,N)\quad .
\label{keyequation}
\end{equation}
$|g.s.>$ is a linear combination of all the possible states with 
$\frac{N}{2}$ spins up and $\frac{N}{2}$ spins down. The 
coefficients are not explicitly known for general $N$. In section 2.2 we
 exhibited $|g.s.>$ explicitly
for finite size systems of 4, 6 and 8 sites. 
The Heisenberg energy of $|g.s.>$ is known exactly and, in the 
thermodynamic limit, is \cite{b34,b6c}
\begin{equation}
H_{J}|g.s.>=(-N\  \ln\ 2)|g.s.>\quad .
\label{mainenergy}
\end{equation}
Eq.(\ref{mainenergy}) provides the second order correction 
Eq.(\ref{secorder3}) to the vacuum energy, $E_{g.s.}^{(2)}=-4 N\ln 2$.

There exist two kinds of excitations created from 
$|g.s.>$; one kind involves only spin flipping and has lower energy 
since no electric flux is created, the other involves fermion 
transport besides spin flipping and thus has a higher energy. For the 
latter excitations the energy is proportional to the coupling times 
the length of the electric flux: the lowest energy is achieved when the 
fermion is transported by one lattice spacing. 
Of course only the excitations of the first kind can be mapped into 
states of the Heisenberg model. 

In~\cite{b6c} the antiferromagnetic Heisenberg model excitations 
have been classified. There it was shown that any excitation may be 
regarded as 
the scattering state of quasiparticles of spin-$1/2$: every 
physical state contains an even number of quasiparticles and 
the spectrum exhibits only integer spin states. 
The two simplest excitations of lowest energy in the thermodynamic 
limit are a triplet and a singlet \cite{b6c}; they 
have a dispersion relation depending on the momenta of the two 
quasiparticles. 
For vanishing total momentum (relative to the ground state 
momentum $P_{g.s.} =0$ for $\frac{N}{2}$ even, 
$P_{g.s.} =\pi$ for $\frac{N}{2}$ odd) in the thermodynamic limit
they are degenerate with the ground state. 

In section 2.2 we showed that even for finite size systems, the excited
states can be grouped in families corresponding to the classification
given in~\cite{b6c}. We explicitly exhibited all the energy 
eigenstates for $N=4$ and $N=6$. The lowest lying are a triplet and a 
singlet, respectively; they have a well defined relative 
(to the ground state) $P$-parity and $G$-parity
- $1^{-+}$ for the triplet and $0^{++}$ for the singlet.
Since they share the same quantum numbers
these states can be identified, in the limit of vanishing fermion mass, 
with the soliton-antisoliton excitations 
found by Coleman in his analysis of the two-flavor Schwinger model.
A related analysis about the parity of the lowest lying states 
in finite size Heisenberg chains, has been given in~\cite{b70}. 

Moreover in \cite{b6c} a whole class $-$ ${\cal M}_{AF}$ $-$ of 
gapless excitations at zero momentum was singled out 
in the thermodynamic limit; these states are eigenstates of the total 
momentum and consequently have positive G-parity at 
zero momentum. The low lying states of the Schwinger model also contain 
\cite{b69} many massless excitations with positive G-parity; they are 
identified \cite{b7,b8} with the excitations belonging to 
${\cal M}_{AF}$. 
The mass of these states in the Schwinger model can be obtained
from the differences between the excitation
energies at zero momentum 
and the ground state energy. 
The energies of the states $|ex.>$ belonging to the class
${\cal M}_{AF}$ have the same 
perturbative expansion of the ground state. 
Consequently, the states $|ex.>$ 
at zero momentum up to the second order in the strong coupling 
expansion have the same energy 
of the ground state (\ref{secorder}), 
$E^{(2)}_{ex}=-4 N\ln 2$.  To this order the
mass gap is zero. 
Higher order corrections may give a mass gap. 

\subsection{The meson masses}
\label{2fmm}

In this section  we determine the masses 
for the states obtained by fermion transport
of one site on the Heisenberg model ground state.
Our analysis shows that besides the $G$-odd pseudoscalar isosinglet
$0^{--}$ with mass $m_S=e_{L}\sqrt{2/\pi}$, there are also a 
$G$-even scalar isosinglet $0^{++}$ and a pseudoscalar isotriplet $1^{-+}$
and a $G$-odd scalar isotriplet $1^{+-}$ with masses of the order of 
$m_S$ or greater.  
The quantum numbers are relative to those of the ground state
$I_{g.s.}^{PG}=0^{++}$ for $N/2$ even $I_{g.s.}^{PG}=0^{--}$ for $N/2$
odd.

Two states can be created using the spatial
component of the vector $j^{1}(x)$ Eq.(\ref{cc}) and 
isovector $j_{\alpha}^{1}(x)$ Eqs.(\ref{ca},\ref{cb}) 
Schwinger model currents. They are the G-odd pseudoscalar 
isosinglet $I^{PG}=0^{--}$ and the
G-even pseudoscalar isotriplet $I^{PG}=1^{-+}$.  The lattice operators
with the correct quantum numbers creating these states at zero momentum, 
when acting on $|g.s.>$, read 
\begin{eqnarray}
S&=&R+L=\sum_{x=1}^{N}j^{1}(x)\\
T_{+}&=&(T_{-})^{\dagger}=R^{(12)}+L^{(12)}=\sum_{x=1}^{N}j_{+}^{1}(x)
\label{ta}\\
T_{0}&=&\frac{1}{\sqrt{2}}(R^{(11)}+L^{(11)}-R^{(22)}-L^{(22)})=
\sum_{x=1}^{N}j_{3}^{1}(x) \quad .
\label{t0}
\end{eqnarray}
$R^{(ab)}$ and $L^{(ab)}$ in (\ref{ta},\ref{t0}) are the 
right and left flavor changing 
hopping operators ($L^{(ab)}=(R^{(ab)})^{\dagger}$)
$$
R^{(ab)}=\sum_{x=1}^{N}\psi_{a,x+1}^{\dagger}e^{iA}\psi_{b,x}\quad .
$$
The states are given by
\begin{eqnarray}
|S>&=&|0^{--}>=S|g.s.>\\
|T_{\pm}>&=&|1^{-+},\pm1>=T_{\pm}|g.s.>
\label{2tpm}\\
|T_{0}>&=&|1^{-+},0>=T_{0}|g.s.>\ .
\label{to}
\end{eqnarray}
They are normalized as
\begin{eqnarray}
<S|S>&=&<g.s.|S^{\dagger}S|g.s.>=-4<g.s.|H_{J}|g.s.>=4N \ln 2
\label{no1}\\
<T_{+}|T_{+}>&=&\frac{2}{3}(N+<g.s.|H_{J}|g.s.>)=\frac{2}{3}N(1-\ln 2)
\label{no2}
\end{eqnarray}
and 
\begin{equation}
<T_{0}|T_{0}>=<T_{-}|T_{-}>=<T_{+}|T_{+}> \ .
\label{no3}
\end{equation}
In Eqs.(\ref{no1},\ref{no2},\ref{no3}) $<g.s.|g.s.>=1$.

The isosinglet energy, up to the second order in the strong coupling 
expansion, is
$
E_{S}=E_{S}^{(0)}+\epsilon^{2}E_{S}^{(2)}
$ with
\begin{eqnarray}
E_{S}^{(0)}&=&\frac{<S|H_{0}|S>}{<S|S>}=1\quad ,\\
E_{S}^{(2)}&=&\frac{<S|H_{h}^{\dag}\Lambda_{S}H_{h}|S>}{<S|S>}\quad ,
\label{secordex}
\end{eqnarray}
$
\Lambda_{S}=\frac{\Pi_{S}}{E_{S}^{(0)}-H_{0}}
$
and $1-\Pi_{S}$ a projection operator onto $|S>$.
On $|g.s.>$ 
\begin{equation}
[H_{0},(\Pi_{S}H_{h})^n S]=(n+1)(\Pi_{S}H_{h})^n 
,\quad
(n=0,1,\dots),
\end{equation}
holds; Eq.(\ref{secordex}) may then be written in terms of spin correlators as
\begin{equation}
E_{S}^{(2)}=E_{g.s.}^{(2)}+4-\frac{\sum_{x=1}^{N}
<g.s.|\vec{S}_{x}\cdot \vec{S}_{x+2}-\frac{1}{4}|g.s.>}{<g.s.|H_{J}|g.s.>}
\quad .
\label{senes}
\end{equation}
One immediately recognizes that the excitation spectrum is determined 
once $<g.s.|\vec{S}_{x}\cdot \vec{S}_{x+2}|g.s.>$ is known. 
Equations similar to Eq.(\ref{senes}) may be established also at a 
generic order of the strong coupling expansion.    

At the zeroth perturbative order the pseudoscalar triplet is 
degenerate with the isosinglet $E_{T}^{(0)}=E_{S}^{(0)}=1$. 
Following the same procedure as before one may compute the energy of the 
states (\ref{2tpm}) and 
(\ref{to}) to the second order in the strong coupling expansion. To 
this order, the energy is given by
\begin{eqnarray}
&&E_{T}^{(2)}=E_{g.s.}^{(2)}-\Delta_{DS}(T)-\nonumber\\ 
&&\frac{4<g.s.|H_{J}|g.s.>+5\sum_{x=1}^{N}
<g.s.|\vec{S}_{x}\cdot \vec{S}_{x+2}-\frac{1}{4}|g.s.>}{N+<g.s.|H_{J}|g.s.>}
\label{senet}
\end{eqnarray}
where in terms of the vector operator 
$
\vec{V}=\sum_{x=1}^{N}\vec{S}_{x}\wedge \vec{S}_{x+1}
$, one can write $\Delta_{DS}(T)$ as
\begin{eqnarray}
\Delta_{DS}(T_{\pm})&=&12
\frac{<g.s.|(V_{1})^2|g.s.>+<g.s.|(V_{2})^2|g.s.>}
{N+<g.s.|H_{J}|g.s.>}\\
\Delta_{DS}(T_{0})&=&12
\frac{2<g.s.|(V_{3})^2|g.s.>}{N+<g.s.|H_{J}|g.s.>}\ .
\end{eqnarray}
The VEV of each squared component of $\vec{V}$ on 
the rotationally invariant singlet $|g.s.>$ give the same 
contribution $i.e.$ $\Delta_{DS}(T_{\pm})=\Delta_{DS}(T_{0})$: the 
triplet states (as in the continuum theory) have a degenerate mass gap. 
This is easily verified by direct computation on finite size systems; 
when the size of the system is finite one may also show that $\Delta_{DS}$ 
is of zeroth order in N.

The excitation masses are given by 
$m_{S}=\frac{e_{L}^2 a}{2}(E_{S}-E_{g.s.})$ and 
$m_{T}=\frac{e_{L}^2 a}{2}(E_{T}-E_{g.s.})$. 
Consequently, the ($N$-dependent) ground state energy terms 
appearing in $E_{S}^{(2)}$ and $E_{T}^{(2)}$ cancel and what is left 
are only $N$ independent terms.
This is a good check of our computation, being the mass an intensive quantity.
 
In principle one should expect also excitations created acting 
on $|g.s.>$ with the chiral currents, in analogy with the one flavor 
Schwinger model where, as shown in ref.~\cite{b58}, the chiral current 
creates a two-meson bound state. The chiral currents operators for the two 
flavor Schwinger model are given by
\begin{eqnarray}
j^{5}(x)&=&\overline{\psi}(x)\gamma^{5}\psi(x)\\
j^{5}_{\alpha}(x)&=&\overline{\psi}_{a}(x)\gamma^{5}
(\frac{\sigma}{2})_{ab}\psi_{b}(x)\quad .
\end{eqnarray}
The corresponding lattice operators at zero momentum are 
\begin{eqnarray}
S^{5}&=R-L=\sum_{x=1}^{N}j^{5}(x)
\label{s5}\\
T_{+}^{5}&=(T_{-}^{5})^{\dagger}=R^{(12)}-L^{(12)}=
\sum_{x=1}^{N}j_{+}^{5}(x)
\label{ta2}\\
T_{0}^{5}&=\frac{1}{\sqrt{2}}(R^{(11)}-L^{(11)}-R^{(22)}+L^{(22)})=
\sum_{x=1}^{N}j_{3}^{5}(x)\quad .
\label{t02}
\end{eqnarray}
The states created by (\ref{s5},\ref{ta2},\ref{t02}) when acting on $|g.s.>$,
are
\begin{eqnarray}
|S^{5}>&=&|0^{++}>=S^{5}|g.s.>\\
|T_{\pm}^{5}>&=&|1^{+-},\pm1>=T_{\pm}^{5}|g.s.>
\label{tpm2}\\
|T_{0}^{5}>&=&|1^{+-},0>=T_{0}^{5}|g.s.>\quad .
\label{to2}
\end{eqnarray}
They are normalized as
\begin{eqnarray}
<S^{5}|S^{5}>&=&<g.s.|S^{5\dagger}S^{5}|g.s.>=-4<g.s.|H_{J}|g.s.>=4N\log 2\\
<T^{5}_{+}|T^{5}_{+}>&=&\frac{2}{3}(N+<g.s.|H_{J}|g.s.>)=\frac{2}{3}
N(1-\log 2)
\end{eqnarray}
and 
\begin{equation}
<T^{5}_{0}|T^{5}_{0}>=<T^{5}_{-}|T^{5}_{-}>=<T^{5}_{+}|T^{5}_{+}>\quad .
\end{equation}

Following the computational scheme used to study $|S>$ and $|T>$, 
one finds for the state $|S^{5}>$
\begin{eqnarray}
E_{S^5}^{(0)}&=&1\\
E_{S^5}^{(2)}&=&E_{g.s.}^{(2)}+12-3\frac{\sum_{x=1}^{N}
<g.s.|\vec{S}_{x}\cdot \vec{S}_{x+2}-\frac{1}{4}|g.s.>}{<g.s.|H_{J}|g.s.>}\quad .
\label{senes5}
\end{eqnarray}
For the triplet $|T^{5}>$ one gets
\begin{eqnarray}
E_{T^5}^{(0)}&=&1\\
E_{T^5}^{(2)}&=&E_{g.s.}^{(2)}+\frac{\sum_{x=1}^{N}<g.s.|\vec{S}_{x}
\cdot\vec{S}_{x+2}-
\frac{1}{4}|g.s.>-4<g.s.|H_{J}|g.s.>}{N+<g.s.|H_{J}|g.s.>}\quad .
\label{senet5}
\end{eqnarray} 

Now we can compute the mass spectrum up to the second order in the 
strong coupling expansion. 
Using Eq.(\ref{corrd2n}), the isosinglet mass reads as
\begin{equation}
\frac{m_{S}}{e^2 a}=\frac{1}{2}+1.9509 \quad \epsilon ^2\quad .
\label{ms} 
\end{equation}
For what concerns the isotriplet mass, since the double sum in 
Eq.(\ref{senet}) is given by
\begin{equation}
\Delta_{DS}(T)=8\frac{<g.s.|\vec{V}\cdot \vec{V}|g.s.>}
{N+<g.s.|H_{J}|g.s.>}\quad ,
\end{equation}
using Eq.(\ref{vvapp}), one gets
\begin{equation}
\frac{m_{T}}{e_{L}^2 a}=\frac{1}{2}+0.0972\quad \epsilon ^2\quad .
\label{mtt} 
\end{equation}

The existence of massive isotriplets was already noticed in \cite{b69}, 
and their mass in the continuum theory was numerically computed 
for various values of the fermion mass. 
In particular there is a $G$-parity even isotriplet with mass 
approximately equal to the mass of the isosinglet $0^{--}$.  

The masses of the $|S_{5}>$ isosinglet and the $|T_{5}>$ isotriplet are
\begin{eqnarray}
\frac{m_{S^5}}{e^2a}&=&\frac{1}{2}+5.85 \epsilon^2 
\label{ms5}\\
\frac{m_{T^5}}{e^2a}&=&\frac{1}{2}+4.4069\epsilon^2\ . 
\label{mt5}
\end{eqnarray}

Equations (\ref{ms}), (\ref{mtt}), (\ref{ms5}) and (\ref{mt5}) 
provide the values of $m_{S}$, $m_{T}$, $m_{S^5}$ and 
$m_{T^5}$ for small values of 
$z=\epsilon^2=\frac{t^2}{e_{L}^4a^4}$ up to the second order 
in the strong coupling expansion. 
Whereas (\ref{mtt}) is only approximate (\ref{ms}), (\ref{ms5}) and (\ref{mt5}) 
are exact at the second order in the $\epsilon$ expansion.
We extrapolated these masses to the continuum limit using the 
standard technique of the Pad\'e approximants~\cite{b8} and we got results in good agreement with the continuum.  

\section{The multiflavor lattice Schwinger models}

We now study the ${\cal N}$-flavor lattice Schwinger models in the hamiltonian formalism using staggered fermions. 
The ${\cal N}$-flavor Schwinger models have many features in in common with four dimensional $QCD$: at the classical level they have a 
symmetry group $U_{L}({\cal N})\otimes U_{R}({\cal N})= SU_{L}({\cal N})\otimes SU_{R}({\cal N}) \otimes U_{V}(1)\otimes U_{A}(1)$ that is broken down 
to $SU_{L}({\cal N})\otimes SU_{R}({\cal N}) \otimes U_{V}(1)$ by the axial anomaly exactly like in $QCD$~\cite{b74}. The massless ${\cal N}$-flavor 
Schwinger models describe no real interactions between their particles as one can infer by writing the model action in a bosonized form. 
The model exhibits one massive and ${\cal N}^{2} -1$ massless pseudoscalar ``mesons"~\cite{b75}. 

On the lattice we shall prove that $-$ at the second order in the strong coupling expansion $-$ 
the lattice Schwinger models are effectively described by $SU({\cal N})$ quantum 
antiferromagnetic spin-$1/2$ Heisenberg Hamiltonians with spins in a particular fundamental representation of the $SU({\cal N})$ 
Lie algebra and with nearest neighbours couplings. The features of the model are very different depending on if ${\cal N}$ is odd or even.
 When ${\cal N}$ is odd, the ground state energy in the strong coupling limit 
is proportional to $e_{L}^2$, the square of the electromagnetic coupling constant. In contrast, when ${\cal N}$ is even the ground state energy 
in the strong coupling limit is of order 1. This difference arises from the proper definition of the charge density
\begin{equation}
\rho(x)=\sum_{a=1}^{\cal N} \psi_{a,x}^{\dagger}\psi_{a,x}-\frac{\cal N}{2}
\label{ron}
\end{equation}
where the constant ${\cal N}/2$ has been subtracted from the charge density operator in order to make it odd under 
the charge conjugation transformation. 
As a consequence, when ${\cal N}$ is even, $\rho(x)$ admits zero eigenvalues and the ground state does 
not support any electric flux, while when ${\cal N}$ is odd the ground 
state exhibits a staggered configuration of the charge density and electromagnetic fluxes. 

In the continuum the Coleman theorem~\cite{b39}  prevents the formation of either an isoscalar chiral condensate 
$<\overline{\psi}\psi>$ or an isovector chiral condensate $<\overline{\psi}T^{a}\psi>$ $-$ where $T^{a}$ is an $SU({\cal N})$ generator $-$
 for every model with an internal $SU({\cal N})$-flavor symmetry. This feature is reproduced on the lattice also 
 for this class of models~\cite{b9}.

\subsection{The continuum  ${\cal N}$-flavor Schwinger models}
\label{cnsm}

The continuum $SU({\cal N})$-flavor Schwinger models are defined by the action 
\begin{equation}
S = \int d^{2} x(\sum_{a=1}^{\cal N} \overline{\psi}_{a}(i\gamma_{\mu}\partial^{\mu}+\gamma_{\mu}
A^{\mu})\psi_{a}-\frac{1}{4e^{2}_{c}}F_{\mu\nu}F^{\mu\nu})
\label{na1n}
\end{equation}
where the ${\cal N}$ fermions have been introduced in a completely symmetric way. Although the theory described by (\ref{na1n}) strictly parallels 
what has been shown in the previous section for the $SU(2)$ model, we shall now report a detailed analysis both for the sake of clarity and 
to show that some difference appears between ${\cal N}$ even and odd. 

The Dirac fields are an ${\cal N}$-plet, $i.e.$ transform according to the fundamental representation of 
the flavor group while the 
electromagnetic field is an $SU({\cal N})$ singlet. The flavor symmetry of the theory cannot be spontaneously broken for the same reasons as 
in the $SU(2)$ case. 
The particles of the theory belong to $SU({\cal N})$ multiplets. The action is invariant under the symmetry 
\beq
SU_{L}({\cal N})\otimes SU_{R}({\cal N})\otimes U_{V}(1) \otimes U_{A}(1) \nonumber
\eeq
The symmetry generators act as follows
\begin{eqnarray}
SU_{L}({\cal N}) &:& \psi_{a}(x)\longrightarrow (e^{i\theta_{\alpha}T^{\alpha}P_{L}})_{ab}\ \psi_{b}(x)\   , \  
\overline{\psi_{a}}(x)\longrightarrow \overline{\psi_{b}}\ (x)(e^{-i\theta_{\alpha}T^{\alpha}P_{R}})_{ba} \\
SU_{R}({\cal N}) &:& \psi_{a}(x)\longrightarrow (e^{i\theta_{\alpha}T^{\alpha}P_{R}})_{ab}\ \psi_{b}(x)\   ,\   
\overline{\psi_{a}}(x)\longrightarrow \overline{\psi_{b}}(x)\ (e^{-i\theta_{\alpha}T^{\alpha}P_{L}})_{ba} \\
U_{V}(1) &:& \psi_{a}(x)\ \longrightarrow (e^{i\theta(x){\bf 1}})_{ab}\ \psi_{b}(x)\  ,\  
\psi_{a}^{\dagger}(x)\longrightarrow \psi_{b}^{\dagger}(x)\ (e^{-i\theta(x){\bf 1}})_{ba} \\
U_{A}(1) &:& \psi_{a}(x)\longrightarrow (e^{i\alpha \gamma_{5}{\bf 1}})_{ab}\ \psi_{b}(x)\  ,\  
\psi_{a}^{\dagger}(x)\longrightarrow \psi_{b}^{\dagger}(x)\ (e^{-i\alpha \gamma_{5}{\bf 1}})_{ba} 
\end{eqnarray}
where $T^{\alpha}$ are the generators of the $SU({\cal N})$ group, $\theta_{\alpha}$, $\theta(x)$ and $\alpha$ are real coefficients and
\begin{equation}
P_{L}=\frac{1}{2}(1-\gamma_{5})\ ,\ P_{R}=\frac{1}{2}(1+\gamma_{5})\quad .
\end{equation}
At the classical level the above symmetries lead to conservation laws for the isovector, vector and axial currents
\begin{eqnarray}
j_{\alpha}^{\mu}(x)_{R}&=&\overline{\psi}_{a}(x)\gamma^{\mu}P_{R}(T_{\alpha})_{ab}\psi_{b}(x)\quad ,
\label{can}\\
j_{\alpha}^{\mu}(x)_{L}&=&\overline{\psi}_{a}(x)\gamma^{\mu}P_{L}(T_{\alpha})_{ab}\psi_{b}(x)\quad ,
\label{cbn}\\
j^{\mu}(x)&=&\overline{\psi}_{a}(x)\gamma^{\mu}{\bf 1}_{ab}\psi_{b}(x)\quad ,
\label{ccn}\\
j^{\mu}_{5}(x)&=&\overline{\psi}_{a}(x)\gamma^{\mu}\gamma ^{5}{\bf 1}_{ab}\psi_{b}(x)\quad .
\end{eqnarray}
At the quantum level the vector and axial currents cannot be simultaneously conserved. If the regularization is gauge invariant, so that 
the vector current is conserved, then the axial current acquires the anomaly which breaks the symmetry $U_{A}(1)$~\cite{b74} 
\beq
\partial_{\mu} j_5^{\mu}(x)={\cal N}\frac{e_{c}^{2}}{2\pi}\epsilon_{\mu \nu}F^{\mu \nu}(x)\quad .
\label{anomalyn}
\eeq
The isoscalar $<\overline{\psi}\psi>$ and isovector 
$<\overline{\psi}T^{\alpha}\psi>$ chiral condensates are zero due to the Coleman theorem~\cite{b39}, in fact they would break not only the $U_A(1)$ symmetry 
of the action but also the continuum internal symmetry $SU_L({\cal N})\otimes SU_R({\cal N})$ down to $SU_V({\cal N})$. 
There is an order parameter just for the breaking of the $U_A(1)$ symmetry~\cite{b57,b67}, the operator
\begin{equation} 
<\overline{\psi}_{L}^{({\cal N})}\ldots \overline{\psi}_{L}^{(1)}\psi_{R}^{(1)}\ldots 
\psi_{R}^{({\cal N})}>=(\frac{e^{\gamma}}{4\pi})^{\cal N} (\sqrt{\frac{\cal N}{\pi}} e_{c})^{\cal N}\quad .
\label{chicu}
\end{equation}
Under a discrete chiral rotation
\begin{equation}
\psi_{L}\rightarrow \gamma_{5} \psi_{L}=-\psi_{L}\quad ,\quad \psi_{R}\rightarrow \gamma_{5} \psi_{R}=\psi_{R}
\label{dga5}
\end{equation}
the operator (\ref{chicu}) of course transforms as
\begin{equation}
 <\overline{\psi}_{L}^{({\cal N})}\ldots \overline{\psi}_{L}^{(1)}\psi_{R}^{(1)}\ldots 
\psi_{R}^{({\cal N})}>\rightarrow (-1)^{{\cal N}} <\overline{\psi}_{L}^{({\cal N})}\ldots \overline{\psi}_{L}^{(1)}\psi_{R}^{(1)}\ldots 
\psi_{R}^{({\cal N})}>
\end{equation}
The operator (\ref{chicu}) is even under (\ref{dga5}) when ${\cal N}$ is even and this implies that notwithstanding the fact that the continuous chiral 
rotations $U_{A}(1)$ are broken by the non-zero VEV (\ref{chicu}), the discrete chiral symmetry (\ref{dga5}) is unbroken. 
When ${\cal N}$ is odd also the discrete chiral symmetry (\ref{dga5}) is broken by the non-zero VEV (\ref{chicu}). 

The usual abelian bosonization procedure may again be applied provided that ${\cal N}$ Bose fields are introduced~\cite{b74,b75,b76}
\beq
:\overline{\psi}_{a}\gamma^{\mu}\psi_{a}:=\frac{1}{\sqrt{\pi}}\epsilon^{\mu \nu}\partial_{\nu}\Phi_{a}\  ,\  a=1,\ldots {\cal N}
\eeq
The electric charge density and the action read
\beq
j_{0}=:\sum_{a=1}^{\cal N}\psi^{\dagger}_{a}\psi_{a}:=\frac{1}{\sqrt{\pi}}\partial_{x}(\sum_{a=1}^{\cal N}\Phi_{a})\quad ,
\label{chden}
\eeq
\beq
S=\int d^{2}x (\frac{1}{2}\sum_{a=1}^{\cal N}\partial_{\mu}\Phi_{a}\partial^{\mu}\Phi_{a}+
\frac{e_{c}^{2}}{2\pi}(\sum_{a=1}^{\cal N}\Phi_{a})^{2})
\label{na2n}
\eeq
The mass matrix is determined by the last term in Eq.(\ref{na2n}) and must be diagonalized. 
The field degrees of freedom span the vector space on which the mass matrix is defined. 
The action must be expressed in terms of an orthonormal basis of field vectors, in order to have a properly normalised kinetic energy term. 
The original $\Phi^{a}$ in Eq.(\ref{na2n})  are orthonormal basis vectors, but they are not eigenvectors of the mass matrix. The mass matrix has one 
non-zero eigenvalue $\frac{e_{c}^{2}}{{\cal N}\pi}$  with associated eigenvector $\frac{1}{\sqrt{\cal N}}\sum_{a=1}^{\cal N}\Phi^{a}$ 
and all the other eigenvalues are zero. The remaining eigenvectors can be made orthonormal by the following change of variables
\begin{equation}
\tilde{\Phi}^a=O^{a}_{b}\Phi^{b}
\end{equation}
where the orthogonal matrices $O^{a}_{b}$ are~\cite{b76}
\begin{eqnarray}
O_{b}^{1}&=&\frac{1}{\sqrt{\cal N}}(1,1,\ldots,1)\quad ,\\
O_{b}^{2}&=&\frac{1}{\sqrt{{\cal N}({\cal N}-1)}}(1,1,\ldots, -{\cal N}+1)\quad ,\\
O_{b}^{3}&=&\frac{1}{\sqrt{({\cal N}-1)({\cal N}-2)}}(1,1,\ldots,-{\cal N}+2,0)\quad ,\\
&\vdots&\nonumber\\
O_{b}^{\cal N}&=&\frac{1}{\sqrt{2}}(1,-1,0,\ldots,0)\quad .
\end{eqnarray}
In terms of these new fields $\tilde{\Phi}^{a}$ the action (\ref{na2n}) reads
\begin{equation}
S=\int d^2x(\frac{1}{2}\sum_{a=1}^{\cal N} \partial_{\mu}\tilde{\Phi}_{a}\partial^{\mu}\tilde{\Phi}^{a}-\frac{1}{2}\mu^2 (\tilde{\Phi}^{1})^2)
\label{na3n}
\end{equation}
where $\mu^{2}={\cal N}\frac{e_{c}^2}{\pi}$. 
The action (\ref{na3n}) describes ${\cal N}$ non interacting fields, one massive and ${\cal N}-1$ massless. 
The multiflavor Schwinger model can also be studied in the framework of non abelian bosonization~\cite{b68bis}, where the relationship 
between isovector currents and bosonic excitations appears in a more symmetrical form~\cite{b68tris}.  

The Hamiltonian, gauge constraint and non-vanishing (anti-)commutators
of the continuum ${\cal N}$-flavor Schwinger models are
\begin{eqnarray}
H=\int dx[\frac{e^2}{2}E^2(x)+\sum_{a=1}^{\cal N}&
\psi^{\dagger}_a (x)\alpha\left(i\partial_x +eA(x)\right)\psi_a
(x)]\label{ham1n}\\ \partial_x E(x)\ +&\sum_{a=1}^{\cal N} \psi^{\dagger}_a
(x)\psi_a (x)\sim 0\label{ga1n}\\
\left[ A(x),E(y)\right]=i\delta(x-y) ~,&
 \left\{\psi_a(x),\psi_b^{\dagger}(y)\right\}=\delta_{ab}\delta(x-y)
 \label{commu1n}
\end{eqnarray}

\subsection{The lattice ${\cal N}$-flavor Schwinger models}

On the lattice the Hamiltonian, constraint and (anti-) commutators reducing to (\ref{ham1n},\ref{ga1n},\ref{commu1n}) 
in the naive continuum limit are 
\begin{eqnarray}
H_{S}=\frac{e_{L}^{2}a}{2}\sum_{x=1}^N E_{x}^{2}&-&\frac{it}{2a}\sum_{x=1}^N
\sum_{a=1}^{\cal N} (\psi_{a,x+1}^{\dag}e^{iA_{x}}\psi_{a,x}
-\psi_{a,x}^{\dag}e^{-iA_{x}}\psi_{a,x+1})\label{hamiltonn}\nonumber\\
E_{x}-E_{x-1}&+&\sum_{a=1}^{\cal N}\psi_{a,x}^{\dag}\psi_{a,x}-\frac{\cal N}{2}\sim
0\ ,
\label{gaussn}\\
\left[ A_x,E_y\right]=i\delta_{x,y}~&,&
\left\{\psi_{a,x},\psi_{b,y}^{\dagger}\right\}=\delta_{ab}\delta_{xy}
\nonumber
\end{eqnarray}
The fermion fields are defined on the sites, $x=1,\ldots, N$, gauge and the electric fields, $ A_{ x}$ and
 $E_{x}$,  on the links $[x; x + 1]$, $N$ is an even integer 
and, when $N$ is finite it is convenient to impose periodic boundary conditions.  When $N$ is finite, the continuum limit is the 
${\cal N}$-flavor Schwinger model on a circle~\cite{b62}.
The coefficient $t$ of the hopping term in (\ref{hamiltonn})
plays the role of the lattice light speed. In the naive continuum limit,
$e_L=e_c$ and $t=1$. 
 
The lattice ${\cal N}$-flavor Schwinger model is equivalent to a one 
dimensional quantum Coulomb gas on the lattice with ${\cal N}$ kinds of particles. To see this one can fix the 
gauge, $A_{x} = A$ (Coulomb gauge). Eliminating the non-constant electric 
field and using the gauge constraint, one obtains the effective Hamiltonian
\begin{eqnarray}
H_{S}&=&H_u+H_p
\equiv\left[\frac{e^{2}_{L}}{2 N}E^{2}+\frac{e^{2}_{L}a}{2}
\sum_{x,y}\rho(x) V(x-y)\rho(y)\right]+\nonumber\\
&+&\left[
-\frac{it}{2a}\sum_{x}\sum_{a=1}^{\cal N}(\psi_{a,x+1}^{\dag}e^{iA}\psi_{a,x}-\psi_{a,x}^{\dag}e^{-iA}
\psi_{a,x+1})\right]\ ,
\label{hsnn}
\end{eqnarray}
where $\rho(x)$ is given in Eq.(\ref{ron}) and the Coulomb potential $V(x-y)$ is given in Eq.(\ref{pote}). 
The constant electric field is normalized so that $[ A, E ] = i$ . 
The constant 
modes of the gauge field decouple in the thermodynamic limit 
$ N \rightarrow \infty$.
In the thermodynamic limit the Schwinger Hamiltonian (\ref{hsnn}), rescaled by the
 factor ${e_{L}^{2}a}/{2}$, reads 
 \begin{equation}
H=H_{0}+\epsilon H_{h}
\end{equation}
with
\begin{eqnarray}
H_{0}&=&\sum_{x>y}[\frac{(x-y)^{2}}{N}-(x-y)]\rho(x)\rho(y)\quad ,
\label{hunn}\\
H_{h}&=&-i(R-L)
\label{hpn}
\end{eqnarray}
and $\epsilon=t/e_{L}^{2}a^{2}$.
In Eq.(\ref{hpn}) the right $R$ and left $L$ hopping operators are defined ($L=R^{\dagger}$) as
\begin{equation}
R=\sum_{x=1}^{N}R_{x}=\sum_{x=1}^{N}\sum_{a=1}^{\cal N}R_{x}^{(a)}= 
\sum_{x=1}^{N} \sum_{a=1}^{\cal N} \psi_{a,x+1}^{\dag}e^{iA}\psi_{a,x}\quad .
\end{equation}
On a periodic chain the commutation relation 
\begin{equation}
[R,L]=0\quad 
\end{equation}
is satisfied.

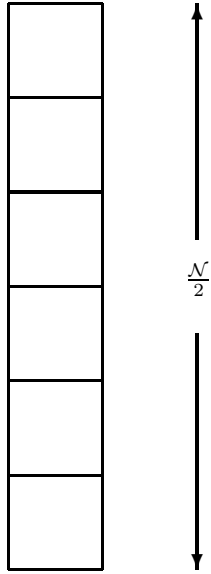
\begin{figure}[htb]
\begin{center}
\setlength{\unitlength}{0.00041700in}%
\begingroup\makeatletter\ifx\SetFigFont\undefined
\def\x#1#2#3#4#5#6#7\relax{\def\x{#1#2#3#4#5#6}}%
\expandafter\x\fmtname xxxxxx\relax \def\y{splain}%
\ifx\x\y   
\gdef\SetFigFont#1#2#3{%
  \ifnum #1<17\tiny\else \ifnum #1<20\small\else
  \ifnum #1<24\normalsize\else \ifnum #1<29\large\else
  \ifnum #1<34\Large\else \ifnum #1<41\LARGE\else
     \huge\fi\fi\fi\fi\fi\fi
  \csname #3\endcsname}%
\else
\gdef\SetFigFont#1#2#3{\begingroup
  \count@#1\relax \ifnum 25<\count@\count@25\fi
  \def\x{\endgroup\@setsize\SetFigFont{#2pt}}%
  \expandafter\x
    \csname \romannumeral\the\count@ pt\expandafter\endcsname
    \csname @\romannumeral\the\count@ pt\endcsname
  \csname #3\endcsname}%
\fi
\fi\endgroup
\begin{picture}(2424,7224)(4789,-6973)
\thicklines
\put(4801,-961){\line( 1, 0){1200}}
\put(4801,-2161){\line( 1, 0){1200}}
\put(4801,-3361){\line( 1, 0){1200}}
\put(4801,-4561){\line( 1, 0){1200}}
\put(4801,-5761){\line( 1, 0){1200}}
\put(4801,239){\line( 1, 0){1200}}
\put(4801,-6961){\line( 1, 0){1200}}
\put(7051,-3361){\makebox(0,0)[lb]{\smash{\SetFigFont{10}{12.0}{rm}$\frac{\cal N}{2}$}}}
\put(4801,239){\line( 0,-1){7200}}
\put(6001,239){\line( 0,-1){7200}}
\put(7201,-2761){\vector( 0, 1){3000}}
\put(7201,-3961){\vector( 0,-1){3000}}
\end{picture}
\end{center}
\caption{The representation of $SU({\cal N})$ at each site when ${\cal N}$ is even}
\label{ne}
\end{figure}

When ${\cal N}$ is even the ground state of the Hamiltonian (\ref{hunn}) is the state $|g.s.>$ with $\rho(x)=0$ on every site, $i.e.$
 with every site half-filled
\begin{equation}
\sum_{a=1}^{\cal N} \psi_{ax}^{\dagger}\psi_{ax}|g.s.>=\frac{\cal N}{2}|g.s.>\quad .
\end{equation}
It is easy to understand that $\rho(x)=0$ on every site in the ground state by observing that the Coulomb Hamiltonian 
(\ref{hunn}) is a non-negative operator and that the states with zero charge density are zero eigenvalues of (\ref{hunn}).  
$|g.s.>$ is an highly degenerate state; in fact at each site $x$ the quantum configuration is 
\begin{equation}
\prod_{a=1}^{\frac{\cal N}{2}}\psi_{ax}^{\dagger}|0>\quad .
\label{qcnp}
\end{equation}
The state (\ref{qcnp}) is antisymmetric in the indices $a=1,\ldots,\frac{\cal N}{2}$; $i.e.$ it takes on any 
orientation of the vector in the representation of the flavor symmetry group $SU({\cal N})$ with Young tableau given in fig.(\ref{ne}).  
The energy of $|g.s.>$ is of order 1, since it is non zero only at the second order in the strong coupling 
expansion.

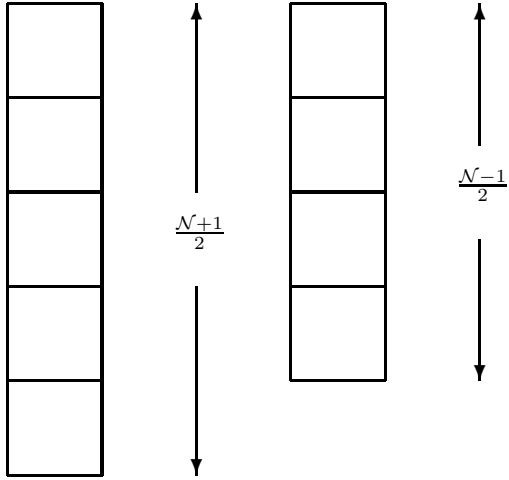
\begin{figure}[htb]
\begin{center}
\setlength{\unitlength}{0.00041700in}%
\begingroup\makeatletter\ifx\SetFigFont\undefined
\def\x#1#2#3#4#5#6#7\relax{\def\x{#1#2#3#4#5#6}}%
\expandafter\x\fmtname xxxxxx\relax \def\y{splain}%
\ifx\x\y   
\gdef\SetFigFont#1#2#3{%
  \ifnum #1<17\tiny\else \ifnum #1<20\small\else
  \ifnum #1<24\normalsize\else \ifnum #1<29\large\else
  \ifnum #1<34\Large\else \ifnum #1<41\LARGE\else
     \huge\fi\fi\fi\fi\fi\fi
  \csname #3\endcsname}%
\else
\gdef\SetFigFont#1#2#3{\begingroup
  \count@#1\relax \ifnum 25<\count@\count@25\fi
  \def\x{\endgroup\@setsize\SetFigFont{#2pt}}%
  \expandafter\x
    \csname \romannumeral\the\count@ pt\expandafter\endcsname
    \csname @\romannumeral\the\count@ pt\endcsname
  \csname #3\endcsname}%
\fi
\fi\endgroup
\begin{picture}(6024,6024)(3589,-5773)
\thicklines
\put(3601,-961){\line( 1, 0){1200}}
\put(3601,-2161){\line( 1, 0){1200}}
\put(3601,-3361){\line( 1, 0){1200}}
\put(3601,-4561){\line( 1, 0){1200}}
\put(3601,-5761){\line( 1, 0){1200}}
\put(3601,239){\line( 0,-1){6000}}
\put(4801,239){\line( 0,-1){6000}}
\put(6001,-2161){\vector( 0, 1){2400}}
\put(6001,-3361){\vector( 0,-1){2400}}
\put(3601,239){\line( 1, 0){1200}}
\put(7201,239){\line( 1, 0){1200}}
\put(9301,-2161){\makebox(0,0)[lb]{\smash{\SetFigFont{10}{12.0}{rm}$\frac{{\cal N}-1}{2}$}}}
\put(7201,-961){\line( 1, 0){1200}}
\put(7201,-2161){\line( 1, 0){1200}}
\put(7201,-3361){\line( 1, 0){1200}}
\put(7201,-4561){\line( 1, 0){1200}}
\put(7201,239){\line( 0,-1){4800}}
\put(8401,239){\line( 0,-1){4800}}
\put(9601,-1561){\vector( 0, 1){1800}}
\put(9601,-2761){\vector( 0,-1){1800}}
\put(5701,-2761){\makebox(0,0)[lb]{\smash{\SetFigFont{10}{12.0}{rm}$\frac{{\cal N}+1}{2}$}}}
\end{picture}
\end{center}
\caption{The representation of $SU({\cal N})$ at each site of the even sublattice and odd sublattice when ${\cal N}$ is odd}
\label{no}
\end{figure}

When ${\cal N}$ is odd the ground states of the Hamiltonian (\ref{hunn}) are characterized by the staggered charge distribution
\begin{equation}
\rho(x)=\pm \frac{1}{2}(-1)^{x}
\label{gsron}
\end{equation}
since (\ref{gsron}) minimizes the Coulomb Hamiltonian (\ref{hunn}); one can have $\rho(x)=+1/2$ on the even sublattice and $\rho(x)=-1/2$ on the odd sublattce 
or viceversa. 
The electric fields generated by the charge distribution (\ref{gsron}) are
 \begin{equation}
E_{x}=\pm \frac{1}{4}  (-1)^{x}
\label{exn}
\end{equation}    
Since now
\begin{equation}
H_{0}|g.s.>=\frac{1}{16}|g.s.>
\end{equation}
the ground state energy is of order $e_{L}^{2}$. 
The states $|g.s.>$ are highly degenerate since they can take up any orientation in the vector space which carries the representation of the 
$SU({\cal N})$ group with the Young tableaux given in fig.(\ref{no}).   

Either when ${\cal N}$ is even or when ${\cal N}$ is odd the ground state degeneracy is resolved at the second order in the strong coupling 
expansion. First order perturbations to the vacuum energy vanish. The vacuum energy at order $\epsilon ^{2}$ reads
\begin{equation}
E_{0}^{(2)}=<H_{h}^{\dagger}\frac{\Pi}{E_{0}^{(0)}-H_{0}}H_{h}>
\label{2vev}
\end{equation}
where the expectation values are defined on the degenerate subspace of ground states and $\Pi$ is a projection operator projecting 
orthogonal to the states of the degenerate subspace. 
Due to the commutation relation
\begin{equation}
\left[H_{0},H_{h}\right]=\frac{N-1}{N}H_{h}-2\sum_{x,y}\left[V(x-y)-V(x-y-1)\right](L_{y}+R_{y})\rho(x)
\end{equation}
Eq.(\ref{2vev}) can be rewritten as
\begin{equation}
E_{0}^{(2)}=-2<RL>\quad .
\label{22vev}
\end{equation}
On the ground state the combination $RL$ can be written in terms of the Heisenberg Hamiltonian  of a generalized $SU({\cal N})$ 
antiferromagnet. By introducing as in the previous section the Schwinger spin operators
\begin{equation}
\vec{S}_{x}=\psi_{ax}^{\dagger}T_{ab}^{\alpha}\psi_{bx}
\end{equation}
where $T^{\alpha}$ are now the generators of the $SU({\cal N})$ group,  the $SU({\cal N})$ Heisenberg Hamiltonian reads
\begin{equation}
H_{J}=\sum_{x=1}^{N}(~\vec{S}_{x}\cdot \vec{S}_{x+1}-\frac{\cal N}{8}+\frac{1}{2\cal N}\rho(x)\rho(x+1)~)=-\frac{1}{2}\sum_{x=1}^{N}L_{x}R_{x}
\label{nhj}
\end{equation}
When ${\cal N}$ is even, on the degenerate ground states one has
\begin{equation}
<H_{J}>=<\sum_{x=1}^{N}(\vec{S}_{x}\cdot \vec{S}_{x+1}-\frac{\cal N}{4})>=<-\frac{1}{2}\sum_{x=1}^{N}L_{x}R_{x}>
\label{nhje}
\end{equation}
while when ${\cal N}$ is odd one has
\begin{equation}
<H_{J}>=<\sum_{x=1}^{N}(\vec{S}_{x}\cdot \vec{S}_{x+1}-\frac{{\cal N}^{2}+1}{8\cal N})>=<-\frac{1}{2}\sum_{x=1}^{N}L_{x}R_{x}>\quad .
\label{nhjo}
\end{equation}
Taking into account that the products of $L_{x}$ and $R_{y}$ at different points have vanishing expectation values on the ground states and using 
Eq.(\ref{nhje}) or Eq.(\ref{nhjo}), Eq.(\ref{22vev}) reads
\begin{equation}
 E_{0}^{(2)}=4<H_{J}>\quad .
\end{equation}
The problem of determining the correct ground state, on which to perform the perturbative expansion, is then reduced again to the diagonalization of the 
$SU({\cal  N})$ Heisenberg spin-$1/2$ Hamiltonian  (\ref{nhj}). As we already pointed out in section (2.4), generalized $SU({\cal N})$ antiferromagnetic 
chains have not been yet analysed in the literature in such a detailed way as the $SU(2)$ chains. Consequentely, 
the study  of the lattice $SU({\cal N})$ flavor lattice Schwinger models become extremely complicated for ${\cal N}>2$. 
Nonetheless, the computational scheme that we developed for the $U(1)$- and $SU(2)$-flavor models~\cite{b10,b8,b9}, should 
work for a generic $SU({\cal N})$-flavor model. 

The ground state of the gauge models is very different depending on if ${\cal N}$ is even or odd. When ${\cal N}$ is even, the ground state 
$|G.S.>$ of 
the spin Hamiltonian (\ref{nhj}) is non-degenerate and translationally invariant, and since it is the ground state of the gauge model in the infinite coupling 
limit, there is 
no spontaneous breaking of the chiral symmetry for any $SU(2{\cal N})$-flavor lattice Schwinger model. 
In contrast, when ${\cal N}$ is odd, the ground state $|G.S.>$ of 
the spin Hamiltonian (\ref{nhj}) is degenerate of order two and is not translationally invariant and consequently any $SU(2{\cal N}+1)$-flavor lattice Schwinger model 
exhibits spontaneous symmetry breaking of the discrete axial symmetry. By translating of one lattice spacing $|G.S.>$ one gets the other one. 
The ${\cal N}$-flavor lattice Schwinger models excitations are also generated from 
$|G.S.>$ by two very different mechanisms, that we already described for the two-flavor model in section (3.2). There are excitations 
involving only flavor changes of the fermions without changing the charge density $\rho(x)$ which corrispond to spin flips in the $SU(2)$ invariant 
model. These excitations are massless. At variance massive excitations involve fermion transport besides flavor changes and are created 
by applying to $|G.S.>$ the latticized currents of the Schwinger models which vary the on site value of $\rho(x)$.      

Very different is the case of the massive multiflavor Schwinger models~\cite{b77}. When ${\cal N}$ is odd, the presence of a non-zero fermionic mass $m$ 
removes the degeneracy and selects one of the two $|G.S.>$ as the non degenerate ground state. When ${\cal N}$ is even the ground state 
remains translationally invariant in the strong coupling limit $e_{L}^{2}\gg m$. In the weak coupling limit $m\gg e_{L}^{2}$ the 
discrete chiral symmetry is broken for every ${\cal N}$. 
\section{Concluding remarks}

In these lectures we analysed the correspondence between the multiflavor strongly 
coupled lattice Schwinger models and the antiferromagnetic Heisenberg 
Hamiltonians to investigate the spectrum of 
the gauge models. 

Using the analysis
of the excitations of the finite size chains,
we showed the equality of the quantum numbers of the states of the Heisenberg 
model and the low-lying excitations of the two-flavor Schwinger model.
We provided also the 
spectrum of the massive excitations of the gauge model; 
in order to extract numerical values for the masses, we explicitly computed  
the pertinent spin-spin correlators of the Heisenberg chain. 
 
The massless and the massive excitations of the gauge 
model are created from the 
spin chain ground state with two very different mechanisms: 
massless excitations involve only spin flipping 
while massive excitations are created by fermion transport 
besides spin flipping and do not belong to the spin chain spectrum.
As in the continuum theory, due to the Coleman 
theorem~\cite{b39}, the massless excitations are not Goldstone bosons,
but may be regarded as the gapless quantum excitations 
of the spin-$1/2$ antiferromagnetic 
Heisenberg chain \cite{haldane}. 

In computing the chiral condensate one can show~\cite{b8,b9} that, also 
in the lattice theory, 
the expectation value of the umklapp operator is 
different from zero, while both 
$<\overline{\psi}\psi>$ and $<\overline{\psi}\sigma^{a}\psi>$ 
are zero to every order in the 
strong coupling expansion. 
This implies that both on the lattice and the continuum 
the $SU(2)$ flavor symmetry is preserved whereas the
$U_A(1)$ axial symmetry is broken. The umklapp operator~\cite{b8,b9} is 
the order parameter for this symmetry, but being 
quadri-linear in the fermi fields, is invariant, in the continuum, under 
chiral rotation of $\pi/2$ and on the lattice under the corresponding 
discrete axial symmetry (translation by one lattice site). 
This shows that the discrete axial 
symmetry is not broken in both cases.
Our lattice computation enhance this result since the ground state of the 
strongly coupled two-flavor Schwinger model is translationally invariant.

The pattern of symmetry breaking of the continuum is 
exactly reproduced even if the Coleman theorem does not 
apply on the lattice and the anomalous symmetry breaking 
is impossible due to the Nielsen-Ninomiya~\cite{nini, b28} theorem. 
At variance with the strongly coupled one-flavor lattice 
Schwinger model, the anomaly 
is not realized in the lattice theory via the spontaneous breaking of 
a residual chiral symmetry~\cite{b10}, but, rather, 
by explicit breaking of the chiral symmetry due to staggered fermions.
The non-vanishing of the VEV of the umklapp operator~\cite{b8,b9} may be regarded as
the only relic, in the strongly coupled 
lattice theory, of the anomaly of the 
continuum two-flavor Schwinger model. 
It is due to the coupling induced by the 
gauge field, between the right and left-movers on the lattice.

When the fermion mass $m$ is different from zero, some 
further difference arises between ${\cal N}$ odd and ${\cal N}$ even. 
When ${\cal N}$ is odd, the mass term induces a translational
non invariant ground state, 
generating a spontaneous chiral symmetry breaking. 
When ${\cal N}$ is even, the ground state remains translationally
invariant in the strong coupling 
limit, $i.e.$ $e^2 \gg m$. 
In the weak coupling limit, $m\gg e^2$, the discrete chiral symmetry
is spontaneously broken 
for every ${\cal N}$. For ${\cal N}=2$, the soliton-antisoliton
excitations \cite{b46} 
acquire a mass.


\begin{thebibliography}{99}
\bibitem{b1}See for instance J. B. Kogut, Rev. Mod. Phys. {\bf 55}, 775 (1983) and references therein.          
\bibitem{b2}T. Banks, S. Raby, L. Susskind, J. Kogut, D. R. T. Jones, P. N. Scharbach and D. K. Sinclair, Phys. Rev. {\bf D15}, 1111 (1977). 
\bibitem{bqcd}E. Langmann and G. W. Semenoff, Phys. Lett. {\bf B297}, 175 (1992).
\bibitem{b3} J. Smit, Nucl. Phys. {\bf B175}, 307 (1980).
\bibitem{b4a} I. Affleck and J. B. Marston, Phys. Rev. {\bf B37}, 3773 (1988); J. B. Marston, Phys. Rev. Lett. {\bf 61}, 1914 (1988). 
\bibitem{b4b} I. Affleck, Z. Zou, T. Hsu and P. W. Anderson, Phys. Rev. {\bf B38}, 745 (1988). 
\bibitem{b4c}P. Wiegmann, Prog. Theor. Phys. Suppl. {\bf 107}, 243 (1992). 
\bibitem{b4d}C. Mudry and E. Fradkin, Phys. Rev. {\bf B49}, 5200 (1994); Phys. Rev. {\bf B50}, 11409 (1994). 
\bibitem{b4e}D. H. Kim and P. A. Lee, cond-mat/9810130.
\bibitem{b5} D. Hofstaeder, Phys. Rev. {\bf B14}, 2239 (1976).
\bibitem{b6} G. W. Semenoff and L. C. Wijewardhana, Phys. Rev. {\bf D45}, 1342 (1992). 
\bibitem{semen} G. W. Semenoff, Mod. Phys. Lett, {\bf A7}, 2811 (1992); M. C. Diamantini, E. Langman, G. W. Semenoff and P. Sodano, 
Nucl. Phys. {\bf B406}, 595 (1993).
\bibitem{schw} J. Schwinger, Phys. Rev. {\bf 125}, 397 (1962); 
Phys. Rev. {\bf 128}, 2425 (1962); 
J. Lowenstein and J. A. Swieca, Ann. Phys. (N.Y.) {\bf  68}, 172 (1971).     
\bibitem{b10}F. Berruto, G. Grignani, G. W. Semenoff and P. Sodano, Phys. Rev. {\bf D57}, 5070 (1998).
\bibitem{b8}F. Berruto, G. Grignani, G. W. Semenoff and P. Sodano, hep-th/9901142.
\bibitem{b9}F. Berruto, hep-th/9902036.
\bibitem{b6b}H.Bethe, Z. Physik {\bf 71}, 205 (1931).
\bibitem{b6c}L. D. Faddeev and L. A. Takhtadzhyan, Phys. Lett. {\bf A85},  375 (1981); 
L. D. Faddeev and L. A. Takhtadzhyan, Zapiski Nauchnych Seminarov LOMI, {\bf 109}, 134 (1981), english translation in J. Sov. Math. {\bf 24}, 241 (1984).
\bibitem{b7}F. Berruto, G. Grignani, G. W. Semenoff and P. Sodano, Phys. Rev. {\bf D59}, 034504(1999).
\bibitem{b31} E. H. Lieb and D. C. Mattis, {\it Mathematical Physics in one dimension}, New York Academic Press (1961); D. C. Mattis, 
{\it The Theory of Magnetism}, Harper \& Row 1965; W. J. Caspers, {\it Spin Systems}, World Scientific 1989; I. Affleck, 
J. Phys. Cond. Mat. {\bf 1}, 3047 (1989); I. Affleck, {\it Field Theory Methods and Quantum Critical Phenomena}, in {\it Fields, Strings and 
Critical Phenomena}, ed. by E. Brezin and J. Zinn-Justin, North Holland (1989); D. C. Mattis, {\it The Many-Body Problem}, World Scientific 1993; 
V. E. Korepin, N. M. Bogoliubov and A. G. Izergin, {\it Quantum Inverse Scattering Method and Correlation Functions}, Cambridge University press 1993; 
A. Auerbach, Interacting Electrons and Quantum Magnetism, Springer Verlag 1994; A. M. Tsvelik, {\it Quantum Field Theory in Condensed Matter Physics}, 
Cambridge University press 1995; L. D. Faddeev, Int. J. Mod. Phys. {\bf A10}, 1845 (1995), hep-th/9605187; 
R. B. Laughlin, D. Giuliano, R. Caracciolo and Olivia L. White, {\it Quantum Number Fractionalization in 
Antiferromagnets}, in {\it Field Theories for Low Dimensional Condensed Matter Systems: Spin Systems and Strongly Correlated Electrons}, ed. by 
R. B. Laughlin, G. Morandi, P. Sodano, A. Tagliacozzo, V. Tognetti, Springer Verlag in press;  
A. Auerbach, F. Berruto and L. Capriotti, {\it Quantum Magnetism Approach to Strongly Correlated Electrons}, 
in {\it Field Theories for Low Dimensional Condensed Matter Systems: Spin Systems and Strongly Correlated Electrons}, ed. by 
R. B. Laughlin, G. Morandi, P. Sodano, A. Tagliacozzo, V. Tognetti, Springer Verlag in press.
\bibitem{b32} P. W. Anderson, Phys. Rev. {\bf B86}, 694 (1952); R. Kubo, Phys. Rev. {\bf 87}, 568 (1952).
\bibitem{b39}N. D. Mermin, H. Wagner, Phys. Rev. Lett. {\bf 22}, 1133 (1966); 
S. Coleman, Commun. Math. Phys. {\bf 31}, 259 (1973).
\bibitem{b34} L. Hulth\'en, Arkiv. Mat. Astron. Fysik {\bf 26A} (11), 1 (1938).
\bibitem{b35}R. B. Laughlin, unpublished (1995).
\bibitem{b38} M. Takahashi, J. Phys. {\bf C10}, 1289 (1977); cond-mat/9708087.
\bibitem{b46} S. Coleman, Ann. Phys. (N.Y.) {\bf 101}, 239 (1976).
\bibitem{b45}J. des Cloizeaux and J. J. Pearson, Phys. Rev. {\bf 128}, 2131 (1962).
\bibitem{b47} V. E. Korepin, A. G. Izergin, F. H. L. Essler and D. B. Uglov, Phys. Lett. {\bf A190}, 182 (1994).
\bibitem{b48} S. Lukyanov, cond-mat/9712314.
\bibitem{b49}I. Affleck, J. Phys. {\bf A31}, 4573 (1998).  
\bibitem{b50} S. V. Tyablikov, Ukrain. Math. Zh. {\bf 11}, 287 (1959); 
D. N. Zubarev, Soviet Physics-Uspekhi {\bf 3}, 320 (1960).
\bibitem{b51} H. Q. Lin and D. K. Campbell, J. Appl. Phys. {\bf 69}, 5947 (1991).
\bibitem{b52} B. Sutherland, Phys. Rev. {\bf B12}, 3795 (1975); P. P. Kulish and Yu. Reshetikhin, Sov. Phys. JETP {\bf 53}, 108 (1981); 
A. Doikou and R. I. Nepomechie, hep-th/9803118.
\bibitem{b53} I. Affleck, Phys. Rev. Lett. {\bf 54}, 966 (1985); 
N. Read and S. Sachdev, Phys. Rev. Lett. {\bf 62}, 1694 (1989);
N. Read and S. Sachdev, Nucl. Phys. {\bf B316}, 609 (1989); 
N. Read and S. Sachdev, Phys. Rev. {\bf B42}, 4568 (1990).
\bibitem{b54}I. Affleck and E. H. Lieb, Lett. Math. Phys. {\bf 12}, 57 (1986).
\bibitem{b55} E. Lieb, T. Schultz and D. Mattis, Ann. Phys. {\bf 16}, 407 (1961).   
\bibitem{b58}T. Banks, L. Susskind and J. Kogut, Phys. Rev. {\bf D13}, 1043 (1976); A. Carrol, J. Kogut, D. K. Sinclair and L. Susskind, Phys. Rev. 
{\bf D13}, 2270 (1976);C. J. Hamer, Z. Weihong and J. Oitmaa, Phys. Rev. {\bf D56}, 55 (1997).
\bibitem{b64b}J. P. Steinhardt, Phys. Rev. {\bf D16}, 1782 (1977).
\bibitem{b58b}R. B. Laughlin, cond-mat/9802180.
\bibitem{b65} Y. Hosotani, J. Phys. {\bf A30}, L757 (1997); hep-th/9809066.
\bibitem{b60}S. Coleman, R. Jackiw and L. Susskind, Ann. Phys. (N.Y.) {\bf 93}, 167 (1975).
\bibitem{b57}J. E. Hetrick and Y. Hosotani, Phys. Rev. {\bf D38}, 2621 (1988); J. E. Hetrick, Y. Hosotani  and S. Iso, 
Phys. Lett. {\bf B350}, 92 (1995); Y. Hosotani, R. Rodriguez, J. E. Hetrick and S. Iso, hep-th/9606129; Y. Hosotani, hep-th/9606167.
\bibitem{b23} S. L. Adler and W. A. Bardeen, Phys. Rev. {\bf 182}, 1517 (1969); 
J. S. Bell and R. Jackiw, Nuovo Cimento {\bf A60}, 47 (1969). 
\bibitem{b62}N. S. Manton, Ann. Phys. (N. Y.) {\bf 159}, 220 (1985).
\bibitem{b65} Y. Hosotani, J. Phys. {\bf A30}, L757 (1997); hep-th/9809066.
\bibitem{b67} C. Gattringer and E. Seiler, Ann. Phys. {\bf 233}, 97 (1994).
\bibitem{b68}S. Coleman, Ann. Phys. (N.Y.) {\bf 101}, 239 (1976).
\bibitem{b68bis}E. Witten, Comm. Math. Phys. {\bf 92}, 455 (1984).
\bibitem{b68tris}D. Gepner, Nucl. Phys. {\bf B252}, 481 (1985).
\bibitem{b69}K. Harada, T. Sugihara, M. Taniguchi and M. Yahiro, Phys. Rev. {\bf D49}, 4226 (1994).
\bibitem{b70}S. Eggert and I. Affleck, Phys. Rev. {\bf B46}, 10866 (1992).
\bibitem{b74} See for example E. Abdalla, M. C. B. Abdalla and K. D. Rothe, {\it Nonperturbative Methods in 2Dimensional Quantum Field Theory}, 
World Scientific 1991. 
\bibitem{b75}C. Gattringer, {\it $QED_{2}$ and the $U(1)$ Problem}, Dissertation University of Graz 1995, hep-th/9503137.
\bibitem{b76}M. Sadzikowski and P. Wegrzyn, Mod. Phys. Lett. {\bf A11}, 1947 (1996).
\bibitem{b77}J. P. Steinhardt, {\it Lattice Theory of $SU(N)$ Flavor Quantum Electrodynamics in $(1+1)$-dimensions}, Ph.D. Thesis, 
Harvard University (1978).  
\bibitem{haldane} F. D. M. Haldane, Phys. Rev. Lett. {\bf 50}, 1153 (1983).
\bibitem{nini}H. B. Nielsen and M. Ninomiya, Nucl. Phys. 
{\bf B185}, 20 (1981); {\bf B193}, 173 (1981); Phys. Lett. {\bf 105B}, 219
(1981); {\bf 130 B}, 389 (1983).
\bibitem{b28}J. B. Kogut, Rev. Mod. Phys. {\bf 51}, 659 (1979); M. Creutz, {\it  Quarks , Gluons and Lattices}, Cambridge University press, 1983; 
I. Montvay and G. M\"{u}nster, {\it Quantum Fields on a Lattice}, Cambridge University press, 1994; H. J. Rothe, {\it  Lattice Gauge Theories: An Introduction}, 
World Scientific, 1997; R. Gupta, {\it  Introduction to Lattice QCD}, Les Houches Lectures 1998, hep-lat/9807028; G. Grignani and G. W. Semenoff, 
{\it Introduction to some common topics in gauge theory and spin systems}, in {\it Field Theories for Low Dimensional Condensed Matter Systems: 
Spin Systems and Strongly Correlated Electrons}, ed. by 
R. B. Laughlin, G. Morandi, P. Sodano, A. Tagliacozzo, V. Tognetti, Springer Verlag in press.   

\end{thebibliography}
\end{document}